\documentclass[varenna]{cimento}

\usepackage{amsmath}
\usepackage{txfonts} 
\usepackage{comment}
\usepackage{pst-node}
\usepackage{listings}
\usepackage[normalem]{ulem}
\usepackage{cite}

\usepackage{enumerate}

\usepackage{lipsum}   
\usepackage{framed}
\usepackage{tikz}
\usetikzlibrary{decorations.pathmorphing,calc}
\pgfmathsetseed{1} 
\pgfdeclarelayer{background}
\pgfsetlayers{background,main}

\tikzset{
  normal border/.style={orange!30!black!10, decorate, 
     decoration={random steps, segment length=2.5cm, amplitude=.7mm}},
  torn border/.style={orange!30!black!5, decorate, 
     decoration={random steps, segment length=.5cm, amplitude=1.7mm}}}

\newenvironment{parchment}[1][Example]{%
  \vskip\baselineskip
  \MakeFramed {\FrameRestore}
  \noindent\tikz\node[inner sep=1ex, draw=black!20,fill=white, 
          anchor=west, overlay] at (0em, 2em) {\sffamily#1};\par}%
{\endMakeFramed}

\title{Models for nuclear reactions with weakly-bound systems}

\author{Antonio M. Moro}

\institute{Departamento de FAMN. Universidad de Sevilla \\
Apartado 1065 \\ 41080 Sevilla, Spain}

\usepackage{graphicx}  
\usepackage{epsfig}

\newcommand{\bi}{\begin{itemize}}
\newcommand{\ei}{\end{itemize}}
\newcommand{\be}{\begin{equation}}
\newcommand{\ee}{\end{equation}}
\newcommand{\asym}{\xrightarrow{R \gg}} 
\def\nuc#1#2{\relax\ifmmode{}^{#1}{\protect\text{#2}}\else${}^{#1}$#2\fi}
\newcommand{\sch}{Schr\"odinger }
\newcommand{\shalfzero}{^{10}{\rm Be(0}^+{\rm )}\otimes \nu 2s_{1/2}}

\newcommand{\dhalftwo}{^{10}{\rm Be(2}^+{\rm )}\otimes \nu 1d_{5/2}}

\newcommand{\bR}{{\vec R}}
\newcommand{\vecR}{{\vec R}}

\newcommand{\bRp}{{\vec R'}}
\newcommand{\br}{{\vec r}}
\newcommand{\vecr}{{\vec r}}
\newcommand{\veck}{{\vec k}}

\newcommand{\brp}{\vec{r}\mkern2mu\vphantom{r}'}
\newcommand{\veckp}{\vec{k}\mkern2mu\vphantom{k}'}

\newcommand{\bK}{{\vec K}}
\newcommand{\bs}{{\vec s}}
\newcommand{\images}{figs}

\newcommand{\miframebox}[1]{\psframebox[linecolor=blue,framearc=0.15]{{  #1 }}}

\newcommand{\mibox}[2]{%
 \vspace{4mm}
  \fcolorbox{black}[HTML]{E9F0E9}{\parbox{0.95\textwidth}{
  \textit{ #1} \\
\noindent 
\rule{0.95\textwidth}{0.5pt} \vspace{2mm}
 #2 }}
}

\begin{document}

\maketitle
\tableofcontents

\begin{abstract}
In this contribution, I present a short overview of the theory of direct nuclear reactions, with special emphasis on the case of reactions induced by weakly-bound nuclei. After introducing some general results of quantum scattering theory, I present specific applications to elastic, inelastic, transfer and breakup reactions. For each of them, I first introduce the most standard framework, followed by some alternative models or extensions suitable for the case of weakly bound nuclei.  A short discussion on semiclassical theory of Coulomb excitation and its application to breakup of halo nuclei is also provided.   
\end{abstract}

\section{Introduction}
Our present knowledge on the properties of atomic nuclei is largely based on the analysis of nuclear reactions. The very existence of the nucleus was inferred by Rutherford in 1905 from his famous $\alpha$ elastic scattering experiment and many  features and phenomena, such as the shell structure, the magic numbers, the  collective and single-particle degrees of freedom, among others, are  investigated  using nuclear reactions. Since the 1980s, thanks to the development of radioactive beams, these studies could be extended to regions of the nuclear chart beyond the stability valley.

In the proximity of the proton and neutron driplines, new exotic structures and phenomena were discovered. Prominent  examples are the popular halo and Borromean nuclei.  It was soon realized that formalisms originally designed to describe the  structure and reactions of ordinary nuclei were not well adapted to describe these new phenomena.  In particular, in the proximity of the driplines, atomic nuclei are weakly bound.  When these fragile systems collide with a stable nucleus they  break up easily due to the Coulomb and nuclear forces exerted by  a target nucleus. Consequently, reaction theories designed to describe these reactions must incorporate the effect of the strong coupling to the breakup channels.

We enumerate some fingerprints of the weak binding on reaction observables:


\begin{itemize} 
\item {\it Large interaction cross sections in nuclear collisions at  high energies.} 
Historically, the first evidence of the unusual properties of halo nuclei came from the pioneering experiments performed by Tanihata and co-workers at Berkeley using very energetic (800 MeV/nucleon)  secondary beams of radioactive species \cite{Tan85a,Tan85b}. At these high energies,  interaction cross sections are approximately proportional to the size of the colliding nuclei. It was found that some exotic isotopes of light isotopes ($^{6}$He, $^{11}$Li, $^{14}$Be) presented much higher interaction cross sections than their neighbour isotopes, which was interpreted as an abnormally large radius. 

\item {\it Narrow momentum distributions of residues following fast nucleon removal}. Momentum distributions of the residual nucleus following the removal of one or more nucleons of a energetic projectile colliding  with a target nucleus are closely related to the momentum distribution of the removed nucleon(s) in the original projectile. Kobayashi {\it et al.} \cite{Kob88} found that the momentum distributions of  $^{9}$Li following the fragmentation process $^{11}$Li+$^{12}$C $\rightarrow$ $^{9}$Li +X were abnormally narrow which, according to the Heisenberg's uncertainty principle, suggested a long tail in the density distribution of the $^{11}$Li nucleus. This result was later found in other weakly bound nuclei. 

\item {\it Abnormal elastic scattering cross sections}. Elastic scattering is affected by the coupling to non-elastic processes. In particular, when coupling to breakup channels is important, elastic scattering cross sections are depleted with respect to the case of tightly bound  nuclei. Some other key signatures are the departure of the elastic cross section with respect to the Rutherford cross section at sub-Coulomb energies and the disappearance of the Fresnel peak at near-barrier energies in reactions induced by halo nuclei on heavy targets \cite{San08,Pie12,Cub12,Pes17}. 

\item {\it Enhanced near-threshold breakup cross section in Coulomb dissociation experiments of neutron-halo nuclei}. When a neutron-halo nucleus, composed of a charged {\it core} and one or two weakly-bound neutrons ($^{11}$Be, $^{6}$He, $^{11}$Li,\ldots) collides with a high-Z target nucleus, the projectile structure is heavily distorted due to the tidal force originated from the uneven action of the Coulomb interaction on the charged core and the neutrons. This produces a stretching which may eventually break up the loosely bound projectile. This gives rise to a large population of the continuum states close to the breakup threshold. 

\end{itemize}

A proper, quantitative understanding of these and other phenomena requires the use of an appropriate reaction theory. But, before we address the features of reactions induced by weakly-bound nuclei, we will review some general concepts and results of quantum scattering theory. 

A remark on the terminology is in order here. In many cases, the word ``exotic'' is used as akin to ``unstable''. Strictly, not all ``unstable nuclei'' show exotic properties (such as weak binding, haloes, etc). Conversely, there are also stable nuclei which exhibit some ``exotic'' (abnormal) features, such as weak binding. This is the case, for instance, of the deuteron system which, albeit not exotic, behaves similarly to halo nuclei due to its relatively small binding energy.

\section{Some general scattering theory}
A  nuclear collision represents a extremely complicated many-body quantum-mechanical scattering problem, whose rigorous solution is not possible in most cases. Therefore, approximate models, usually tailored to specific types of reactions, are used. These models tend to emphasize specific degrees of freedom, those which are most likely activated during the reaction under study. For example, when low-lying collective states are present in either the projectile or target nucleus, the possibility of exciting and populating these states must be somehow (explicitly or effectively) taken into account. For weakly-bound nuclei, such as halo nuclei, the dissociation (``breakup'') of the valence nucleon(s) must be considered.

\begin{figure}
\begin{center}
\includegraphics[width=0.7\columnwidth]{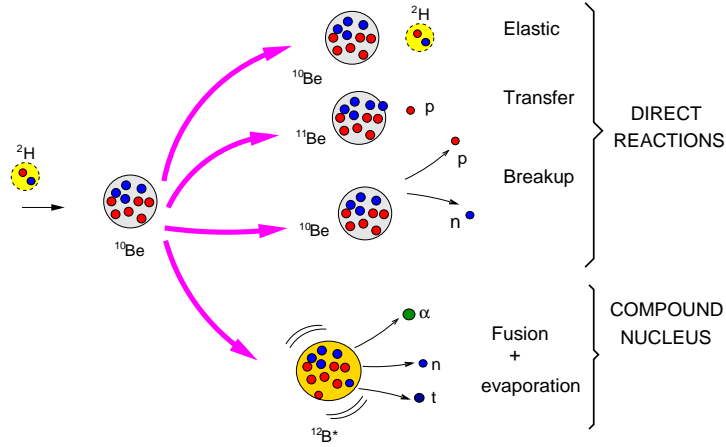}    
\caption{\label{fig:be10d_chans}Direct and compound nucleus reaction channels taking place in a d+$^{10}$Be reaction.}  
\end{center}
\end{figure}

As an example, in fig.~\ref{fig:be10d_chans} we illustrate schematically some of the channels taking place in 
a $d$+$^{10}$Be reaction. These channels can be divided into two categories, according to the characteristic collision time and degrees of freedom involved. On one side, the {\it direct reaction channels}, which are relatively fast ($t\sim 10^{-21}$~s) and peripheral processes, usually involving a few degrees of freedom and small momentum transfer. This is the case of elastic, inelastic scattering and rearrangement (transfer) processes. Angular distributions of the projectile-like fragment usually peak at forward angles. On the other side, the {\it compound nucleus} reactions, which take place over a longer time scale ($t\sim 10^{-18}-10^{-16}$~s), lead to a significant  redistribution of the initial kinetic energy among the nucleons of the collision partners and, hence, a larger number of degrees of freedom involved. The compound nucleus is usually left in a high excited state, which tends to de-excite by particle or gamma-ray emission, whose angular distributions tend to be isotropic in the center-of-mass (CM) frame. 

The very different nature of direct and compound nucleus reactions results also in very different formal treatments. The latter are treated using statistical models, first proposed by Bohr \cite{Boh36}. 
In the remainder of this  contribution, only direct reactions will be discussed.

\subsection{The concept of cross section}

\begin{figure}
\begin{center}
\begin{minipage}{0.62\textwidth}
\includegraphics[width=0.8\columnwidth]{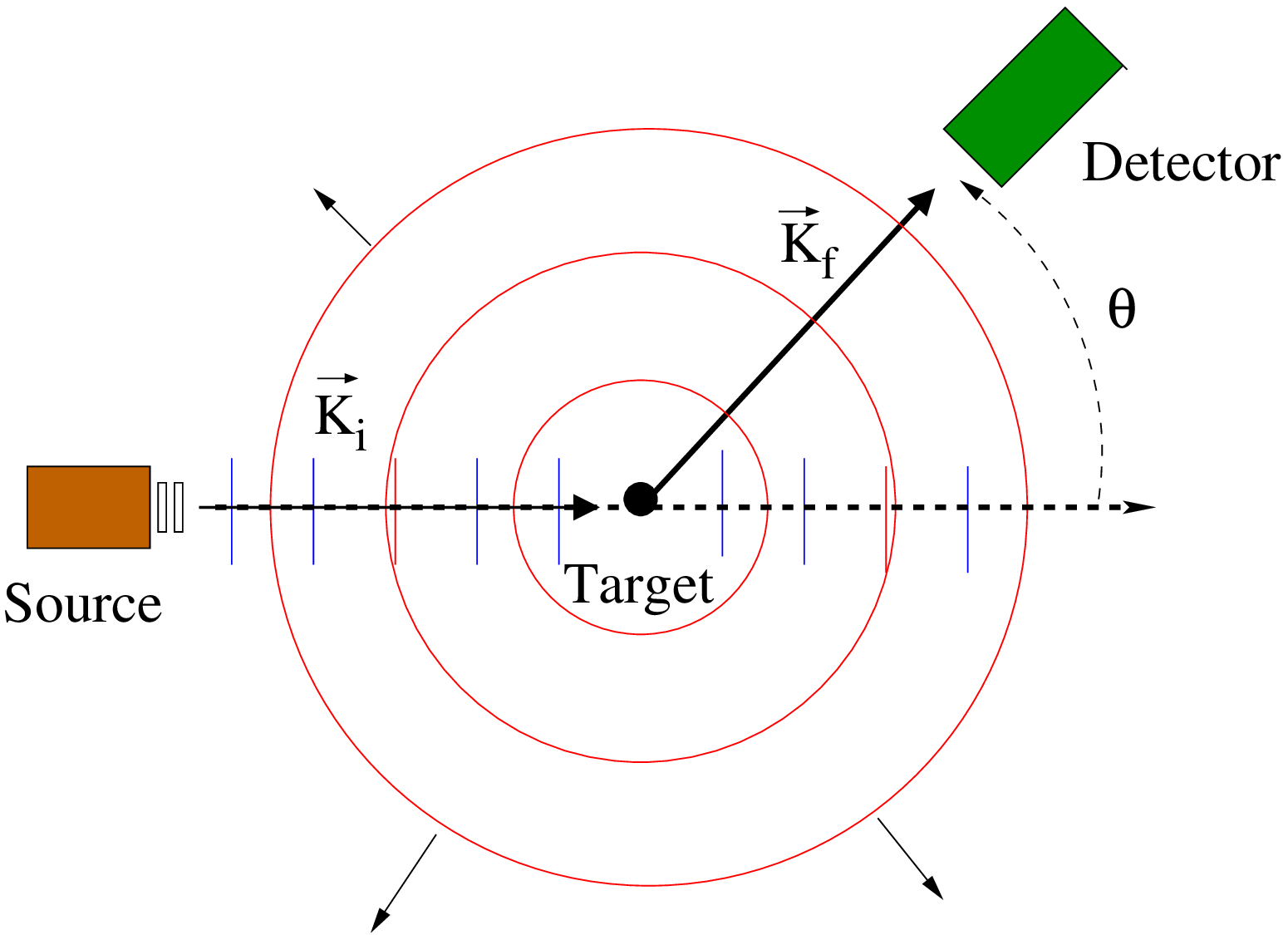}
\end{minipage}
\begin{minipage}{0.35\textwidth}
\includegraphics[width=0.65\columnwidth]{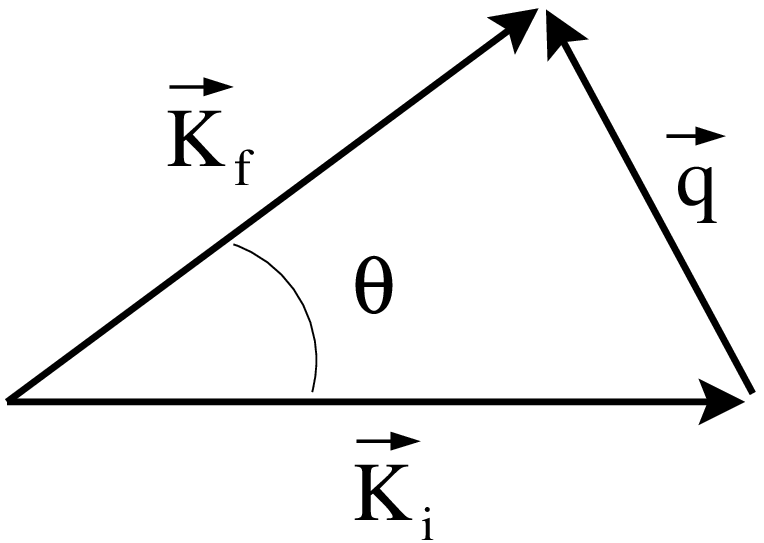}
\end{minipage}
\end{center}
\caption{\label{fig:scat}Left: Schematic representation of a scattering experiment, showing the incident and scattered waves. Right: incident and final momenta, and momentum transfer}
\end{figure}

A typical nuclear reaction experiment (see fig.~\ref{fig:scat}) measures the number of particles, integrated over a given amount of time, of one or more species resulting from a collision between two nuclei, as a function of  its scattering angle and/or its energy. This number of particles will depend on the experimental conditions, such as the beam intensity and the target thickness. To compare with the theoretical predictions, it is convenient to introduce the so-called {\it differential cross section} which is denoted  ${d \sigma}/{d\Omega}$ and is defined  as
the {\it flux of scattered particles through the area $dA=r^2 d\Omega$ in the direction $\theta$, per unit incident flux}, i.e.
\be
\frac{d\sigma}{d\Omega} =  \frac{\textrm{flux of scattered particles through $dA= r^2 d\Omega$} }{ \textrm{incident flux} } ,
\ee
and can be extracted experimentally from the number of recorded events as
\be
\label{eq:dsdw_exp}
\Delta I = I_0 ~ n_t ~  \frac{d \sigma}{d\Omega}  \Delta \Omega ,
\ee
where $I_0$ is the number of incident particles per unit time,
$\Delta \Omega$ is the  solid angle subtended by the detector, $\Delta I$ is the number of detected particles per unit time in $\Delta \Omega$, and  $n_t$ the number of target nuclei per unit surface.

Inspection of eq.~(\ref{eq:dsdw_exp}) shows that the differential cross section has units of area.  The differential cross section is {\it independent} of the experimental conditions, such as the beam intensity, the elapsed time of the measurement and the target thickness. Instead, it depends on the interaction between the projectile and target systems, which is the important quantity that a scattering experiment aims to isolate and probe. 

The final goal of the scattering theory is to develop appropriate models to which compare the measured observables, with the aim of extracting information on the structure of the colliding nuclei as well as understanding the dynamics governing these processes. The measured quantities are typically total or partial cross sections with respect to angle and/or energy of the outgoing nuclei. Therefore, the challenge of reaction theory is to obtain these cross sections by solving the dynamical equations of the system (at non-relativistic energies, the Schr\"odinger equation) with a realistic but manageable structure model of the colliding nuclei. 

\subsection{Model Hamiltonian and scattering wave function}
The mathematical treatment of a scattering problem requires the solution of the time-dependent or time-independent Schr\"odinger equation for the system%
\footnote{For a enlightening discussion on the relation between the time-dependent and time-independent approaches, see Chapter 1 of Ref.~\cite{Aus70}.}. In the second case, this equation reads
\be
\left[ H - E \right]\Psi  = 0 ~ .
\label{eq:sch-gen}
\ee
 This wave function will be  a function of the degrees of freedom (eg.\ internal coordinates) of the projectile and target, denoted generically as $\xi_p$ and $\xi_t$, as well as on the relative coordinate between them ($\bR$). Thus, we will express the total wave function as $\Psi(\bR,\xi_p,\xi_t)$. The Hamiltonian of the system is written in the form
\be
\label{eq:H}
H= \hat{T}_\bR + H_p(\xi_p) + H_t(\xi_t) + V(\bR,\xi_p,\xi_t) ,
\ee
where $\hat{T}_\bR$ is the kinetic energy operator ($\hat{T}=-\frac{\hbar^2}{2 \mu} \nabla^2$),
$H_p(\xi_p$) ($H_t(\xi_t)$) denote the projectile (target) internal Hamiltonian and $V(\bR,\xi_p,\xi_t)$  is the projectile-target interaction. After the collision, the projectile and target may exchange some nucleons, or even break up, so the Hamiltonian (\ref{eq:H}) corresponds actually to the entrance channel. To distinguish between different mass partitions that may arise in a reaction, we will use Greek letters, with $\alpha$ denoting the initial partition. So, the previous Hamiltonian is rewritten as 
\be
\label{eq:Halpha}
H= \hat{T}_\alpha + H_\alpha(\xi_\alpha) + V_\alpha(\bR_\alpha,\xi_\alpha) 
\ee
where $\xi_\alpha$ denotes the projectile and target internal coordinates in partition $\alpha$ and $H_\alpha(\xi_\alpha)\equiv H_p(\xi_p) + H_t(\xi_t)$. 
The total energy of the system is given by the sum of the kinetic energy ($E_\alpha$)  and the internal energy ($\varepsilon_\alpha$) of the projectile and target:
\be
E= E_\alpha + \varepsilon_{\alpha} = \frac{\hbar^2 {K_\alpha}^2}{2 \mu_\alpha} +\varepsilon_{\alpha} ~ ,
\ee
where $\hbar {\bK}_\alpha$ is  the linear momentum and $\varepsilon_{\alpha}$ is the sum of the projectile and target internal energies.

Equation (\ref{eq:sch-gen}) is a second-order differential equation that must be solved subject to  appropriate boundary conditions. The latter must reflect the nature of a scattering process. In our time-independent picture, the incident beam will be represented by a plane wave%
\footnote{This is only true for the case of short-range potentials; in  presence of the Coulomb potential, the incident wavefunction is represented by a Coulomb wave}
\footnote{A more realistic description would be in terms of wave-packets but the formal treatment is much more complicated. To link both pictures, one can bear in mind that a wave packet can be constructed as a superposition of plane waves.}. After the collision with the target, a set of outgoing spherical waves will be formed. The situation is schematically depicted in fig.~\ref{fig:scat}.  So, asymptotically,
\be
\Psi^\mathrm{(+)}_{\bK_\alpha}(\bR,\xi) \rightarrow \Phi_\alpha(\xi) e^{i \bK_\alpha \cdot \bR_\alpha} ~ + ~\textrm{outgoing spherical waves},
\ee
with $\Phi_\alpha(\xi) \equiv \phi^{(p)}_0(\xi_p) \phi^{(t)}_0(\xi_t)$ and 
where the superscript ``+'' indicates that this corresponds to the solution with outgoing boundary conditions (mathematically, one may construct also the solution with incoming boundary conditions).

During the collision, the incident wave will be highly distorted due to the interaction with the target nucleus but, after the collision, at  sufficiently large distances (that is, when $V$ becomes negligible), the projectile and target will emerge in any of their (kinematically allowed) eigenstates of the projectile and target nuclei. So, asymptotically, we may write%
\footnote{Note that we distinguish between $\vecR_\alpha$ and $\vecR_\beta$ since, for a rearrangement process, the coordinates will be different. We will return to this issue later on.}
%
%
%
%
\begin{align}
\label{eq:Psi-asym}
\Psi^\mathrm{(+)}_{\bK_\alpha}   \xrightarrow{R_\alpha \gg} & \Phi_\alpha(\xi_\alpha) e^{i \bK_\alpha \cdot \bR_\alpha} +  \Phi_\alpha(\xi_\alpha) f_{\alpha,\alpha}(\theta) \frac{e^{i K_\alpha R_\alpha}}{R_\alpha} 
+\sum_{\alpha' \neq \alpha} \Phi_{\alpha'}(\xi_\alpha) f_{\alpha',\alpha}(\theta) \frac{e^{i K_{\alpha'} R_\alpha}}{R_\alpha}  \\
\Psi^\mathrm{(+)}_{\bK_\alpha}   \xrightarrow{R_\beta \gg}  &  \sum_{\beta \neq \alpha} \Phi_\beta(\xi_\beta) f_{\beta,\alpha}(\theta) \frac{e^{i K_\beta R_\beta}}{R_\beta} .
\end{align}
The first line corresponds to the elastic and  inelastic channels (hence the coordinate $R_{\alpha}$), whereas the second line is for rearrangement (i.e.\ transfer) channels. 
 The function ${e^{i K_\alpha R_\alpha}}/{R_\alpha}$ represents a spherical outgoing wave. The function  multiplying this outgoing wave, $f_{\alpha,\alpha}(\theta)$,  is the {\it scattering amplitude} for elastic scattering. Its argument, $\theta$, is the CM scattering angle, and corresponds to the angle between the incident and final momenta (see fig.~\ref{fig:scat}). Likewise, the coefficients $f_{\alpha',\alpha}$
and $f_{\beta,\alpha}$ correspond to the scattering amplitudes for inelastic and transfer channels, respectively.  From the definition of flux given above, it turns out that  
(see e.g.\ Chap.~3, Sec.\ G of \cite{Glen83})
\be
\left ( \frac{d\sigma}{d\Omega} \right)_{\alpha \rightarrow \beta} = \frac{v_\beta}{v_\alpha} \left| f_{\beta,\alpha}(\theta) \right|^2 .
\ee 
where $v_\alpha$ and $v_\beta$ are initial and final asymptotic velocities. 

It is customary to define the transition matrix (T-matrix):
\be
{\cal T}_{\beta \alpha} (\theta) = -\frac{2 \pi  \hbar^2}{\mu_\beta} f_{\beta \alpha}(\theta) ~,
\ee
in terms of which
\be
\left ( \frac{d\sigma}{d\Omega} \right)_{\alpha \rightarrow \beta} = 
 \frac{\mu_\alpha \mu_\beta}{(2 \pi  \hbar^2)^2}  \frac{K_\beta}{K_\alpha}\left| {\cal T}_{\beta \alpha}(\theta) \right|^2 .
\ee

\subsection{An integral equation for $f_{\beta,\alpha}(\theta)$}
Consider that we are interested in a particular channel $\beta$. The scattering amplitude corresponding to this particular channel can be obtained from the asymptotic form of the total wavefunction, eq.~(\ref{eq:Psi-asym}), multiplying on the left by the ``internal'' wavefunction $\Phi^{*}_\beta(\xi_\beta)$ corresponding the channel of interest, and integrating over the coordinates $\xi_{\beta}$, i.e.,
\be
( \Phi_\beta | \Psi^\mathrm{(+)}_{\bK_\alpha} \rangle \asym 
\delta_{\beta,\alpha} e^{i \bK_\alpha \cdot \bR_\alpha} 
+  f_{\beta,\alpha}(\theta) \frac{e^{i K_\beta R_\beta}}{R_\beta} ,
\ee
where $(\ldots \rangle$ denotes integration over internal coordinates only. Thus, $( \Phi_\beta | \Psi^\mathrm{(+)}_{\bK_\alpha} \rangle$ remains a function of $\bR_\beta$, so we may define  $X_\beta (\bR_\beta) \equiv (\Phi_\beta | \Psi^\mathrm{(+)}_{\bK_\alpha} \rangle$. 
If we know $\Psi^\mathrm{(+)}_{\bK_\alpha}$ or an approximation to it, we can extract the scattering amplitude from the asymptotics of $X_\beta (\bR_\beta)$. Using this result, it is possible to obtain a formal expression for $f_{\beta,\alpha}(\theta)$. We start by writing the Schr\"odinger equation, using the form of the Hamiltonian appropriate for the channel $\beta$, that is,
\be
H= \hat{T}_\beta + H_\beta({\xi_\beta}) + V_\beta (\bR_\beta, \xi_\beta) .
\ee
Using this form of the Hamiltonian in the Schr\"odinger equation, eq.~(\ref{eq:sch-gen}), multiplying on the left by $\Phi^{*}_\beta(\xi_\beta)$  and integrating over the coordinates $\xi_{\beta}$ we get the {\it projected} equation:
\be
\label{eq:source}
[\hat{T}_\beta + \varepsilon_\beta -E] X_\beta (\bR_\beta) = - ( \Phi_\beta |V_\beta | \Psi^\mathrm{(+)}_{\bK_\alpha} \rangle ,
\ee
where we have used $\varepsilon_\beta =\langle \Phi_\beta(\xi_\beta) | H_\beta | \Phi_\beta(\xi_\beta) \rangle$ and the fact that the kinetic energy operator does not depend on the internal coordinates $\xi_\beta$. This is a second-order  inhomogeneous  differential equation for the function  $X_\beta$. The most general solution is the sum of the solution of the corresponding homogeneous equation, plus a particular solution of the inhomogeneous equation. The homogeneous equation is trivially solved, since it contains only the kinetic energy operator; its solution is just a plane wave with momentum $\bK_\beta$, with modulus $K_\beta=\sqrt{2 \mu_\beta (E-\varepsilon_\beta)}/\hbar$. The particular solution of the inhomogeneous equation can be formally obtained using Green's functions techniques (see, for example, \cite{Mes81,Glen83}) leading to:
%
\be
X_\beta(\bR_\beta) = e^{i \bK_\alpha \cdot \bR_\alpha} \delta_{\alpha,\beta}  
- \frac{\mu_\beta}{2 \pi \hbar^2} \int G_\beta(\bR_\beta, \bR'_\beta)
( \Phi_\beta |V_\beta  \Psi^\mathrm{(+)}_{\alpha} \rangle d\bR'_\beta \, ,
\ee 
where $G_\beta$ is the  Green's function in channel $\beta$. Explicitly:
\be
G_\beta(\bR_\beta, \bR'_\beta) = \frac{e^{i K_\beta |\bR_\beta - \bR'_\beta|}}{|\bR_\beta - \bR'_\beta|} .
\ee
To extract the scattering amplitude, we must take the asymptotic limit, $R_\beta \gg R'_\beta$. In this limit, the Green's function reduces to%
\footnote{For $R_\beta \gg R'_\beta$, $|\bR_\beta - \bR'_\beta| \approx R_\beta - \hat{R}_\beta \cdot \vec{R'}_\beta =
 R_\beta - \hat{K}_\beta \cdot \vec{R'}_\beta$   .}
\be
G_\beta(\bR_\beta, \bR'_\beta) \rightarrow \frac{e^{iK_\beta R_\beta }}{R_\beta} e^{-i \vec{K}_{\beta} \cdot \bR'_{\beta}}, 
\ee
and the function $X_\beta(\bR_\beta)$ tends to
\be
\label{eq:X}
X_\beta(\bR_\beta) \xrightarrow{R_\beta \gg} e^{i \bK_\alpha \cdot \bR_\alpha} \delta_{\alpha,\beta}   -
 \frac{\mu_\beta}{2 \pi \hbar^2}  \frac{e^{i K_\beta R_\beta}}{R_\beta}  \int e^{-i \vec{K}_{\beta} \cdot \bR'_{\beta}} ( \Phi_\beta |V_\beta  \Psi^\mathrm{(+)}_{\bK_\alpha} \rangle   d\bR'_\beta.
\ee

Comparing with the asymptotic form (\ref{eq:Psi-asym}), and recalling the definition of the scattering amplitude, we have
\begin{align}
f_{\beta,\alpha}(\theta) &= - \frac{\mu_\beta}{2 \pi \hbar^2} \langle  e^{i \bK_\beta \bR_\beta}  \Phi_\beta |V_\beta  \Psi^\mathrm{(+)}_{\bK_\alpha} \rangle  
\nonumber \\
 & = -   \frac{\mu_\beta}{2 \pi \hbar^2} \int \int e^{-i \bK_\beta \bR_\beta} 
  \Phi^{*}_\beta (\xi_\beta) V_\beta(\bR_\beta,\xi_\beta)  \Psi^\mathrm{(+)}_{\bK_\alpha}(\bR_\alpha,\xi_\alpha) \, d\xi_\beta d\bR_\beta .
\end{align}
Or, in terms of the T-matrix,
\be
\label{eq:Tpw}
{\cal T}_{\beta,\alpha} = \int \int e^{-i \bK_\beta \bR_\beta} 
  \Phi^{*}_\beta (\xi_\beta) V_\beta(\bR_\beta,\xi_\beta)  \Psi^\mathrm{(+)}_{\bK_\alpha}(\bR_\alpha,\xi_\alpha) d\xi_\beta d\bR_\beta .
\ee

\subsection{Gell-Mann--Goldberger transformation (aka two-potential formula) \label{sec:twopot} }
%
A more general expression for eq.~(\ref{eq:Tpw}) can be obtained introducing an auxiliary (and by now arbitrary) potential $U_\beta(\bR_\beta)$ on both sides of eq.~(\ref{eq:source}),
\be
\label{eq:sourceU}
[\hat{T}_\beta + U_\beta + \varepsilon_\beta -E] X_\beta(\bR_\beta) = - ( \Phi_\beta |V_\beta -U_\beta | \Psi^\mathrm{(+)}_{\bK_\alpha} \rangle.
\ee

The solution of (\ref{eq:sourceU}) is given by a general solution of the homogeneous equation, plus a particular solution of the full equation. 
The homogeneous equation is given by
\be
[\hat{T}_\beta + U_\beta + \varepsilon_\beta -E] \chi^{(+)}_\beta(\bR_\beta) =0 .
\ee
This equation represents the scattering of the particles in channel $\beta$ under the potential $U_\beta$. The solution is of the form
\be
\chi^{(+)}_\beta(\bR_\beta) = e^{i \bK_\beta \cdot \bR_\beta} + \textrm{outgoing spherical waves} \, .
\ee
In the next section, we shall discuss in more detail how this equation is solved in practical situations, making use of the partial wave expansion.

Finally, the full equation (\ref{eq:sourceU}) is solved adding a particular solution of the inhomogeneous equation. This is done using again Green's functions techniques. Details are given in \cite{Aus70}. The full solution (which generalizes eq.~(\ref{eq:X})) is given by
\be
X_\beta(\bR_\beta) \equiv ( \Phi_\beta | \Psi^\mathrm{(+)}_{\bK_\alpha} \rangle = \chi^{(+)}_\beta(\bR_\beta) \delta_{\alpha \beta} +
       \int G_\beta^{(+)}(\bR_\beta,\bR'_\beta)  ( \Phi_\beta |V_\beta -U_\beta | \Psi^\mathrm{(+)}_{\bK_\alpha} \rangle d \bR'_\beta .
\ee

The scattering amplitude (or the T-matrix) is extracted from the asymptotics of the outgoing waves. But note that we have now outgoing waves in both terms of the RHS of the previous equation giving rise also to two contributions to the scattering amplitude,
\be
\label{eq:Tdist}
\miframebox{
{\cal T}_{\beta,\alpha} = {\cal T}^{(0)}_{\beta,\alpha} \delta_{\alpha \beta} +
\int \int \chi^{(-)*}_{\beta}(\bK_\beta, \bR_\beta) 
  \Phi^{*}_\beta (\xi_\beta) [V_\beta-U_\beta]  \Psi^\mathrm{(+)}_{\bK_\alpha}  d\xi_\beta d\bR_\beta ,
}
\ee
The first term is the scattering amplitude due to the potential $U_\beta$ and is present only for $\beta=\alpha$ (i.e.\ elastic scattering). The function  $\chi_{\beta}^{(-)}$ is the time-reversed of $\chi^{(+)}$ and corresponds to the solution of a homogeneous equation consisting on a plane wave with momentum $\bK_\beta$ and ingoing spherical waves. It can be readily obtained from  $\chi^{(+)}$ using the relationship $\chi^{(-)*}(\bK,\bR)$=  $\chi^{(+)}(-\bK,\bR)$. 

The result (\ref{eq:Tdist}) is known as the {\it Gell-Mann--Goldberger transformation} or {\it two-potential formula}. This expression is exact but it cannot be solved as such, since it contains the exact wave function of the system. However, it provides a very useful starting point to derive approximate expressions, as we will see later on.

\section{Defining the modelspace \label{sec:pq} }
%
We have seen that the dynamics of the system in a scattering process is encoded in the full wave function, $\Psi^{(+)}$. Formally, it can be obtained by solving the Schr\"odinger equation of the system. In our time-independent approach, this wavefunction consists asymptotically on an incident plane, and outgoing spherical waves in all possible channels. Practical calculations require as a first step reducing the full space to a tractable modelspace. This is motivated by two facts: (i) the channels of interest to analyze a particular experiment and (ii) the numerical/computational complexity of the problem. For example, if we are interested in analyzing some inelastic scattering experiment, our model space might  consist on the ground state of the projectile and target, plus the excited states more strongly populated in the experiment. 

A formal procedure to reduce the problem from the full space to a selected modelspace was developed by Feshbach \cite{Fes58,Fes62}.  The idea is to separate the full space into two parts, denoted respectively as P and Q. The P space comprises the channels of interest and will therefore be taken into account  explicitly in the model wave function $\Psi^{(+)}$. The Q space is composed by the remaining channels. So, following Feshbach (see also \cite{Aus70} and \cite{Glen83}, Chapter 8G), we may write $\Psi^{(+)}=\Psi_P + \Psi_Q$. The components $\Psi_P$ and  $\Psi_Q$ obey a  system coupled equations, with the  deceptively simple form
\begin{align}
(E-H_{PP}) \Psi_P & = H_{PQ} \Psi_Q \\
(E-H_{QQ}) \Psi_Q & = H_{QP} \Psi_P 
\end{align}
where $H_{PP}=P H P$, $H_{PQ}=P H Q$, and so on. The projected Hamiltonian $H_{PP}$  contains the coupling among the states of the P space, and likewise for $H_{QQ}$. The terms $H_{PQ}$ and $H_{QP}$ describe couplings between the states of P and those of Q. Since we are interested only in $\Psi_P$, we eliminate $\Psi_Q$ from the RHS of the first equation, using the second equation:%
\footnote{The $i\epsilon$ guarantees the outgoing boundary condition. The limit $\epsilon \rightarrow 0$ is understood in these expressions.}
\be
\left [ E - H_{PP} - H_{PQ} \frac{1}{E- H_{QQ} + i \epsilon} H_{QP} \right ] \Psi_P =0 
\ee
This equation can be also written as 
\be
\left [ E - H_\alpha - T_\alpha - {\cal V} \right ] \Psi_P =0 ,
\label{eq:HVP}
\ee
with 
\be
\label{eq:veff}
{ \cal V } = V_{PP} + V_{PQ} \frac{1}{E- H_{QQ} + i \epsilon} V_{QP} .
\ee

The first term of the RHS ($V_{PP}$) is the potential operator acting only among the states of the P space, and the second term describes the coupling with the omitted (Q) states. This term is found to be complex, energy-dependent and non-local. Because of the presence of the Q operator, it involves the coupling to all the possible channels and so it cannot be exactly evaluated in practice. 
 Yet, this formal solution provides an useful  guidance on how to replace such a complicated object by a more manageable one. Direct reaction theories replace (\ref{eq:HVP}) by an approximated one of the form
\be
\label{eq:Sch_heff}
(E-H_\mathrm{eff}) \Psi_\mathrm{model} = 0,
\ee
where $H_\mathrm{eff}$ is an effective Hamiltonian which contains an approximation of the complicated object ${\cal V}$, usually involving some phenomenological forms. 

\section{Single-channel scattering: the optical model \label{chap:elas}}


The simplest approximation to the P space is to reduce the physical space to just the ground states of the projectile and target. This gives rise to the optical model formalism. In this case, the effective Hamiltonian acquires the form
\be
H_\mathrm{eff} = H_\alpha + U_\alpha(\bR) ,
\ee
where $U_\alpha(\bR)$ is meant to represent the effective potential (\ref{eq:veff}) when P is reduced to the projectile and target ground states. Note that this potential does not contain any explicit dependence on the internal degrees of freedom $\xi$. Thanks to that, the total wave function of the system can be written in the factorized form%
\footnote{The subscript $\alpha$ is omitted here when implicitly understood.}
\be
\label{eq:Psi_1chan}
\Psi^{(+)}_\textrm{model}(\xi,\bR) = \Phi_{0}(\xi) \chi^{(+)}_{0}(\bR).
\ee
Using the fact that, by construction, $H_\alpha \Phi_{0}(\xi)= \varepsilon_0 \Phi_{0}(\xi)$, replacement of the previous equation on the \sch (\ref{eq:Sch_heff}) gives 
\be
[T_\alpha +  U_\alpha(\bR) -E_0] \chi^{(+)}_{0} (\bR) =0  ,
\ee
where $E_0 = E - \varepsilon_0$, i.e., the kinetic energy associated with the relative motion between the projectile and target.

If the effective Hamiltonian, $H_\mathrm{eff}$, is to represent the complicated Feshbach operator, describing not only the interaction in the P space, but also the couplings between the P and Q spaces (all non-elastic channels in this case), then the effective interaction  $U_\alpha(\bR)$ will be complex, non-local and energy-dependent. The imaginary part accounts for the flux leaving the elastic channel (P space) to the channels not explicitly included (the Q space). The energy dependence is usually  taken into account phenomenologically, by parametrizing $U$ with some suitable form and adjusting the parameters to the experimental data over some energy region. Finally, non-locality is rarely taken into account, or it is simply taken into account approximately, by including its effect in the effective local potential \cite{Per62}. Recently, however, this topic  has received  renewed attention \cite{Tit14,Tia15,Lov17}. The effective interaction $U_\alpha$ is referred to as {\it optical potential}.

\subsection{Partial wave expansion}
As an additional simplification, we consider the case in which the spins of the colliding particles are ignored and the optical potential is assumed to be a function only of the projectile-target separation, $R=|\bR|$.  In this case, the wave function can be expanded in spherical harmonics as,
\be
\label{eq:part-wave}
\chi^{(+)}_0 (\bK,\bR) =  \frac{1}{K R} \sum_{\ell } i^\ell (2 \ell +1) \chi_{\ell}(K,R) P_{\ell}(\cos \theta) ,
\ee
where the constant factors are introduced for convenience. The radial functions $\chi_{\ell}(K,R)$ are a solution of 
\begin{equation}
\left[- \frac{\hbar^{2}}{2\mu}\frac{d^{2}}{dR^{2}} + \frac{\hbar^{2}}{2\mu}\frac{\ell(\ell+1)}{R^{2}}+U(R)-E_0 \right]\chi_{\ell}(K,R)=0.
\label{Eq:radScheq}
\end{equation}
%
%
In the case of $U_\alpha=0$, the solution $\chi^{(+)}_0 (\bK,\bR)$ must reduce to a plane wave, whose partial wave expansion is known
\begin{align}
\label{eq:pw}
e^{i \bK \cdot \bR} & =  \frac{4 \pi}{K R} \sum_{\ell,m} i^\ell F_\ell(K R) Y_{\ell m}(\hat{R}) Y^{*}_{\ell m}(\hat{K})
=  \frac{1}{K R} \sum_{\ell} i^\ell (2 \ell +1) F_\ell(K R) P_{\ell}(\cos \theta) ,
\end{align}
where  $F_\ell(KR)=(KR) j_{\ell}(K,R)$ with  $j_{\ell}(K,R)$ a spherical Bessel function. Comparing this expression with (\ref{eq:part-wave}), we see that, in the $U_\alpha=0$ case, $\chi_{\ell}(K,R) \rightarrow F_\ell(KR)$. 


For non-zero potential, we can still say that $\chi^{(+)}_0 (\bK,\bR)$  must verify the following equation at large distances,
\begin{equation}
\left[- \frac{\hbar^{2}}{2\mu}\frac{d^{2}}{dR^{2}} + \frac{\hbar^{2}}{2\mu}\frac{\ell(\ell+1)}{R^{2}}-E_0 \right]\chi_{\ell}(K,R)=0 \quad \textrm{(for large R)} ,
\label{Eq:rad-asym}
\end{equation}
whose most general solution is a combination of two independent solutions for this equation. One of them can be taken as the regular solution $F_\ell(K R)$. The other can be the irregular solution,
\be
G_\ell(K R)= - (K R) n_\ell (K R) 
\ee
or any combination of $G$ and $F$, that is,
\be
\label{eq:XasymFG}
\chi_{\ell}(K,R) \asym A F_\ell(K R) + B G_\ell (K R) .
\ee
%

The  combination appropriate for our purposes is suggested by the known asymptotic behavior of our physical scattering wavefunction, i.e.
\be
\label{eq:om-asym}
\chi^{(+)}_0 (\bK,\bR)  \asym  e^{i \bK \cdot \bR} +  f(\theta) \frac{e^{i K R}}{R} .
\ee
The exponential part of the outgoing wave, $e^{i K R}$, turns out to be just a definite combination of the $F$ and $G$ functions, because
\begin{align}
G_\ell(\rho)  + i F_\ell (\rho) \equiv  H^{(+)}_{\ell}(\rho) \rightarrow e^{i (\rho - \ell \pi/2)}   .
\end{align}

So, returning to the partial wave expansion, the appropriate boundary condition consistent with the behavior (\ref{eq:om-asym}) is given by
\be
\label{eq:asymT}
\chi_{\ell}(K,R) \rightarrow   F_\ell(K R) +  T_\ell H^{(+)}_\ell (K R) ,
\ee
where the  coefficients $T_\ell$ are to be determined by numerical integration of the differential equation.   It is usual to write $T_\ell$ in terms of the so-called {\it phase-shifts},
\be
T_\ell = e^{i \delta_\ell} \sin (\delta_\ell) 
\ee
or, in terms of the {\it reflection coefficient}, $S_\ell$, or {\it S-matrix},%
\footnote{When these expressions are generalized to the multiple channel case, the quantity $S_\ell$ becomes a matrix and is referred to as {\it scattering or collision matrix} (the name is also used in single-channel case, but the terminology is less obvious).} 
\be
S_\ell  = 1 + 2 i T_\ell = e^{2 i \delta_\ell} .
\ee

The condition (\ref{eq:asymT}) can be also written as,
\be
\label{eq:asymS}
\chi_{\ell}(K,R) \rightarrow  \frac{i}{2} \left [ H^{(-)}_\ell(K R) -  S_\ell H^{(+)}_\ell (K R) \right ] \, ,
\ee
where 
\be
H^{(-)}_{\ell}(\rho)  =  G_\ell(\rho)  - i F_\ell (\rho)   \rightarrow e^{-i (\rho -\ell \pi/2)}  .
\ee

The S-matrix $S_\ell$ is therefore the coefficient of the outgoing wave ($H^{(+)}_\ell$) for the partial wave $\ell$.  It reflects the effect of the potential on this particular wave in the sense that,
\begin{itemize}
\item If no potential is present, there is no outgoing wave. Then, $T_\ell=0$ or, equivalently, $S_\ell=1$ and $\delta_\ell=0$. 
\item As a consequence of the previous result, for large values of $\ell$ the centrifugal barrier keeps the projectile well apart from the target, and thus the effect of the (short-ranged) potential $U_\alpha$ will be negligible. Consequently, for $\ell \rightarrow \infty$ $\Rightarrow$ $S_\ell \rightarrow 1$.   
\item If the scattering potential is real, the overall outgoing flux for a given partial wave must be conserved, and hence $|S_\ell|=1$. 
\item On the other hand, for a complex potential (with negative imaginary part), we have $|S_\ell| < 1$, which reflects the fact that part of the incident flux has left the elastic channel in favor of other channels.
\end{itemize}

In the accompanying  shaded box, we give some basic guidelines on how the wave functions and  phase-shifts are actually computed (single-channel case). 

\begin{parchment}[Numerical calculation of the scattering wave function and phase-shifts]
For a single-channel case, the wave functions and phase-shifts can be computed as follows:
\begin{enumerate}
\item Integrate the radial differential equation from the origin outwards, with the initial value $\chi_\ell (K,0)=0$ and some finite (arbitrary) slope.
\item At a sufficiently large distance, $R_\mathrm{max}$, beyond which the nuclear potentials have become negligible, the numerically obtained solution is matched to the asymptotic form 
$$
N \chi_{\ell}(K,R_\mathrm{max}) \rightarrow   F_\ell(\eta,K R_\mathrm{max}) +  T_\ell H^{(+)}_\ell (\eta,K R_\mathrm{max})
$$
\item This equation contains two unknowns, $T_\ell$ and the normalization $N$. Thus, it is supplemented with the condition of continuity of the derivative
$$
N \chi'_{\ell}(K,R_\mathrm{max}) \rightarrow   F'_\ell(\eta,K R_\mathrm{max}) +  T_\ell (H^{(+)}_\ell (\eta,K R_\mathrm{max}))'
$$
From these two conditions, one obtains the $T$-matrix (or, equivalently, the S-matrix) and the phase-shifts. 

\item The procedure is repeated for each $\ell$, from $\ell=0$ to $\ell_\mathrm{max}$, such that  $S_{\ell_\mathrm{max}} \approx 1$.
\end{enumerate}
\end{parchment}

\subsection{Scattering amplitude}
To get the scattering amplitude, we substitute the asymptotic radial function $\chi_{\ell}(K,R)$ from (\ref{eq:asymS}) into the full expansion (\ref{eq:part-wave}):
\begin{align}
\chi^{(+)}_0 (\bK,\bR) &  \rightarrow  \frac{1}{K R} \sum_{\ell} i^\ell (2 \ell +1)
    \left \{  F_\ell(K R) +  T_\ell H^{(+)}_\ell (K R)  \right \}
    P_{\ell}(\cos \theta)  
\nonumber \\
                         & =  \frac{1}{K R} \sum_{\ell} i^\ell  (2 \ell +1) F_\ell(K R) P_{\ell}(\cos \theta) + 
 \frac{1}{K} \sum_{\ell } i^\ell (2 \ell +1)   T_\ell    \frac{e^{i (K R-\ell \pi/2)}}{R}   P_{\ell}(\cos \theta) 
\nonumber \\
                         & =  e^{i \bK \cdot \bR}  + 
 \frac{1}{K} \sum_{\ell }   (2 \ell +1)  e^{i \delta_\ell} \sin{\delta_\ell}     P_{\ell}(\cos \theta) \frac{e^{i K R}}{R} 
\end{align}
The elastic scattering amplitude is the coefficient of  $e^{i K R}/{R}$ in the last line, i.e.,
\be 
\label{eq:fam-om}
f(\theta)   = \frac{1}{K} \sum_{\ell} (2 \ell +1)  e^{i \delta_\ell} \sin{\delta_\ell}  P_{\ell}(\cos \theta) 
           = \frac{1}{2 i K} \sum_{\ell}(2 \ell +1) (S_\ell -1 ) P_{\ell}(\cos \theta) .
\ee 
%

The differential elastic cross section will be given by 
\be
\frac{d\sigma}{d\Omega} = | f(\theta) | ^ 2  .
\ee 

In principle, the  sum in (\ref{eq:fam-om}) runs from $\ell=0$ to infinity. However, remember that, for large values of $\ell$, the S-matrix tends to 1 so, in practice, the sum can be safely truncated at a maximum value $\ell_\mathrm{max}$, determined by some convergence criterion of the cross section.

\subsection{Coulomb  case}
The Coulomb case deserves a special consideration because
the expressions derived in the previous section are strictly applicable to the case of short-range potentials, for which the asymptotic form (\ref{eq:om-asym}) is appropriate. For a pure Coulomb case, we can perform a partial wave expansion of the scattering wavefunction $\chi_C(\bK,\bR)$ of the form
\be
\chi_C(\bK,\bR) = \frac{1}{K R} \sum_{\ell} (2 \ell +1)  i^\ell \chi^C_\ell(K R)  P_{\ell}(\cos (\theta)) \, ,
\ee
with the radial functions $\chi^C_\ell(K R)$ obeying the equation
\be
\label{eq:radC}
\left [ \frac{d^2}{dR^2} + K^2 - \frac{2 \eta K}{R} + \frac{\ell (\ell +1)}{R^2} \right ]  \chi^C_\ell(K R) = 0 \, ,
\ee
where
\be
\label{eq:eta}
\eta= \frac{Z_p Z_t e^2}{\hbar v } = \frac{Z_p Z_t e^2 \mu}{\hbar^2 K} 
\ee
is the so-called Coulomb or {\it Sommerfeld parameter}.

The solution of (\ref{eq:radC}) must be regular at the origin. Asymptotically, it behaves as
\be
\chi^C_\ell(K R) \asym e^{i \sigma_\ell }  F_\ell(\eta,KR)
\ee
where $F_\ell(\eta,KR)$ is the regular Coulomb function and $\sigma_\ell$ is the Coulomb phase-shift for a partial wave $\ell$,
\be
\sigma_\ell= \arg \Gamma(\ell + 1 + i \eta) .
\ee 
The Coulomb function behaves asymptotically as \cite{Abr72}
\be
F_\ell(\eta,\rho) \rightarrow \sin (\rho  - \eta \ln (2 \rho) - \ell \pi/2 + \sigma_\ell) ,
\ee
which in the case $\eta=0$ ($\sigma_\ell=0$) reduces to the  regular   function $F_\ell(KR)$ introduced in the case of short-range potentials
\be
F_\ell(\eta=0,\rho) =  F_\ell(\rho) = \rho j_\ell (\rho)   .
\ee
Analogously, an irregular solution of (\ref{eq:radC}) can be found, which reduces to $G_\ell(\rho)$ in the no Coulomb case
\be
G_\ell(\eta,\rho) \rightarrow \cos (\rho  - \eta \ln (2 \rho) - \ell \pi/2 + \sigma_\ell) \xrightarrow{\eta=0}  G_\ell(\rho) = -\rho n_\ell (\rho) ,
\ee
as well as the ingoing and outgoing functions,
\begin{align}
H^{(+)}_\ell(\eta,\rho) & = G_\ell(\eta,\rho)  + i F_\ell(\eta,\rho) \\
H^{(-)}_\ell(\eta,\rho) & = G_\ell(\eta,\rho)  - i F_\ell(\eta,\rho) \, .
\end{align}

For the pure Coulomb case, the scattering amplitude will be given by
\be
f_C(\theta) = \frac{1}{2 K} \sum_{\ell}  (2 \ell +1) (e^{2 i \sigma_\ell} -1 ) P_{\ell}(\cos \theta) \, .
\ee
This integral is not convergent (cannot be truncated at a finite $\ell$) but the full result is known analytically and is given by 
\be
f_C(\theta) = - \frac{\eta}{2 K \sin^2 (\frac{1}{2}\theta) } e^{-i \eta \ln(\sin^2(\frac{1}{2}\theta)  + 2 i \sigma_0) } .
\ee
The differential cross section yields  the well-known Rutherford formula
\be
\frac{d\sigma_R}{d \Omega} = |f_C(\theta)|^2 = \frac{\eta^2}{4 K^2 \sin^4(\frac{1}{2}\theta) } = 
\left ( \frac{Z_p Z_t e^2}{4 E}  \right )^2  \frac{1}{\sin^4(\frac{1}{2}\theta)} \, .
\ee

\subsection{Coulomb plus nuclear case}
If both Coulomb and nuclear potentials are present, the scattering function $\chi^{(+)}_{0}(\bK,\bR)$  will never reach the asymptotic form of a plane wave plus outgoing waves, due to the presence of the $1/R$ term in  the Schr\"odinger equation. Nevertheless, it can be written as
\be
\label{eq:cn}
\chi_{0}^{(+)}(\bK,\bR) \rightarrow \chi^{(+)}_C (\bK,\bR) + \textrm{outgoing spherical waves} \, ,
\ee
where the {\it outgoing waves} are now proportional to the functions $H^{(+)}_\ell (\eta, KR)$. Of course, when only the Coulomb potential is present, this term vanishes, and the scattering wave function reduces to $\chi^{(+)}_C (\bK,\bR)$. 

If we write, as usual, the $\chi_{0}^{(+)}(\bK,\bR)$ function as a partial wave expansion, the corresponding  radial coefficients $\chi_\ell(K,R)$ verify the asymptotic condition
\begin{align}
\chi_\ell (K,R) & \rightarrow e^{i \sigma_\ell} \left [ F_\ell (\eta,KR) + T_\ell H^{(+)}_\ell (\eta, KR)  \right ]  \nonumber \\
                & = e^{i \sigma_\ell}\frac{i}{2} \left [  H^{(-)}_\ell  (\eta,KR) -S_\ell H^{(+)}_\ell (\eta, KR)  \right ] ,
\end{align}
which is very similar to (\ref{eq:asymT}) and (\ref{eq:asymS}), except for the additional Coulomb phase $e^{i \sigma_\ell}$ and the replacement of the functions $F(KR)$, $H^{(+)}$, etc by their Coulomb generalizations. 

The scattering amplitude results
\be
f(\theta) = f_C (\theta) + \frac{1}{2 i K} \sum_{\ell} (2 \ell +1) e^{2 i \sigma_\ell} (S_\ell -1) P_\ell(\cos \theta) ,
\ee
where the first term corresponds to the pure-Coulomb amplitude, which arises from the outgoing waves in the first term of (\ref{eq:cn}) and the second term is the so-called  Coulomb modified nuclear amplitude.

\subsection{Parametrization of the phenomenological optical potential}
The effective optical optical potential is usually taken as the sum of Coulomb and nuclear central potentials $U(R)= U_{C}(R) +  U_N(R)$, with the Coulomb part taken as the potential corresponding to a uniform distribution of charge of radius $R_c$:
\be
U_C(R)=\left\{ \begin{array}{ll}
\frac{Z_p Z_t e^2}{2 R_c} \left( 3- \frac{R^2}{R_c^2}\right) & \textrm{if $R \leq R_c$} \\
\frac{Z_p Z_t e^2}{R}  & \textrm{if $R \geq R_c$} 
\end{array} \right .
\ee

As for the nuclear part, it contains in general real and imaginary parts. The most popular parametrization is the so-called Woods-Saxon form,
\be
\label{eq:WS}
U_N(R)=V(R) + i W(R) = -\frac{V_0}{1+\exp\left(\frac{R-R_0}{a_0}\right)}- i~\frac{W_0}{1+\exp\left(\frac{R-R_i}{a_i}\right)} .
\ee

The parameters $V_0$, $R_0$ and $a_0$ are the depth, radius and diffuseness (likewise for the imaginary part). They are usually determined from the analysis of elastic scattering data so, strictly, this potential might account also for $\lambda>0$ Coulomb multipoles. The radius $R_x$ is sometimes parametrized using the projectile ($A_p$) and target ($A_t$) mass numbers introducing the so-called reduced radius, $R_x = r_x (A_p^{1/3} +A_t^{1/3})$. For ordinary nuclei $r_x \approx 1.1-1.3$~fm.  

If the spin of the projectile (or target) is considered, the potential will contain also spin-dependent terms. The most common one is the spin-orbit term, which is usually parametrized as
\be
U_{so}(R) = (V_{so} + i W_{so}) \left ( \frac{\hbar}{m_\pi c } \right )^2 \frac{1}{R}\frac{df(R,R_{so},a_{so})}{dR} (2  {\bf \ell} \cdot {\bf s}) ,
\ee
where the radial function $f(R,R_{so},a_{so})$ is again a Woods-Saxon form, and $ \left ( {\hbar}/{m_\pi c } \right )^2 = 2~\mathrm{fm}^2$, is just introduced in order $U_{so}$ has dimensions of energy.

\subsection{Microscopic optical potentials}
\begin{figure}[bt]
\begin{center}
\includegraphics[width=0.35\columnwidth]{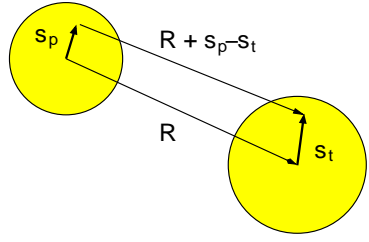}
\end{center}
\caption{\label{fig:dfold}Relevant coordinates for nucleus-nucleus  folding model calculations.}
\end{figure}

The optical potential or, at least, part of it, can be also calculated microscopically, starting from some effective nucleon-nucleon (NN) interaction. For example, the bare potential can be computed microscopically, by means of a folding procedure in which an effective nucleon-nucleon interaction  (JLM, M3Y, etc) is convoluted with the projectile and target densities (see fig.~\ref{fig:dfold}). 
\be
V_\mathrm{fold}(\bR) = \int \rho_p(s_p) \rho_t(s_t) v_{NN}(|\bR +\bs_p- \bs_t|) d\bs_p d\bs_t  \, ,
\ee
where  $\rho_p(s_p)$ and $\rho_t(s_t)$ are the projectile and target densities, respectively. 
Since the latter are g.s.\ densities, $V_\mathrm{fold}(\bR)$ accounts only for the bare potential $V_{PP}$ (P-space part) and ignores the effect of non-elastic channels.  The remaining part of the effective potential (second term in eq.~(\ref{eq:veff})) must be supplied, using some phenomenological or microscopic prescription.

\section{Elastic scattering phenomenology}

\subsection{Elastic scattering in presence of strong absorption}
When the projectile and target nuclei are composite systems, the scattering is largely dominated by the absorptive part of the nucleus-nucleus potential. This means that the effects of the coupling to nonelastic channels is dominant. Absorption introduces quantal effects, such as diffraction, which are analogous to those observed in optical phenomena. Heavy-ion collisions are also characterized by large angular momenta and small de Broglie wavelengths associated to the relative motion, in comparison with the dimensions of the nuclei. In this situation, the projectile-target motion can be interpreted in terms of classical trajectories, which is a useful concept to assist our intuition. 

Resorting to this idea of classical trajectories, an important concept in strong-absorption scattering is that of grazing collision, which refers to those trajectories for which the colliding nuclei begin to experience the strong nuclear interaction. Associated to this, one may introduce also the concepts of grazing angle, $\theta_{gr}$, (the scattering angle for a grazing collision) and grazing angular momentum $\ell_\text{gr}$.  Trajectories with angular momentum $\ell < \ell_\text{gr}$  will be strongly absorbed and the corresponding elastic scattering cross section will be largely suppressed.   

The specific features of a given reaction in presence of strong absorption are largely determined by the  grazing angular momentum $\ell_\text{gr}$ and the Sommerfeld parameter ($\eta$). 
 In particular, for large values of $\ell_\text{gr}$, where many partial waves are involved, one observes characteristic diffraction patterns analogous to the Fresnel and Fraunhofer patterns encountered in optics.

Recalling the definition of the Sommerfeld parameter, it can be regarded as a measurement of the energy of the system relative to the Coulomb barrier. The latter is defined as the top of the real (nuclear + Coulomb) potential. This is exemplified in fig.~\ref{fig:veff} for the $^4$He+$^{58}$Ni system, whose Coulomb barrier is about 10 MeV. 

Depending on the values of $\eta$ and $\ell_\text{gr}$, we may distinguish three distinct regimes:

\begin{figure}[bt]
\begin{center}
\includegraphics[width=0.6\columnwidth]{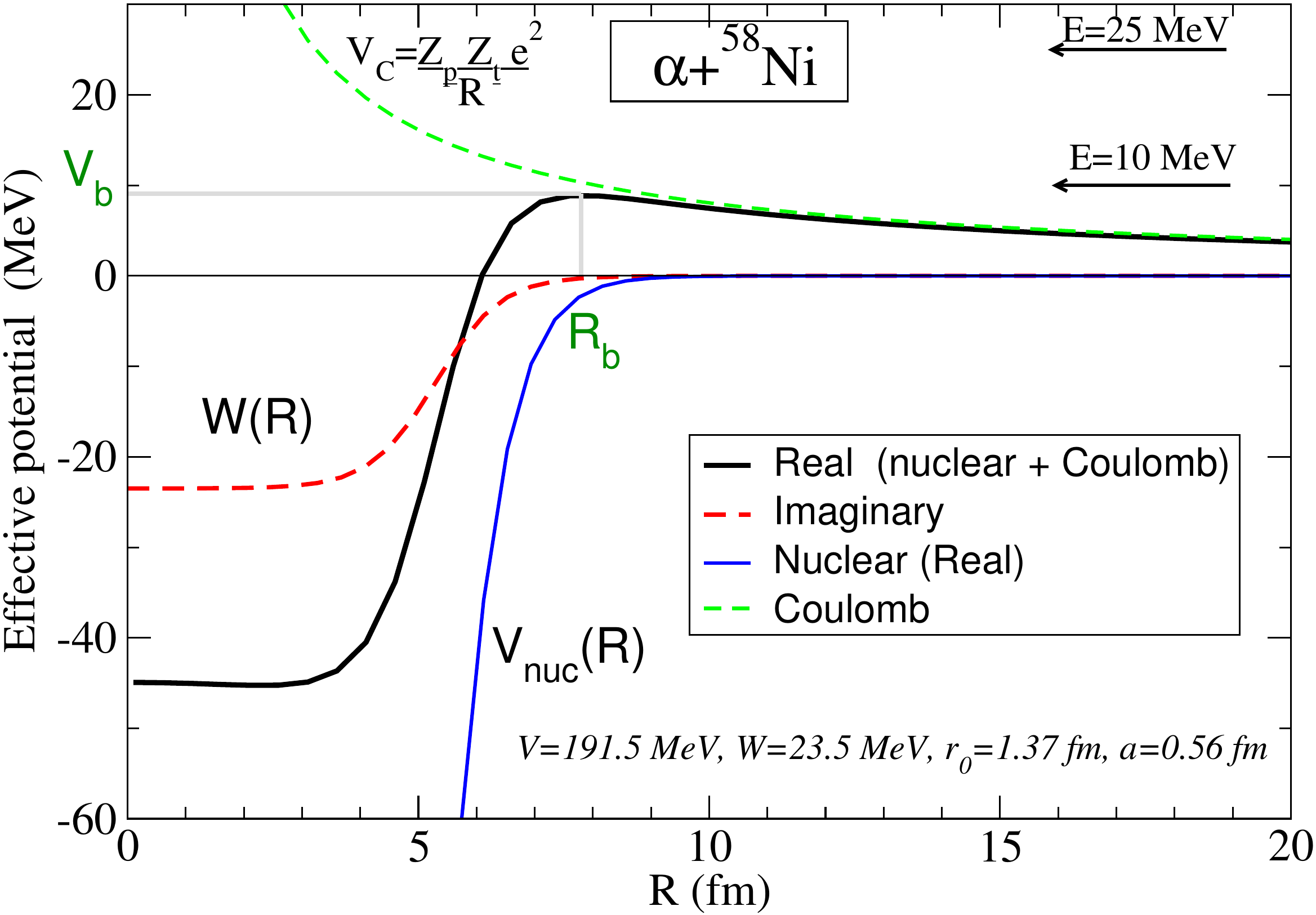}
\end{center}
\caption{\label{fig:veff}Effective potential for a $^4$He+$^{58}$Ni system. The labels $R_b$ and $V_b$ indicate the distance and top of the Coulomb barrier.}
\end{figure}

\begin{figure}
\begin{center}
\begin{minipage}[c]{.32\textwidth}
{\par \resizebox*{0.9\textwidth}{!}
{\includegraphics{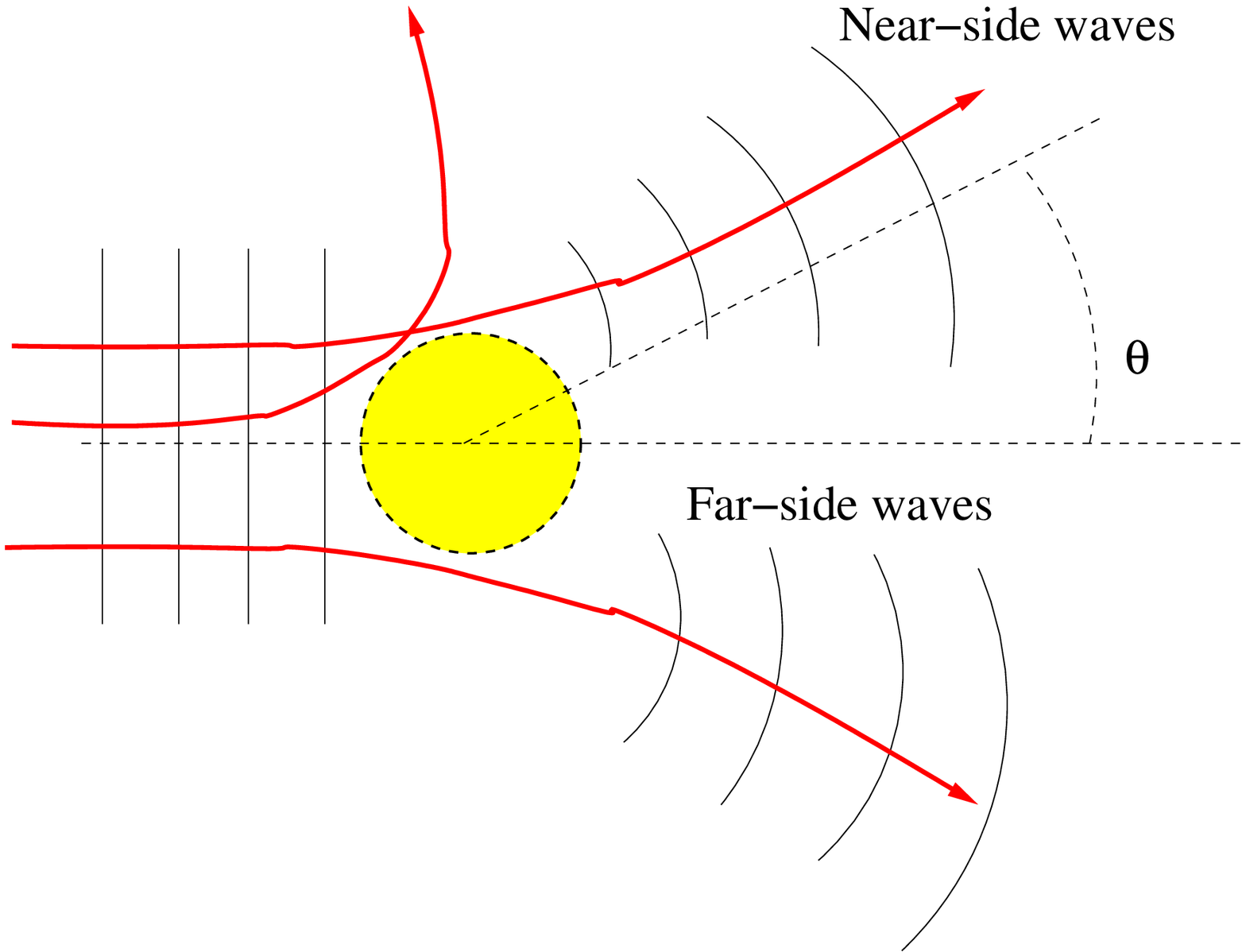}} \par}
\end{minipage}
\begin{minipage}[c]{.32\textwidth}
{\par \resizebox*{0.85\textwidth}{!}
{\includegraphics{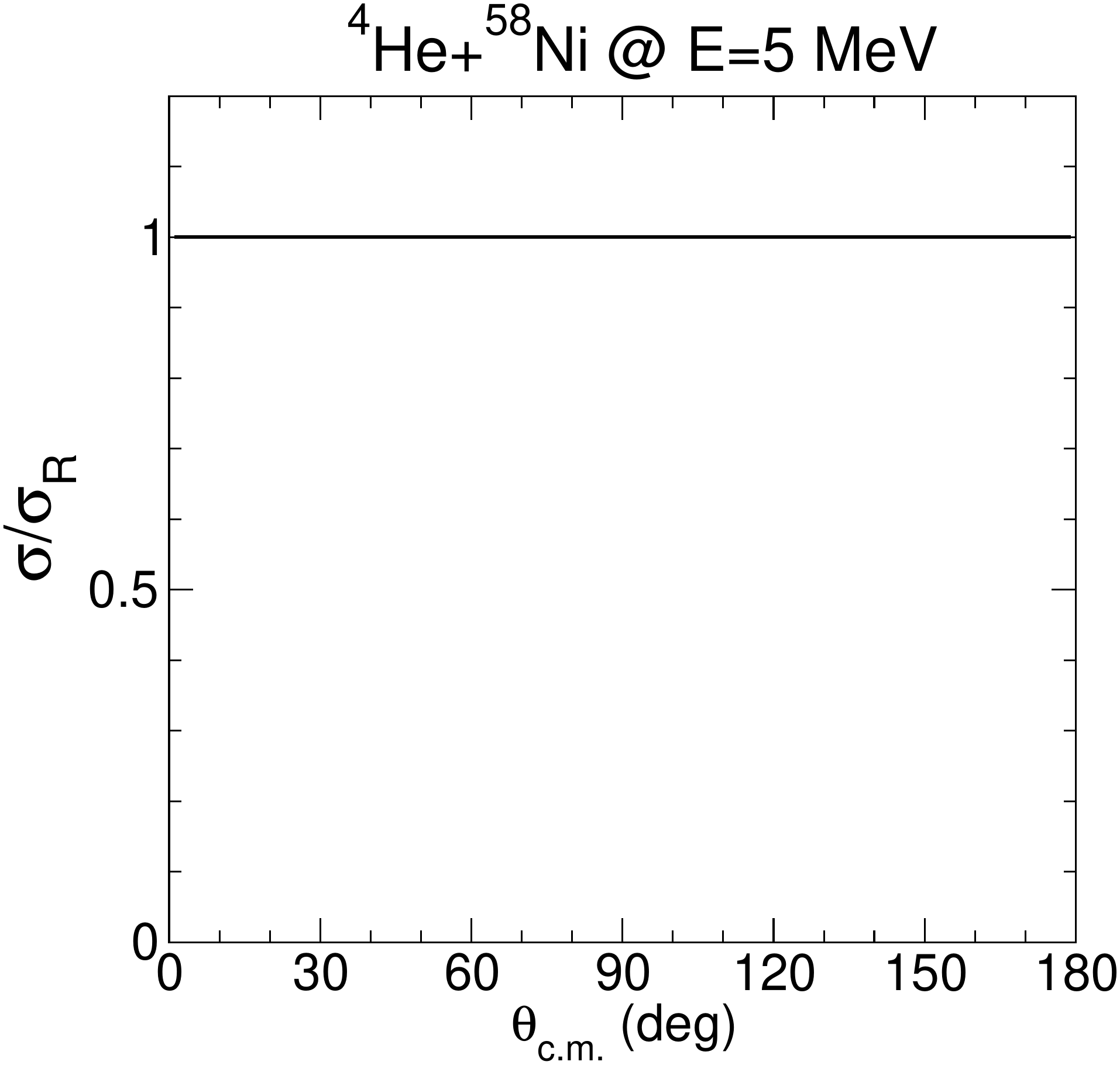}} \par}
\end{minipage}
\begin{minipage}[c]{.32\textwidth}
{\par \resizebox*{0.85\textwidth}{!}
{\includegraphics{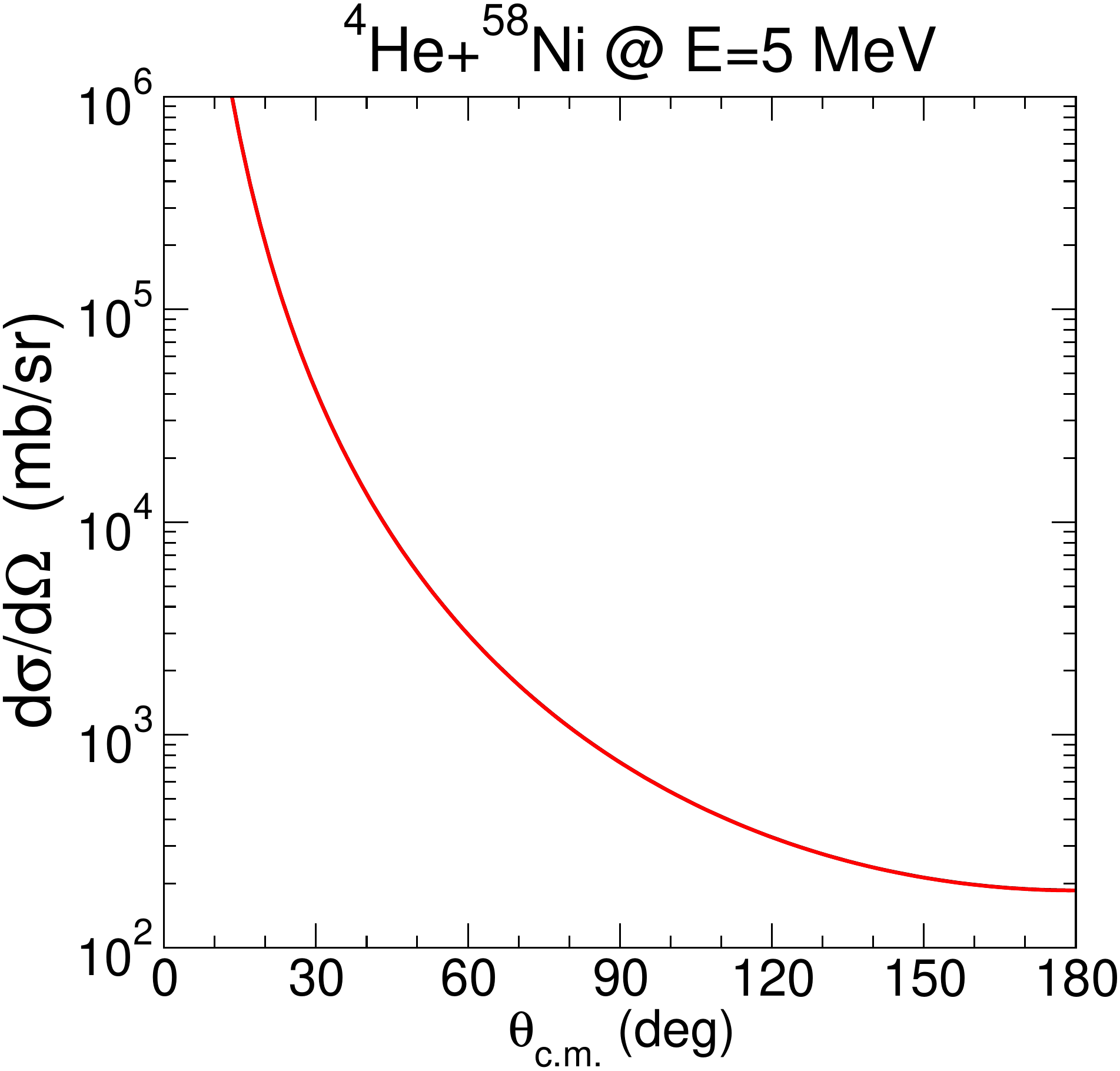}} \par}
\end{minipage}

\end{center}
\begin{center}
\begin{minipage}[c]{.32\textwidth}
{\par \resizebox*{0.95\textwidth}{!}
{\includegraphics{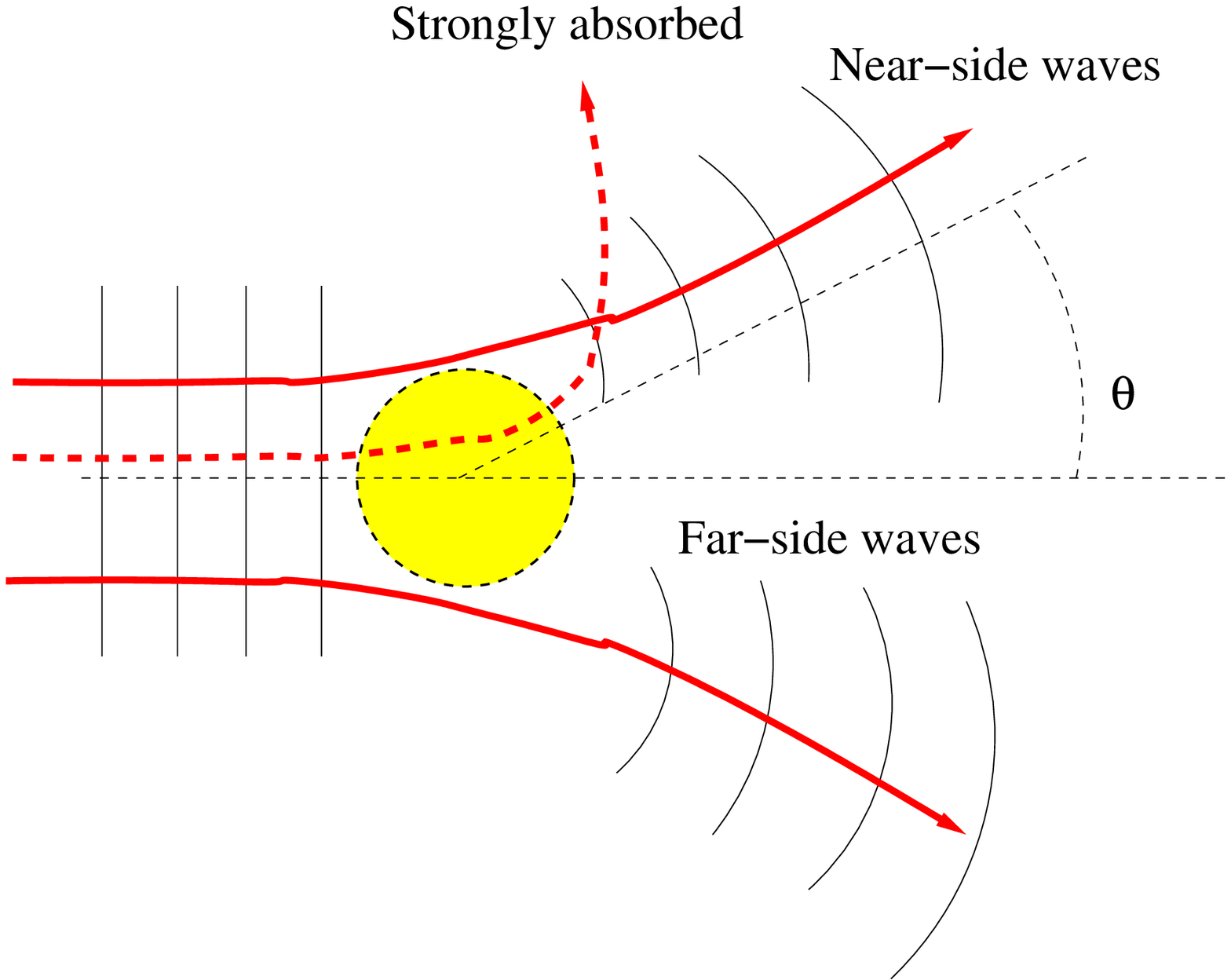}} \par}
\end{minipage}
\begin{minipage}[c]{.32\textwidth}
{\par \resizebox*{0.85\textwidth}{!}
{\includegraphics{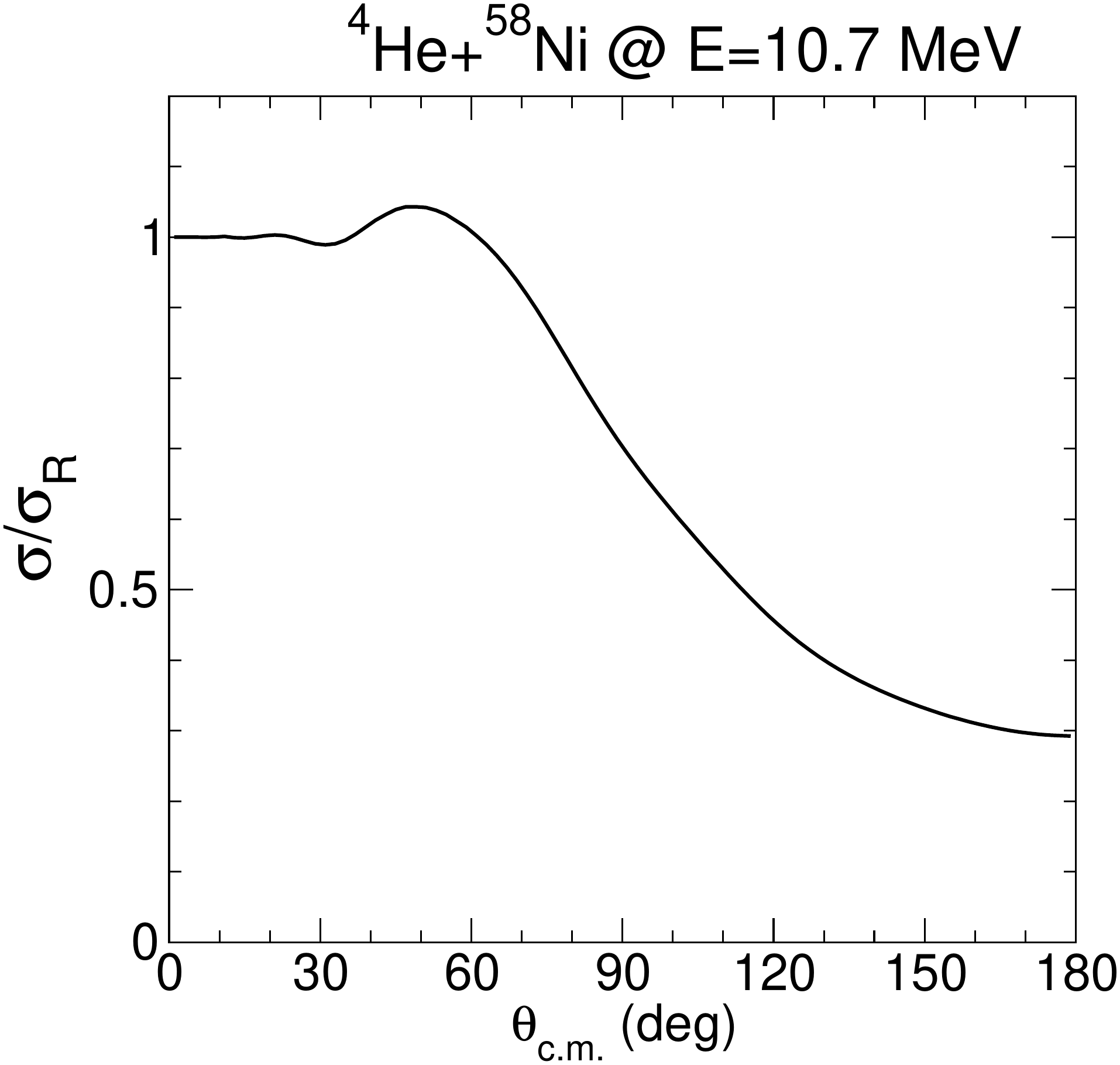}} \par}
\end{minipage}
\begin{minipage}[c]{.32\textwidth}
{\par \resizebox*{0.85\textwidth}{!}
{\includegraphics{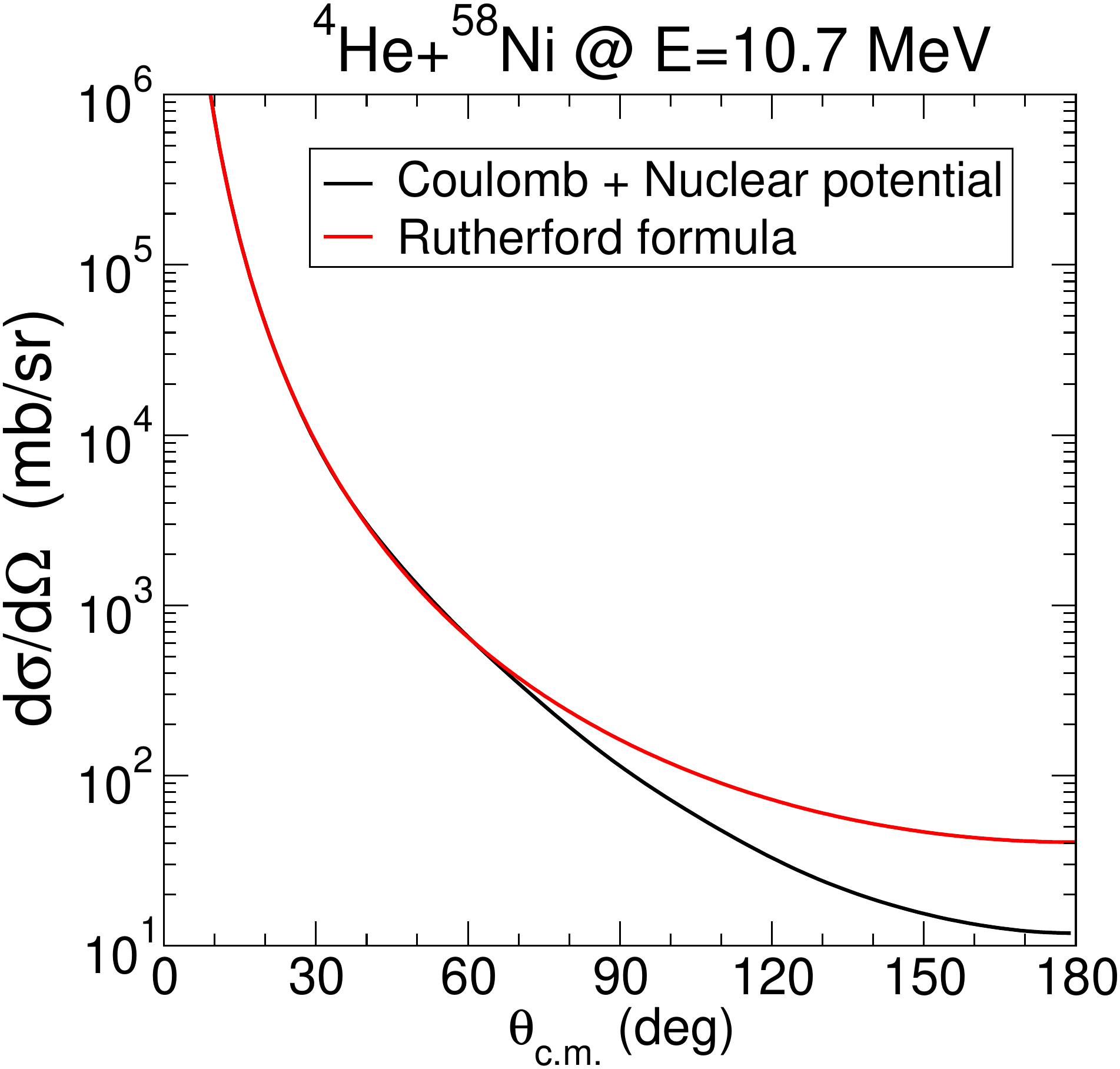}} \par}
\end{minipage}
\end{center}
\begin{center}
\begin{minipage}[c]{.32\textwidth}
{\par \resizebox*{0.95\textwidth}{!}
{\includegraphics{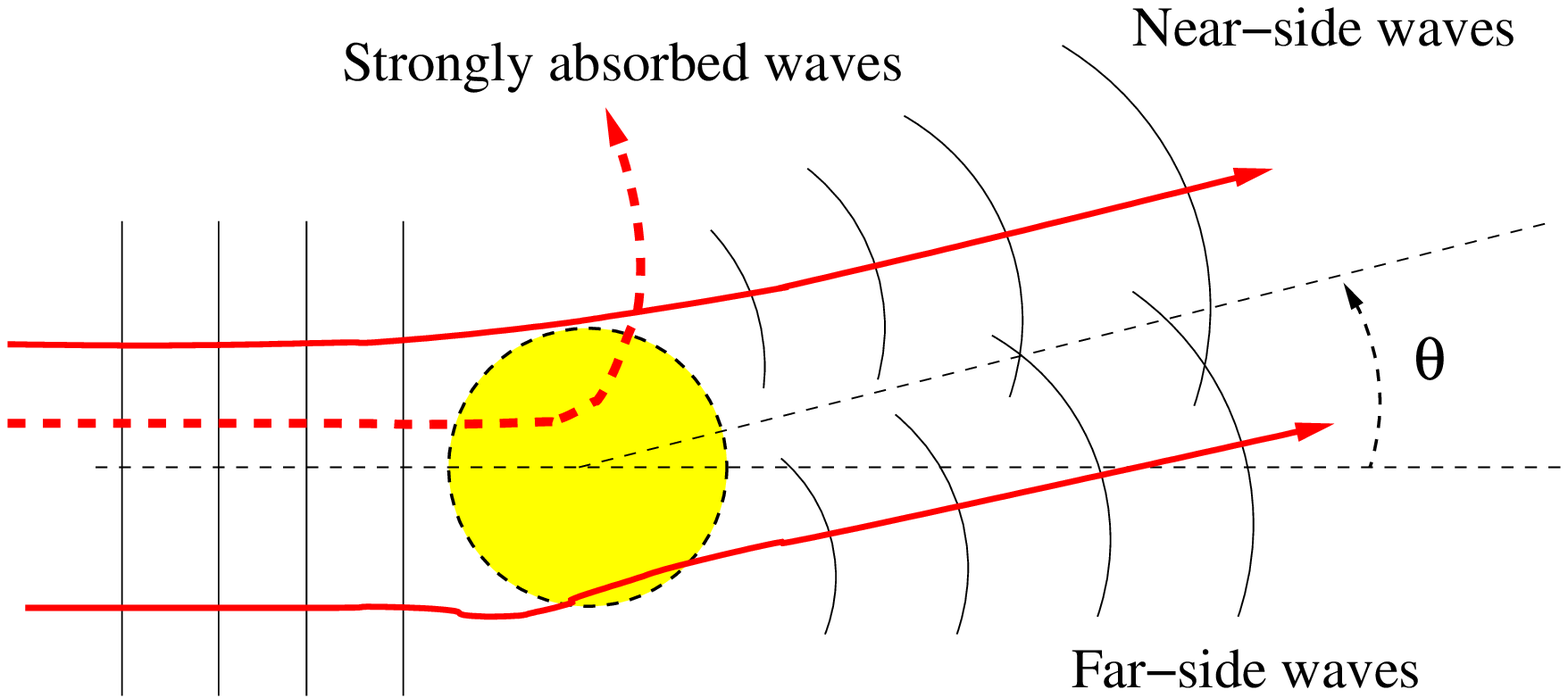}} \par}
\end{minipage}
\begin{minipage}[c]{.32\textwidth}
{\par \resizebox*{0.85\textwidth}{!}
{\includegraphics{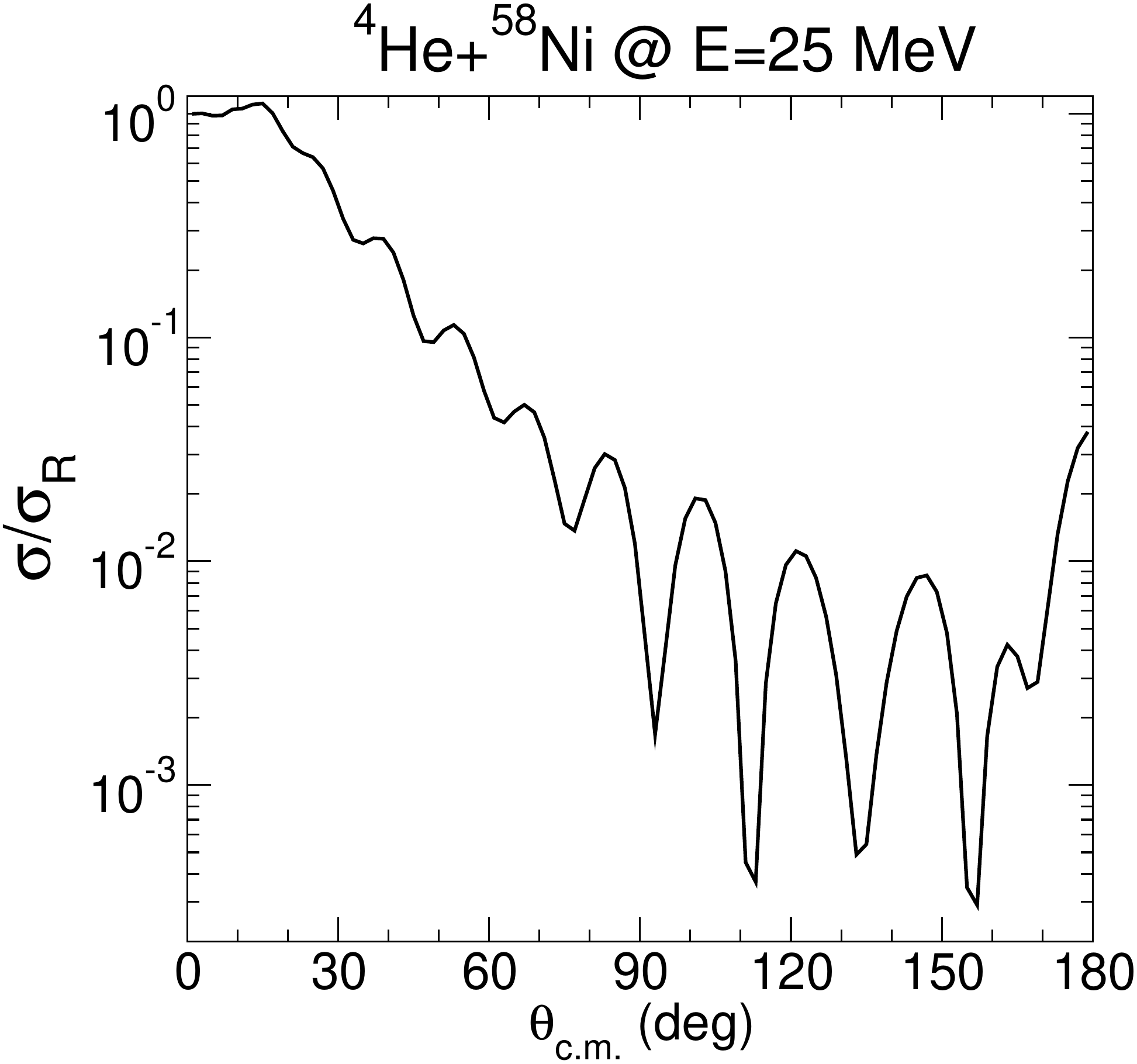}} \par}
\end{minipage}
\begin{minipage}[c]{.32\textwidth}
{\par \resizebox*{0.85\textwidth}{!}
{\includegraphics{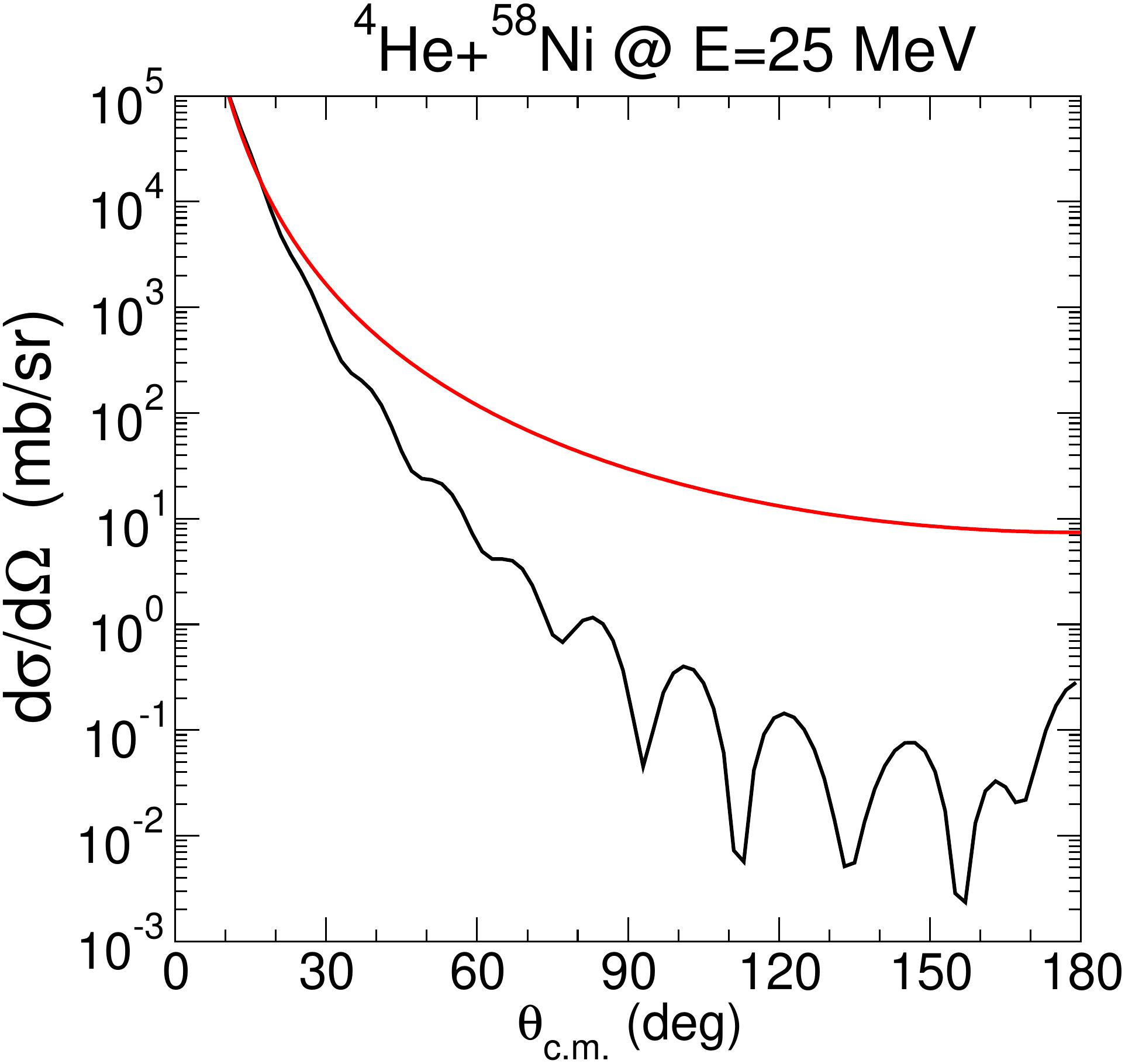}} \par}
\end{minipage}
\caption{\label{fig:strong_abs} Three characteristic regimes encountered in reactions dominated by strong-absorption, illustrated for the reaction $^{4}$He+$^{58}$Ni at $E=5$~MeV (top panels), 10.7~MeV (middle) and 25 MeV (bottom). On the left panels, a pictorial representation in terms of classical trajectories. Dashed trajectories get inside the nuclear range and are likely to be absorbed (i.e.\ leave the elastic channel).}
\end{center}
\end{figure}

\begin{itemize}
\item {\it Rutherford scattering:} When the CM energy is well below the Coulomb barrier, the colliding partners feel only the Coulomb interaction. In absence of strong $\lambda >0$ Coulomb couplings, the projectile--target motion is dictated by the monopole term $Z_p Z_t e^2/R$, and the differential cross sections follows the Rutherford formula. These situations are characterized by large Sommerfeld parameters ($\eta \ggg 1$). In terms of classical trajectories (see LHS of the first row of fig.~\ref{fig:strong_abs}), the repulsive Coulomb interaction acts as a diverging lens, preventing the trajectories to enter into the inner region (dominated by the nuclear interaction). 

\item {\it Fraunhofer scattering:} A very different scenario occurs when the incident energy is much higher than the Coulomb barrier.  The nuclear potential gains importance with respect to the Coulomb potential ($\eta \lesssim 1$) which affects in two ways. First, due to its attractive character, far-side orbits (orbits scattered on the opposite side with respect the incoming projectile) are deflected inward and are allowed to interfere with  {\it near-side}  trajectories scattered at the same angle (see bottom panels of fig.~\ref{fig:strong_abs}). This produces a characteristic oscillatory pattern in the angular distribution, with maxima and minima corresponding to the constructive and destructive interference. Second, due to absorption, some of the trajectories entering into the range of the nuclear potential will be absorbed (i.e.\ will be removed from the elastic channel due to non-elastic processes). This produces an overall reduction of the elastic cross section with respect to the Rutherford formula.  

\item {\it Fresnel scattering:}
Fresnel scattering takes place at incident energies  slightly above the top of the Coulomb barrier and so it can be considered an intermediate situation between Rutherford and Fraunhofer scattering. Distant trajectories (as the one labeled as 1 in the middle panel of fig.~\ref{fig:strong_abs}) are  scattered by the Coulomb potential and hence undergo pure Rutherford scattering. However, closer trajectories experience grazing collisions with the target. Some of them, like the one labeled as 2, can be scattered at the same scattering angle as some more distant Coulomb trajectories. These accumulation of trajectories entering within a narrow range of impact parameters and exiting at about the same scattering angle are responsible for the prominent peak observed in the middle panel, and characteristic of Fresnel scattering. These grazing trajectories divide the angular range into two regions, usually called ``illuminated'' and ``shadow'' regions. The former, corresponding to trajectories more distant than the grazing ones, do not feel the nuclear potential and hence   do not experience absorption. Conversely, trajectories entering with impact parameters smaller than the grazing ones will experience strong absorption. These are the trajectories with larger deflections and, hence, for scattering angles larger than the grazing ones, a drastic reduction of the elastic cross section is found, as seen in the middle panel in the second row of fig.~\ref{fig:strong_abs}. 

The three scenarios described in this section (Rutherford, Fresnel and Fraunhofer) are typical of ordinary, tighly-bound nuclei. In the following sections, we will see how these features are modified in the case of weakly-bound nuclei. 

\end{itemize}

\subsection{Elastic scattering of weakly bound nuclei \label{sec:elas_weakly} }
We have stressed that the elastic scattering is affected by the coupling with non-elastic channels (inelastic, transfer, breakup, fusion...). The relative importance of these channels will depend on the participant nuclei as well and on the energy regime. In the case of weakly-bound projectiles, which is the core topic of this contribution, we have to pay particular attention to the role of the breakup channels since the weak binding usually translates into a large dissociation probability. What modifications should we expect in optical potential, as compared to {\it normal} nuclei? 

\begin{figure}
\begin{center}
\begin{minipage}[t]{.49\textwidth}
{\par \resizebox*{0.8\textwidth}{!}
{\includegraphics{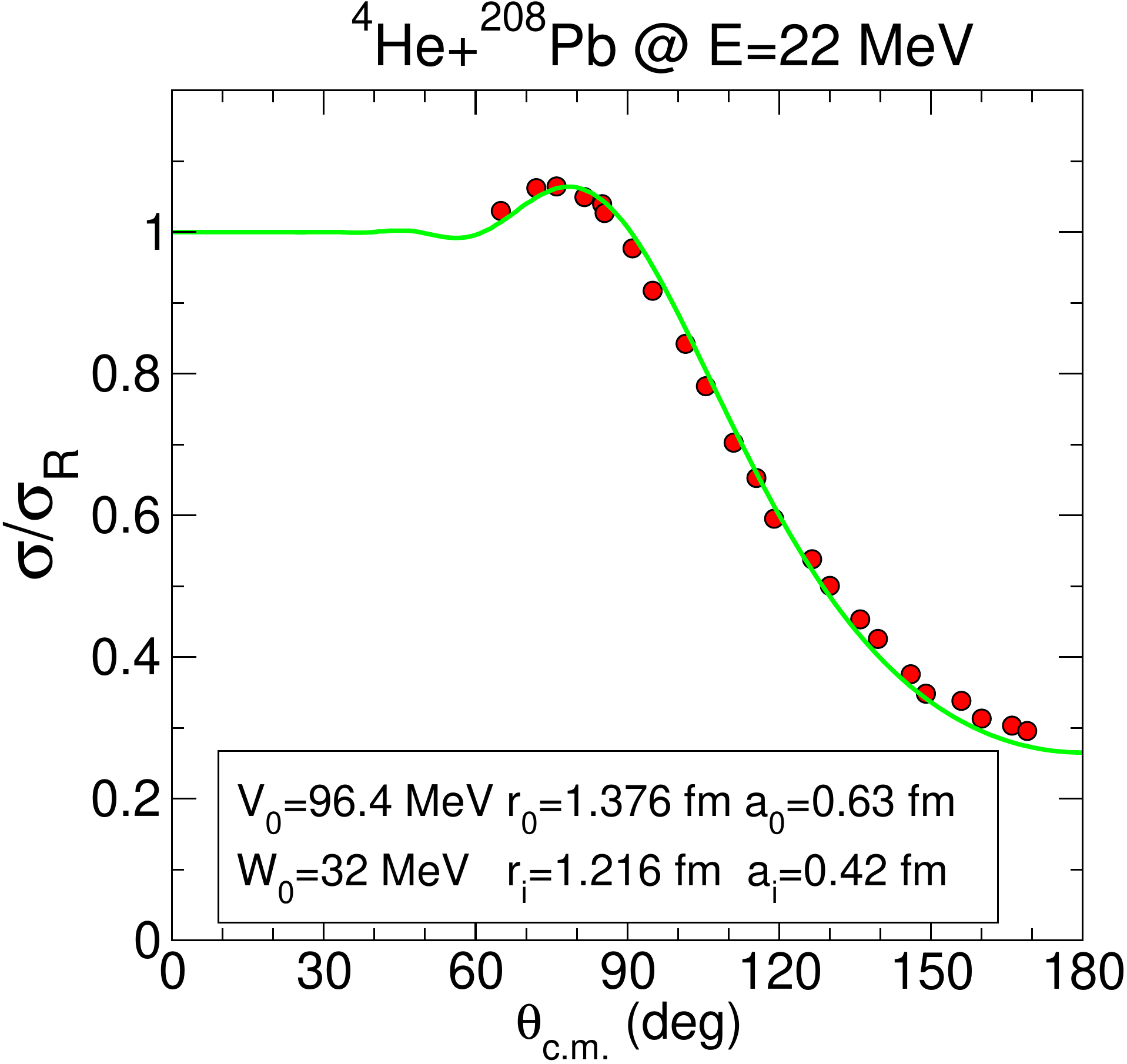}} \par}
\end{minipage}
\begin{minipage}[t]{.49\textwidth}
{\par \resizebox*{0.8\textwidth}{!}
{\includegraphics{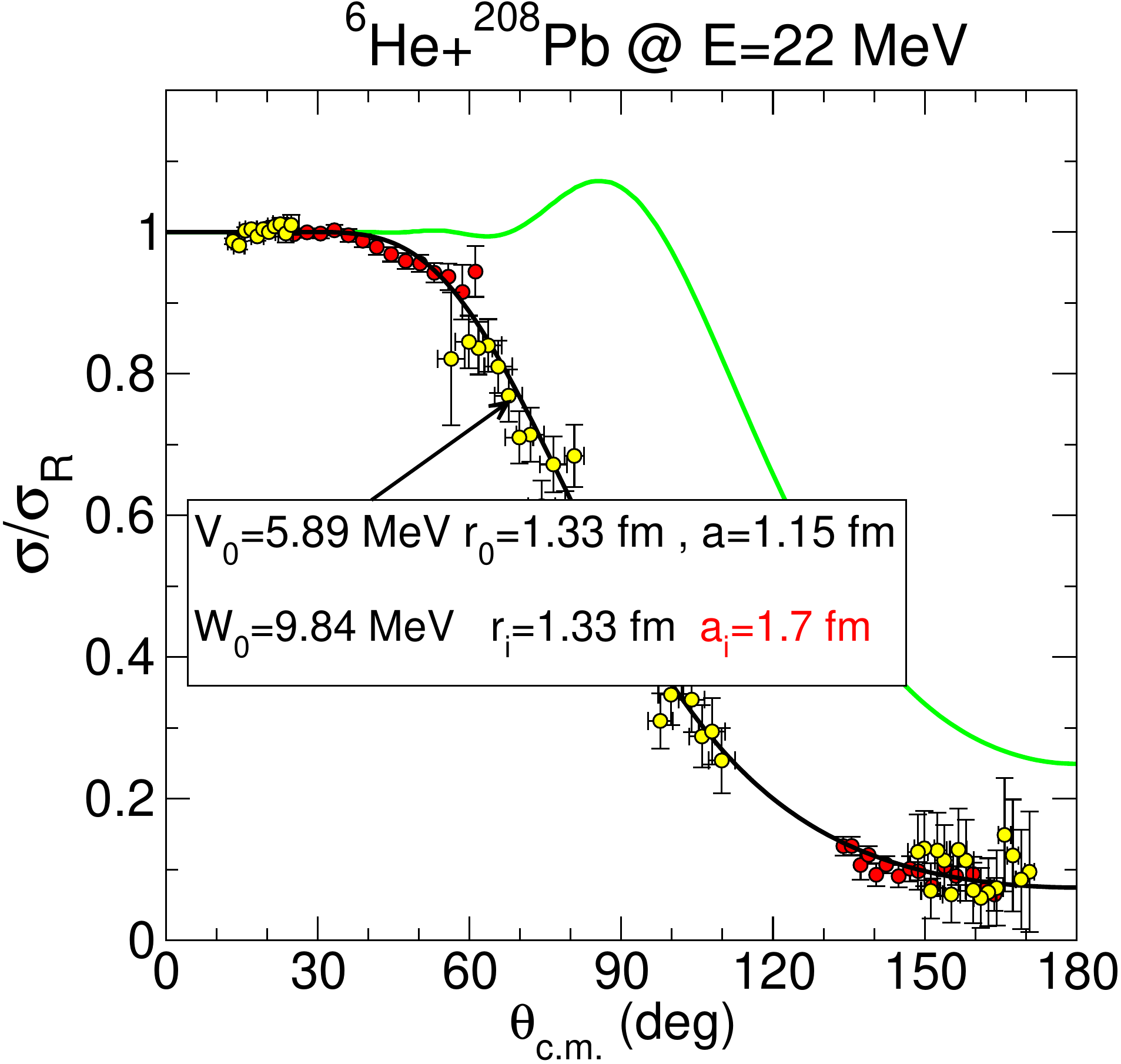}} \par}
\end{minipage}
\caption{\label{fig:hepb_om} Elastic scattering of $^{4,6}$He+$^{208}$Pb at $E_\mathrm{lab}=22$~MeV. The $^{4}$He and $^{6}$He data are, respectively, from refs.~\cite{Bar74} and \cite{San08,Aco11}.}
\end{center}
\end{figure}

Let us consider as examples the $^{4,6}$He+$^{208}$Pb reactions at $E_\mathrm{lab}=22$~MeV (see fig.~\ref{fig:hepb_om}). In the $^4$He case, the measured differential cross section shows a typical Fresnel pattern, with a maximum around the grazing angle and a rapid decrease at larger angles. The  $^6$He case is markedly different. The cross section is largely suppressed with respect to the Rutherford formula and the Fresnel peak is completely absent. The reduction with respect to the Rutherford cross section starts at relatively small angles which, classically, correspond to large impact parameters (i.e.\ distant trajectories). This suggests the existence of a long-range  non-elastic mechanism, which removes a significant part of the flux from the elastic channel. Since $^{6}$He is bound by  only $\sim$1~MeV, a natural candidate is of course the breakup of the projectile but, other mechanisms, such as neutron transfer, can contribute as well. These features can be also seen in the optical model potentials describing these data. The curves shown in  fig.~\ref{fig:hepb_om} are optical model calculations using phenomenological WS forms [eq.~(\ref{eq:WS})].   In the case of the $^4$He projectile, the radius and diffuseness parameters of the real and imaginary parts follow closely the densities of normal, well-bound nuclei  ($a\sim$0.56~fm). If this potential is used for the $^{6}$He+$^{208}$Pb case (scaling the radii according to the mass number of $A$ or,  equivalently, using the same reduced radii), we get the dashed line of the right panel in which, as can be seen, the Fresnel behaviour persists, in clear disagreement with the data. If the potential parameters are varied to reproduce the data (keeping the radii $r_0=r_i=1.33$~fm to reduce the number of free parameters) one obtains the values listed in the figure, and the corresponding differential cross section (solid line). Its more salient feature is the large value of the real and imaginary diffuseness parameters ($a_i > 1$~fm). This is a clear indication of the influence of the non-elastic channels (possibly transfer and breakup) which, in the Feshbach formalism, would be embedded in the polarization potential [c.f.\ eq.~(\ref{eq:veff})]. 

Understanding and disentangling the nature of these non-elastic channels requires going beyond the optical model. This can be done, for example, using approximate forms of the polarization potential or within the coupled-channels method, described below.

\subsection{\label{sec:cdp}Coulomb dipole polarization potentials}
The effect of Coulomb dipole polarizability (CDP) on the elastic scattering can be included by means of a polarization potential.  From physical arguments, we may expect  this potential to be complex, whose real and imaginary parts can be understood as follows:
\begin{enumerate}
\item The strong Coulomb field will produce a polarization (``stretching'') of the projectile, giving rise 
to a dipole contribution on the {\it real} potential. 
\item The weakly bound nucleus can eventually break up, leading to a loss of flux of the elastic channel, which corresponds to the imaginary part of the polarization potential.
\end{enumerate}

The CDP acquires a particularly simple form in the so-called adiabatic limit, in which one assumes that the excitation energies are
high enough so the characteristic time for a transition to a state $n$ ($\tau_{ex} \approx \hbar /(\varepsilon_n- \varepsilon_0)$)
is small compared to the characteristic time for the collision ($\tau_{coll}\approx a_0 /v$, where $a_0$ is the distance of closest approach in a head-on collision and $v$ is the projectile velocity). Applying
second-order perturbation theory, one gets the following expression for this adiabatic
dipole polarization potential \cite{Ald75}:
\be
\label{eq:adpol}
V_\mathrm{ad} (R) = -
\sum_{n=1} \frac{ |\langle n| V_\mathrm{dip} |0 \rangle |^2}{\varepsilon_n - \varepsilon_0} 
= - \alpha \frac{ (Z_t e)^2}{R^4} ,
\ee
where $\alpha$ is the {\it dipole polarizability parameter}, defined as
$$
\alpha= \frac{8\pi}{9} \frac{B(E1; gs \rightarrow n)}{\varepsilon_n - \varepsilon_0 }  ,
$$
with $B(E1; gs \rightarrow n)$ the dipole strength for the coupling to the dipole excited state
$|n \rangle$.

We see that the adiabatic  polarization potential is purely
real and does not depend on the collision energy. When the average excitation energies are small, as it is the case of weakly bound nuclei (such as halo nuclei), the adiabatic approximation is questionable. A expression for a non-adiabatic CDP will be presented in sec.~\ref{sec:AW} in the context of the semiclassical theory of Alder and Winther.

\section{Inelastic scattering: the coupled-channels method }

Nuclei are not inert or {\em frozen} objects; they do have an internal structure of protons and neutrons
that can be modified (excited), for example, in collisions with other nuclei. In fact, an important and common process that may occur in a collision  between two nuclei  is the  excitation of one (or both) of the nuclei. 
Inelastic scattering is an example of {\it direct reaction}  and, as such, the colliding nuclei preserve their collision after the collision. 

The energy required to  excite a nucleus is {\it taken} from the kinetic energy  
of the projectile-target relative motion. This means that, if one  of the colliding nuclei is excited, the final 
kinetic energy of the system is reduced by an amount equal to the excitation energy of the excited state populated in the reaction. So, by measuring the kinetic energy of the outgoing fragments, one can infer the excitation energy of the projectile and target. This has been indeed a common technique to measure and identify such excited states. 

\begin{figure}
\begin{center}
{\par \resizebox*{0.6\textwidth}{!}{\includegraphics[angle=0]{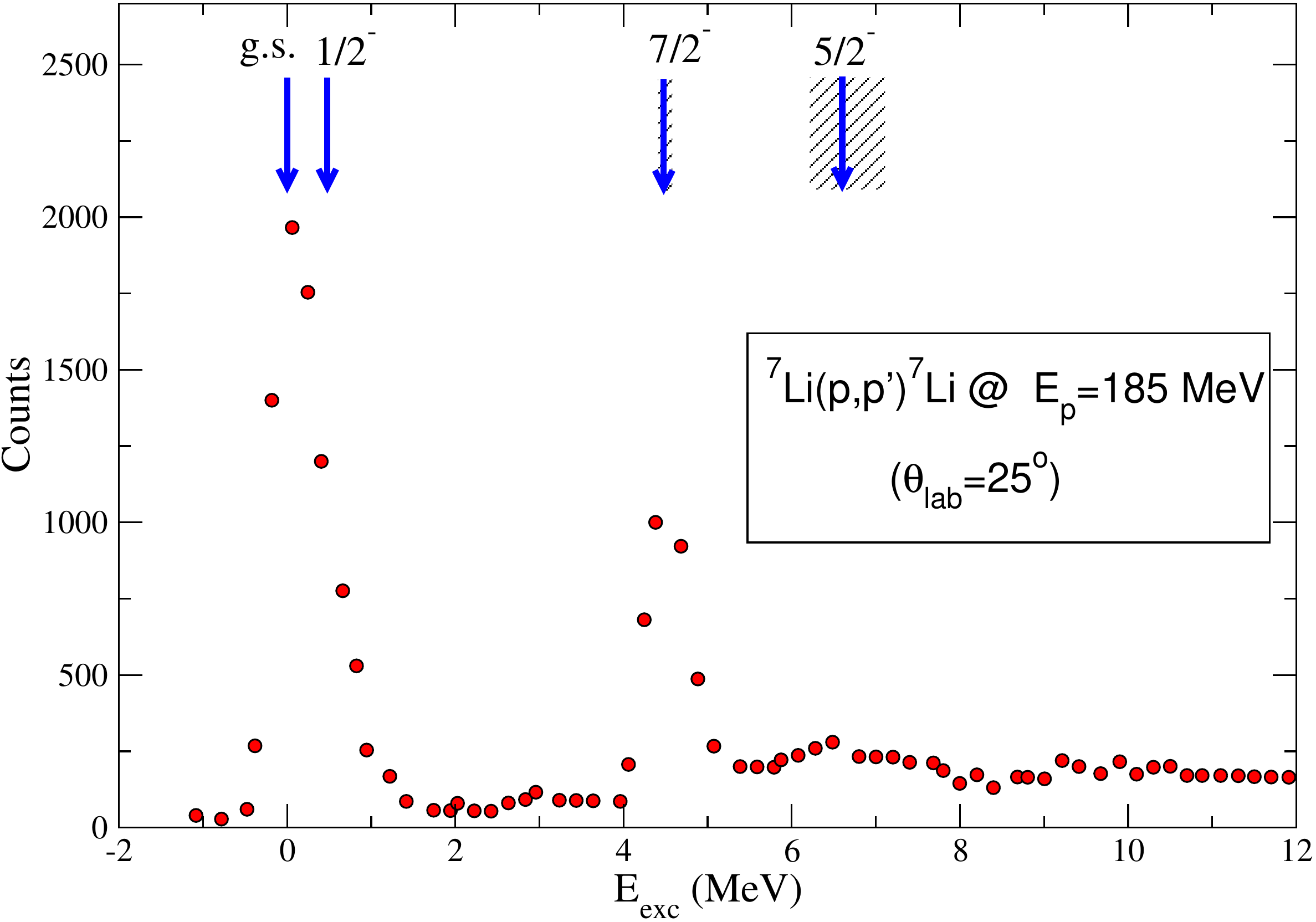}} \par}
\end{center}
\caption{\label{fig:pli7} Energy spectrum of detected outgoing protons of 185 MeV scattered from a \nuc{7}{Li} target. The vertical arrows indicate the position of the $^{7}$Li bound states and two lowest resonances. Experimental data are from 
ref.~\cite{Has65}.}
\end{figure}

As an example, let us consider the scattering of a proton beam off a \nuc{7}{Li} target. Figure \ref{fig:pli7} shows  
the experimental excitation energy spectrum inferred from the energy of the outgoing protons detected at an scattering angle of $25^\circ$, for a proton incident energy of 185 MeV \cite{Has65}. We have superimposed the position of the low-lying levels of \nuc{7}{Li} to highlight the correspondence between the observed peaks 
and these states.  The peak at $E_x=0$ (corresponding to $Q=0$) corresponds to the \nuc{7}{Li} ground state. Thus, it is 
just elastic scattering. At $E_x=0.48$~MeV, we should see a second peak corresponding to the first excited state of 
\nuc{7}{Li}. However, due to the energy resolution, this peak is not resolved in these data from the elastic peak. At $E_x=4.6$~MeV there is a prominent peak corresponding to a $7/2^-$ state in \nuc{7}{Li}. This state is above the \nuc{4}{He}+$^{3}$H threshold placed at 2.47~MeV and does actually correspond to a continuum resonance. This threshold corresponds to the energy necessary to 
dissociate the \nuc{7}{Li} nucleus into $\alpha+t$. Therefore, for excitation energies above this value, we have a continuum of 
accessible energies, rather than a discrete spectrum,  and any value of $E_x$ is possible. This explains the {\it background} observed at these excitation energies. 

Note that the information provided by these data is not enough to determine other properties of the energy spectrum, such as as the spin/parity assignment or their collective/single-particle character. Further information can be obtained from the shape and magnitude of the angular distribution of the emitted ejectile.  To do that, one needs to compare the data with a suitable reaction calculation, as we will see in the next section. 

\subsection{Formal treatment of inelastic reactions}

\subsubsection{The coupled-channels (CC) method}

Remember from Sec.~\ref{sec:pq} that any practical solution of the scattering problem starts with a reduction of the full physical space into P and Q subspaces, the former corresponding to the channels that are to  be explicitly included. In an inelastic process, this P space will comprise the elastic channel, plus some excited states of the projectile and/or target, those more strongly coupled in the process or, at least, those that will be compared with the experimental data.  

Let us consider the scattering of a projectile $a$ by a target $A$, and let us assume for simplicity that 
only the projectile is excited during the process, the target remaining in its ground state.
We denote this mass partition by the index $\alpha$, i.e., $\alpha \equiv a+A$.  Our model Hamiltonian will describe a set of states of the projectile and possible couplings between them during the collision.  This model  Hamiltonian will be expressed as [c.f.\ eq.~(\ref{eq:Halpha})]:
\begin{equation}
 H=-\frac{\hbar^2}{2\mu}\nabla^2_{\bR} + H_a(\xi) +V_{\alpha}(\xi,\bR) ,
\label{eq:hinel}
\end{equation}
where $H_a(\xi)$ is the projectile internal Hamiltonian and $\xi$ its internal coordinates.

Let us denote by $\{\phi_n(\xi)\}$ the internal states of the projectile. These will be the eigenstates of the Hamiltonian 
$H_{a}(\xi)$:
$
H_{a}\phi_n = \varepsilon_n \phi_n .
$
The idea of the CC method is to expand  the total wave function of the system in 
  the set of internal states $\{\phi_n(\xi)\}$,
\begin{equation}
\label{eq:Psi-cc}
 \Psi^{(+)}(\bR,\xi)=\phi_{0}(\xi)\chi_{0}(\bR)+ \sum_{n>0}^{N} \phi_{n}(\xi)\chi_{n}(\bR) ,
\end{equation}
with $\phi_0(\xi)$ representing the ground-state wave function and $N$ the number of excited states included. 

The unknown coefficients  $\chi_{n}(\bR)$  describe
the relative motion between the projectile and target in the corresponding
internal states. They tell us the relative 
``probability'', as a function of $\bR$, for the projectile  being in state $n$. The different possibilities 
for $n$ are frequently referred to as ``channels''. The total wave function $\Psi(\bR,\xi)$ verifies 
the \sch equation: $[E-H]\Psi^{(+)}(\bR,\xi)=0$.  We now proceed as follows: (i) insert the expansion (\ref{eq:Psi-cc}) and the Hamiltonian (\ref{eq:hinel}) in this equation; (ii) multiply on the left by each of the basis functions $\phi^{*}_{n}(\xi)$, and (iii) integrate over the internal coordinates $\xi$. For each $n$, we get a differential equation of the form:
\begin{equation}
\label{eq:cc}
\left[E-\varepsilon_{n}-\hat{T}_\bR -V_{n,n}(\bR) \right] \chi_{n}(\bR)  = 
\sum_{n' \neq n} V_{n,n'}(\bR) \chi_{n'}(\bR) ,
\end{equation}
where  $V_{n,n'}$  are the so called {\em coupling potentials}, defined as:
\begin{equation}
V_{n,n'}(\mathbf{R})=\int d\xi \phi_{n}^{*}(\xi)V(\xi,\bR)\phi_{n'}(\xi) .
\label{eq:transpot}
\end{equation}
So, for example, $V_{0,m}$ is the
potential responsible for the excitation from the ground state ($n=0$)
to a given final state $m$. We have not yet defined the form of the effective potential $V(\xi,\bR)$ and the internal states $\phi_{n}$, that is, the model Hamiltonian. These potentials are constructed within a certain model, as we will see later.

Note that the equation associated with a given value of $n$ contains not only the unknown $\chi_n(\bR)$, but also  $\chi_{n'}(\bR)$ with $n' \neq n$. Consequently, eq.~(\ref{eq:cc}) represents a set of coupled differential equations for the set of functions $\{\chi_n(\bR)\}$. 

\subsubsection{Boundary conditions}
Similarly to the OM case, the CC equations must be solved with  appropriate boundary conditions. 
These boundary conditions correspond to the physical situation in which the projectile is initially in the ground-state  ($\phi_0$) and impinges with momentum $\bK_0$. The projectile-target relative motion is represented by a plane wave with momentum $\bK_0$.  As a result of the collision with the target, a series of outgoing spherical waves  is created (fig.~\ref{fig:scat}). Recalling the general asymptotic behaviour of the total wave function, eq.~(\ref{eq:Psi-asym}), for the case of inelastic scattering we will have
\begin{eqnarray}
\label{Psi_asym}
\Psi^{(+)}_{\bK_0}(\bR,\xi) & \xrightarrow{R \gg} & \left\{ e^{i \bK_0 \cdot \bR}  +  
          f_{0,0}(\theta) \frac{e^{i K_0 R}}{R} \right\} \phi_0(\xi) +
          \sum^{N}_{n>0}f_{n,0}(\theta) \frac{e^{i K_n R}}{R} \phi_n(\xi).
\end{eqnarray}

Comparing with (\ref{eq:Psi-cc}) we see that the functions $\chi_n(\bR)$ must verify the following boundary conditions:
\begin{align}
\label{boundary-cc}
\chi_0^{(+)}(\bK_0,\bR) & \rightarrow  e^{i \bK_0 \cdot \bR}  +  f_{0,0}(\theta) \frac{e^{i K_0 R}}{R} & n= 0
\quad \textrm{(elastic)} \\
\chi_n^{(+)}(\bK_n,\bR) & \rightarrow                           f_{n,0}(\theta) \frac{e^{i K_n R}}{R}, & n\neq 0
\quad \textrm{(non-elastic)}
\end{align}
from which the elastic and inelastic differential cross sections are to be obtained from the coefficient of the corresponding outgoing wave:
\begin{equation}
\left ( {d \sigma(\theta) \over d \Omega} \right )_{0 \to n} = \frac{K_n}{K_0} |{f_{n,0}(\theta)}|^2 .
\end{equation}

Note that: 
\begin{itemize}
\item Plane waves are present only in the $\chi_0$ component (that is, the elastic component) but outgoing waves appear in all components. 
 
\item  The scattering angle in the CM frame, $\theta$, is determined by the direction of the momenta $\bK_0$ and  $\bK_n$. Defining the {\it momentum transfer}  as ${\vec q}={\vec K}_n - {\vec K}_0$, we have (see fig.~\ref{fig:scat}):
\begin{equation}
q^2 = K^2_0 + K^2_n -2 K_0 K_n \cos (\theta)
\end{equation}
 
\item  The modulus of $\bK_n$ is obtained from energy conservation\footnote{For $\varepsilon_n > E$, the kinetic energy is negative and the corresponding momentum $K_n$ becomes imaginary. Consequently, the asymptotic solutions $\chi_n$ of  eq.~(\ref{boundary-cc}) vanish exponentially and then these channels do not contribute to the outgoing flux.}:
\be
 E=\varepsilon_0 + \frac{\hbar^2 K_0^2}{2 \mu} = \varepsilon_n + \frac{\hbar^2 K_n^2}{2 \mu}
\ee
\end{itemize}

\subsubsection{The DWBA method for inelastic scattering}
If the number of states is large, the solution of the coupled equations can be a difficult task. In many situations, however, some of
the excited states are very weakly coupled to the ground state and can be treated perturbatively.  In this case, the set of equations (\ref{eq:cc}) can be solved iteratively, starting from the elastic channel equation, and setting to zero the source term (the RHS of the equation). This allows the calculation of the distorted wave $\chi_0(\bK_0,\bR)$. This solution is then inserted into the equation corresponding to an excited state $n$, thus providing a first order approximation for  $\chi_n(\bK_0,\bR)$. If the process is stopped here, then the method is referred to as {\it distorted wave Born approximation} (DWBA). 

We provide here an alternative derivation of the DWBA method, which leads to a more direct connection with  the scattering amplitude. We make use of the {\it exact} scattering amplitude (\ref{eq:Tdist}) derived in subsection  \ref{sec:twopot} using the Gell-Mann--Goldberger transformation.
%
%
To particularize this general result to our case, we consider  a transition between an initial state $i$ (typically, the g.s.) and a final state $f$. Since these states belong to the same partition ($\alpha$) we do not need to specify explicitly the subscripts $\alpha$ and $\beta$. Then, the general amplitude (\ref{eq:Tdist}) reduces to:
\be
{\cal T}_{f,i} = \int \int \chi_{f}^{(-)*}(\bK_f, \bR) 
  \phi_{f}^{*}(\xi) [V_f - U_f]  \Psi^\mathrm{(+)}_{\bK_i} d\xi  d\bR \, ,
\ee
where, within the CC method, $\Psi^\mathrm{(+)}_{\bK_i}$ is given by the expansion (\ref{eq:Psi-cc}). Recall that, in this expression,  $\chi_{f}^{(-)}(\bK_f, \bR)$ is the time reversal of $\chi_{f}^{(+)}(\bK_f, \bR)$, which is a solution of 
\be
[\hat{T}_\bR + U_f(\bR) + \varepsilon_f -E] \chi^{(+)}_{f}(\bK_f,\bR) =0 ,
\ee
for some auxiliary potential $U_f(\bR)$. Typically, $U_f(\bR)$ is chosen as a phenomenological potential that describes the elastic scattering of the $a+A$ system at the energy of the exit channel ($E_f=E-\varepsilon_f$). 


The DWBA formula is obtained by approximating the total wave function $\Psi^\mathrm{(+)}_{\bK_i}$ by the factorized form:
\begin{equation}
\Psi^{(+)}(\bR,\xi) \simeq \chi^{(+)}_{i}(\bK_i,\bR) \phi_i(\xi),
\end{equation}
where $\chi^{(+)}_{i}(\bK,\bR)$ is the distorted wave describing the projectile--target motion in the entrance channel,
\begin{equation}
\left[\hat{T}_{\bR}+ U_{i}(\bR) +\varepsilon_{i} - E \right] \chi^{(+)}_{i}(\bK_i,\bR)  =  0 \, ,
\label{chii}
\end{equation}
where $U_{i}(\bR)$ is the average potential in the initial channel, and is usually taken as the potential that describes the elastic scattering in this channel. With this choice, one hopes to include effectively some of the effects of the neglected channels. 

In DWBA, the scattering amplitude corresponding to the inelastic excitation of the projectile from the initial 
state $\phi_i(\xi)$ and momentum $\bK_i$ to a final state $\phi_f(\xi)$ and momentum $\bK_f$ is given by 
\begin{equation}
f^\mathrm{DWBA}_{f,i}(\theta)=-\frac{\mu}{2 \pi \hbar^2} \int d\bR \, \chi_{f}^{(-)*}(\bK_f,\bR)  W_{if} (\bR) \chi_{i}^{(+)}(\bK_i,\bR)
\label{eq:dwba}
\end{equation}
where $W_{if}(\bR)$ is the coupling potential 
\begin{equation}
W_{if}(\bR) \equiv  \langle \phi_{f} | V_f - U_f| \phi_i \rangle  = \int d\xi \, \phi_{f}^{*}(\xi)  (V_f - U_f)\phi_{i}(\xi) .
\label{eq:transpot2}
\end{equation}

In actual calculations,  the internal states have definite angular momentum (spin) so, we may introduce this dependence explicitly using the following notation:
$$
 \phi_{i}(\xi) = | I_i  M_i \rangle 
\quad \quad 
\textrm{and}
 \quad \quad
\phi_{f}(\xi) = | I_f  M_f \rangle .
$$
To exploit this property, one usually expands the projectile-target interaction in multipoles:
\be
V(\bR,\xi)= \sqrt{4\pi} \sum_{\lambda, \mu}  V_{\lambda \mu}(R,\xi) Y_{\lambda \mu}(\hat{R}) .
\ee

DWBA calculations require the matrix elements:
\be
 \langle I_f  M_f | V(\bR,\xi) | I_i  M_i \rangle =
   \sqrt{4 \pi} 
  \sum_{\lambda,\mu}   \langle I_f M_f  | V_{\lambda \mu}(R,\xi) | I_i M_i  \rangle  Y_{\lambda \mu}(\hat{R}) .
\ee

The dependence on the spin projection can be singled out using the {\it Wigner-Eckart theorem}%
\footnote{We assume thorough  this contribution the Bohr and Mottelson convention of reduced matrix elements \cite{BM1}.}
\be
 \langle I_f M_f | V_{\lambda \mu}(R,\xi) | I_i M_i \rangle = 
 (2I_f +1)^{-1/2}   \langle I_f M_f |   I_i M_i \lambda \mu  \rangle 
  \langle I_f \| V_{\lambda}(R,\xi) \| I_i  \rangle 
\ee
where the quantities $\langle I_f \| V_{\lambda}(R,\xi) \| I_i  \rangle$ are called {\it reduced matrix elements}. These are independent of the spin projections, as the notation implies.  

Actual applications of the DWBA amplitude (\ref{eq:dwba}) require the specification of the structure model (that will determine the functions $\phi_{i}(\xi)$) as well as the projectile--target interaction $V_f$. We give some examples in the following section.

\subsection{Specific models for inelastic scattering \label{sec:inelmodel}  } 

\subsubsection{Macroscopic (collective) models}

Ignoring spin-dependent forces, the nuclear interaction between spherical, static nuclei is a function of the distance between the nuclear surfaces of the colliding nuclei ($V_N=V_N(R-R_0)$, with $R_0\simeq R_p+R_t$). However, if one (or both) of the colliding nuclei is deformed (rotor) the nucleus-nucleus interaction will depend on the orientation of the deformed nucleus in space (because, depending on this orientation, the distance between the surfaces will vary accordingly). This introduces a dependence on the angles of the relative coordinate, and the nucleus-nucleus potential will not be central any more. A similar situation occurs when one of the nuclei undergoes surface vibrations. 

If the deviation from the spherical shape is small, one may use a Taylor expansion of the potential around this spherical shape, 
\be
V_N(\bR, \xi) = V_N(R - R_0) - \sum_{\lambda,\mu} \hat{\delta}_{\lambda \mu}  {d V(R-R_0) \over d R} Y_{\lambda \mu}(\theta, \phi) + \ldots
\ee
where $\hat{\delta}_{\lambda,\mu}$ are the so-called {\it deformation length operators}. For a nucleus with a permanent deformation (rotor) they are related to the intrinsic shape of the nucleus. For a spherical vibrational nucleus, a formally analogous expression can be obtained, but in this case the operators $\hat{\delta}_{\lambda,\mu}$ are to be understood as dynamical quantities, which produce surface vibrational excitations under the action of the potential exerted by the other nucleus.

Similarly, for the Coulomb interaction, we make use of the multipole expansion of the electrostatic interaction between the charge distribution of the projectile nucleus and that for the target (assumed here to be represented by a point-charge $Z_t e$ for simplicity):
\be
V_C(\bR, \xi)= \frac{Z_t Z_p e^2}{R} +  \sum_{\lambda> 0,\mu}   
\frac{4 \pi}{2 \lambda +1} \frac{Z_t e}{R^{\lambda+1}}  {\cal M}(E \lambda, \mu)  Y_{\lambda \mu}(\hat R)  ,
\ee
where ${\cal M}(E\lambda, \mu) \equiv e \sum_i^{Z_p} r_i^\lambda  Y_{\lambda \mu}^{*}(\hat r_i)$ is the {\it electric multipole operator}.

We include in the auxiliary potentials $U_i$ and $U_f$ the monopole parts of the nuclear and Coulomb interactions (those which cannot excite the nuclei), i.e. 
$$
U_i=U_f=V_N(R-R_0) + \frac{Z_t Z_p e^2}{R} ,
$$
and incorporate the $\lambda >0$ terms in the residual interaction $W= V_f - U_f$.  
The transition potentials for the nuclear and Coulomb parts of this residual interaction are, respectively,
\be
W^N_{if}(\bR)   =  - {d V_{N}(R-R_0) \over d R} 
 \sum_{\lambda > 0, \mu} \langle f; I_f M_f|  \hat{\delta}_{\lambda \mu} |i; I_i M_i \rangle  Y_{\lambda \mu}(\hat R) ,
\ee
and 
\be 
W^C_{if}(\bR)  =
 \sum_{\lambda > 0, \mu} 
\frac{4 \pi}{2 \lambda +1} \frac{ Z_t e}{  R^{\lambda+1} } 
\langle f; I_f M_f|{\cal M}(E \lambda, \mu)|i; I_i M_i \rangle 
Y_{\lambda \mu}(\hat R) .
\ee

The dependence on the spin projections of the structure matrix elements can be singled out using the Wigner-Eckart theorem. For example, for the Coulomb matrix element:
\be
{\langle f; I_f M_f|{\cal M}(E \lambda, \mu)|i; I_i M_i \rangle } =
(2I_f +1)^{-1/2} \langle  I_i M_i \lambda \mu | I_f M_f \rangle 
\langle f; I_f \|{\cal M}(E \lambda) \| i; I_i  \rangle , 
\ee
and the reduced matrix element is related to the electric reduced probabilities:
\be
B(E\lambda; I_i \rightarrow I_f ) =  (2I_i+1)^{-1} ~| \langle f; I_f \|{\cal M}(E \lambda, \mu) \| i; I_i  \rangle |^2 
\quad
(I_i \neq I_f)
\ee

Likewise, for the nuclear matrix elements,
\be
{\langle f; I_f M_f|{\delta}_{\lambda \mu}|i; I_i M_i \rangle } =
(2I_f +1)^{-1/2} \langle  I_i M_i \lambda \mu | I_f M_f \rangle 
\langle f; I_f \|{\delta}_{\lambda} \| i; I_i  \rangle , 
\ee
which are proportional to the reduced matrix elements of the deformation length operator. For a $I_i=0 \to I_f$ transition in a  even-even nucleus characterized by a deformation parameter $\beta_\lambda$, this reduced matrix element is simply given by $\langle f; I_f \|{\delta}_{\lambda} \| i; I_i  \rangle = \beta_\lambda R_0$, where $R_0$ is the average radius.

For a purely nuclear excitation process, with multipolarity $\lambda$, the DWBA amplitude  is  
\be
f^N_{f,i}(\theta)=\frac{\mu}{2 \pi \hbar^2}
 \langle f; I_f M_f|\hat{\delta}_{\lambda \mu}|i; I_i M_i \rangle 
\int d\bR ~ \chi_{f}^{(-)*}(\bK_f,\bR)
 \frac{d V_{N}}{d R} Y_{\lambda \mu}(\hat{R})
\chi_{i}^{(+)}(\bK_i,\bR) \, ,
\ee
whereas for a purely Coulomb excitation, also of multipolarity $\lambda$,
\be
f^C_{f,i}(\theta)=-\frac{\mu}{2 \pi \hbar^2}
\frac{4 \pi Z_t e}{2 \lambda +1}  
\langle f; I_f M_f|{\cal M}(E \lambda, \mu)|i; I_i M_i \rangle 
\int d\bR ~ \chi_{f}^{(-)*}(\bK_f,\bR)
\frac{ 1}{  R^{\lambda+1} }  Y_{\lambda \mu}(\hat{R})
\chi_{i}^{(+)}(\bK_i,\bR) .
\ee

\begin{figure}
\begin{center}
{\par \resizebox*{0.6\textwidth}{!}{\includegraphics[angle=0]{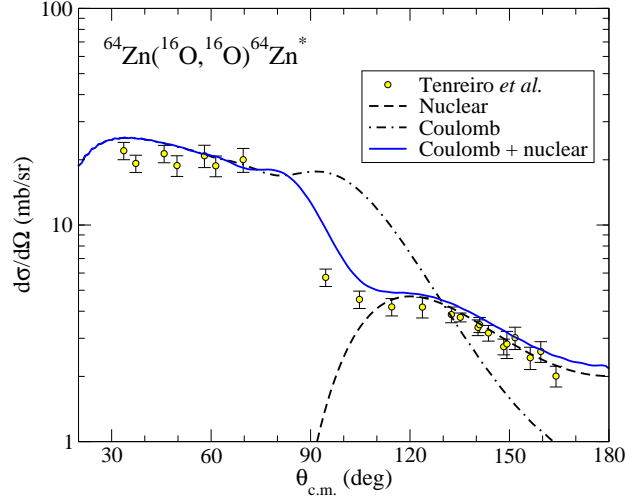}} \par}
\end{center}
\caption{\label{fig:o16zn}Differential inelastic cross section for the excitation of the first excited state of $^{64}$Zn ($E_x=0.992$~MeV, $I=2^+$) in the reaction $^{16}$O+$^{64}$Zn at $E=44$~MeV. The data from  ref.~\cite{Ten96} are compared with DWBA calculations based on a collective model of the target nucleus including nuclear, Coulomb and nuclear+Coulomb couplings. The structure parameters and potentials are those from \cite{Ten96}.}
\end{figure}

In general, both the nuclear and Coulomb potentials may contribute to the excitation mechanism and so the total scattering amplitude will be the coherent sum of both amplitudes and the differential cross section will be%
\footnote{Note that this expression correspond to definite initial and final spin projections. For unpolarized projectile and target, the actual cross section would correspond to a sum over the spin projections of the final nuclei, and an average over the initial ones \cite{Glen83}.} 
\be
  \left( \frac{d\sigma(\theta)}{d\Omega} \right)_{i\to f}= \frac{K_f}{K_i} 
 \left | f^N_{f,i} + f^C_{f,i} \right | ^2 .
\ee
This means that interference effects will arise at those angles for which the nuclear and Coulomb amplitudes are of the same order. These interference effects are illustrated in fig.~\ref{fig:o16zn} for the $^{64}$Zn($^{16}$O,$^{16}$O)$^{64}$Zn$^*$ reaction  at $E=44$~MeV, where a clear destructive interference  between the Coulomb and nuclear couplings is observed around 90$^\circ$.

\subsubsection{Few-body model}
Some nuclei exhibit a marked cluster structure. This is trivially  the case of the deuteron ($d=p+n$) but also of other nuclei, particularly in the light region of the nuclear chart. Some examples are \nuc{6}{Li}=$\alpha$+$d$, \nuc{7}{Li}=$\alpha$+$t$ and $^9$Be=$\alpha$+$\alpha$+n, among many others.  

If the separation energy between the clusters is small  compared to the cluster excitation energies, it is plausible to treat these clusters as inert objects and consider only possible excitations between them.  Using the Feshbach terminology, we include in the P space only the inter-cluster excitations.  In this way, we convert the many-body structure (and reaction) problem into a few-body problem. In this Feshbach reduction,  cluster-target interactions are described by effective potentials (complex in general) evaluated at the corresponding energy per nucleon. Additionally, the inter-cluster interaction is described with an effective potential tuned to describe the known properties of the projectile, such as the separation energy, spin-parity, rms radius, etc. 

Considering as an example the case of a two-body projectile, the projectile-target interaction will be described by the effective potential
\be 
 V(\bR,\xi) \equiv V(\bR,\br)= U_{1}(\br_1) + U_2(\br_2) ,
\ee
where $\br$ is the inter-cluster coordinate and $\vecr_i$ the cluster-target coordinates. 
Note that, in this model, the internal variables $\xi$ are represented by the relative coordinate $\vecr$. 

To apply the CC or DWBA methods we need to evaluate the coupling potentials
\be
 V_{n,n'}(\vecR)=\int d\vecr \, \phi_{n}^{*}(\br)\left[U_{1}(\br_1) + U_2(\br_2)\right] \phi_{n'}(\br) .
\ee

\begin{figure}
\begin{center}
\begin{minipage}[c]{.35\linewidth}
\includegraphics[width=0.95\columnwidth]{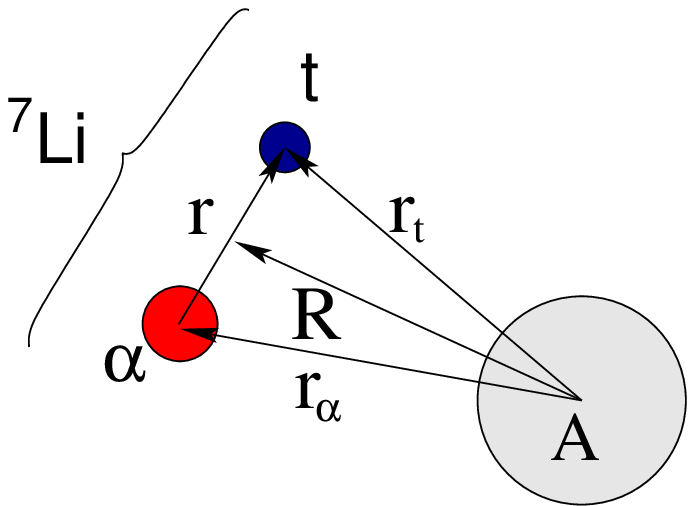} 
\end{minipage}
\begin{minipage}[c]{.61\linewidth}
\begin{center}
\includegraphics[width=0.89\columnwidth]{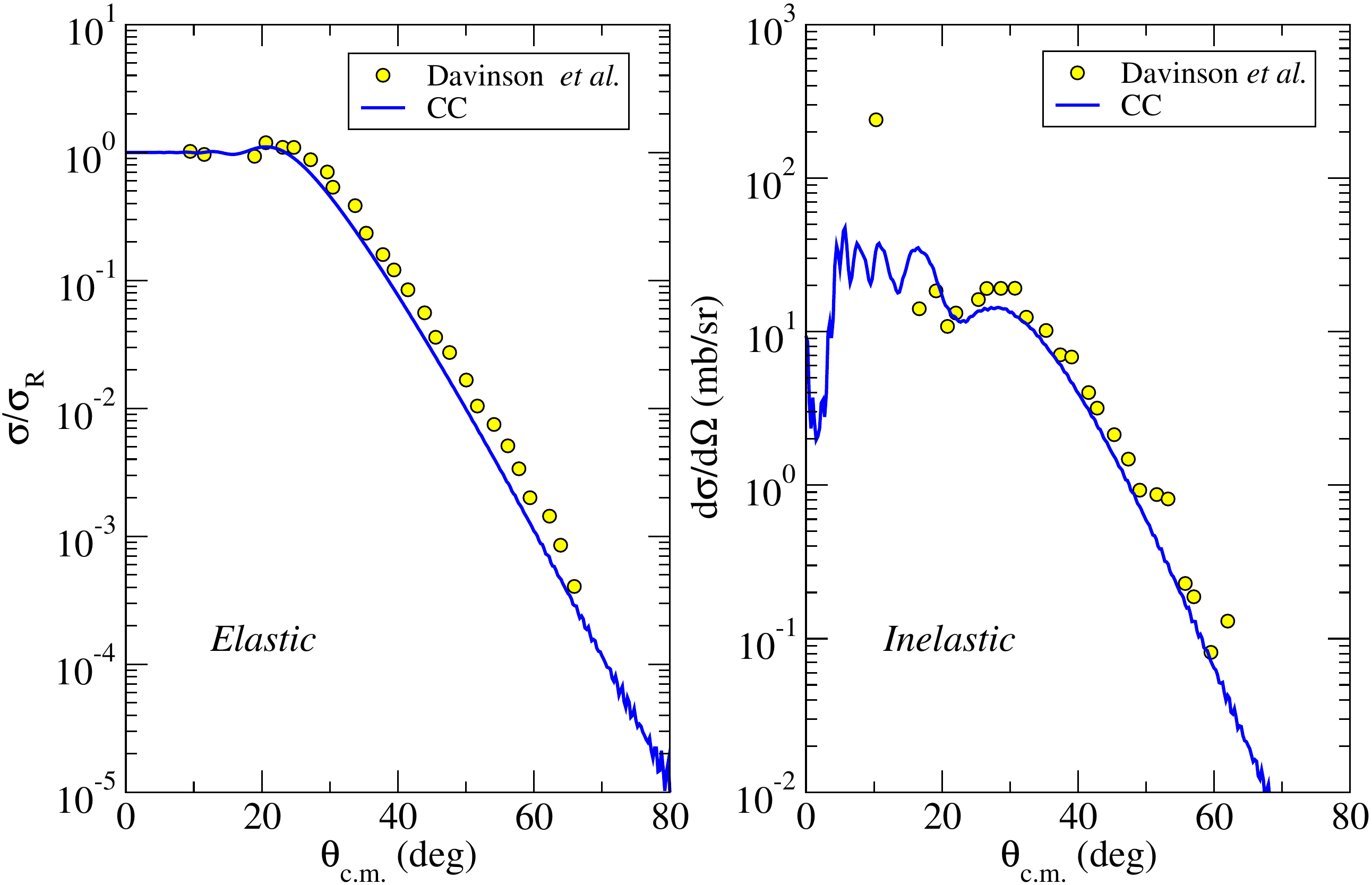} 
\end{center}
\end{minipage}
\end{center}
\caption{\label{fig:li7pb_cc} Left:  Relevant coordinates for a $^{7}$Li+$A$ reaction, using a two-body cluster model of $^{7}$Li$ (\alpha$+$t$). Right:  Application of the CC formalism to the \nuc{7}{Li}+\nuc{208}{Pb} reaction at $E_\text{lab}=68$~MeV, the two-body cluster model for \nuc{7}{Li}.  The left and right cross sections are for  the elastic and inelastic (excitation of $1/2^-$ excited state of $^{7}$Li) differential cross sections. Experimental data are from ref.~\cite{Dav84}.}
\end{figure}

As an example, consider the scattering of $^7$Li, treated as $\alpha$+$t$, by a target $A$ (see fig.~\ref{fig:li7pb_cc}).  In this model, the $^7$Li ground-state  ($3/2^-$) can be interpreted assuming that the $\alpha$ ($0^+$) and $t$ ($1/2^+$) clusters are in a $\ell=1$ state of relative motion. Analogously, the first excited state ($1/2^-$) can be interpreted also  assuming a $\ell=1$ configuration and so it would correspond to a spin-orbit partner of the ground-state level. The  wave functions for these states would be obtained from a single-channel \sch  equation
\be
[\hat{T}_\br + V_{\alpha-t}(\br) - \varepsilon_n ]\phi_n(\br)=0 ,
\ee
where, for the ground-state ($n=0$), $\varepsilon_0=-2.47$~MeV. 

A successful application of such a model to elastic and inelastic scattering of \nuc{7}{Li}+\nuc{208}{Pb} at 68 MeV is shown in fig.~\ref{fig:li7pb_cc}, where the calculations are compared with the data from ref.~\cite{Dav84}. In this case, the model space was restricted to the two bound states of $^{7}$Li. In fact, the application of the CC method to unbound  states (resonant or non-resonant) requires appropriate extensions of the method, as explained in the following section. 

\section{Breakup reactions (I): quantum-mechanical approach}
\subsection{The CDCC method}

If one of the colliding partners is weakly-bound and is excited above its breakup threshold, the system will become unbound and will eventually dissociate into two or more fragments (recall the $^{7}$Li($p$,$p'$) example of fig.~\ref{fig:pli7}). This will be the case of halo nuclei, an example of which is the $^{6}$He nucleus already discussed. If we are interested in the description of these breakup channels, our modelspace must be augmented to include, at least, part of them. A way of doing that is by means of the coupled-channels (CC) method. However, direct application of this method, as introduced in the previous section, is not possible because (i) the breakup states are continuous in energy, thereby leading to an infinite number of states and (ii) the positive-energy wave functions, unlike those for bound states, do not vanish at large distances, presenting an oscillatory asymptotic behaviour. Consequently, they can not be normalized. Coupling potentials calculated with this kind of functions will also oscillate at large distances, posing severe problems to the standard methods of solution of the coupled equations.

\begin{figure}
\begin{center}
\begin{minipage}[t]{.48\linewidth}
\includegraphics[width=0.85\columnwidth]{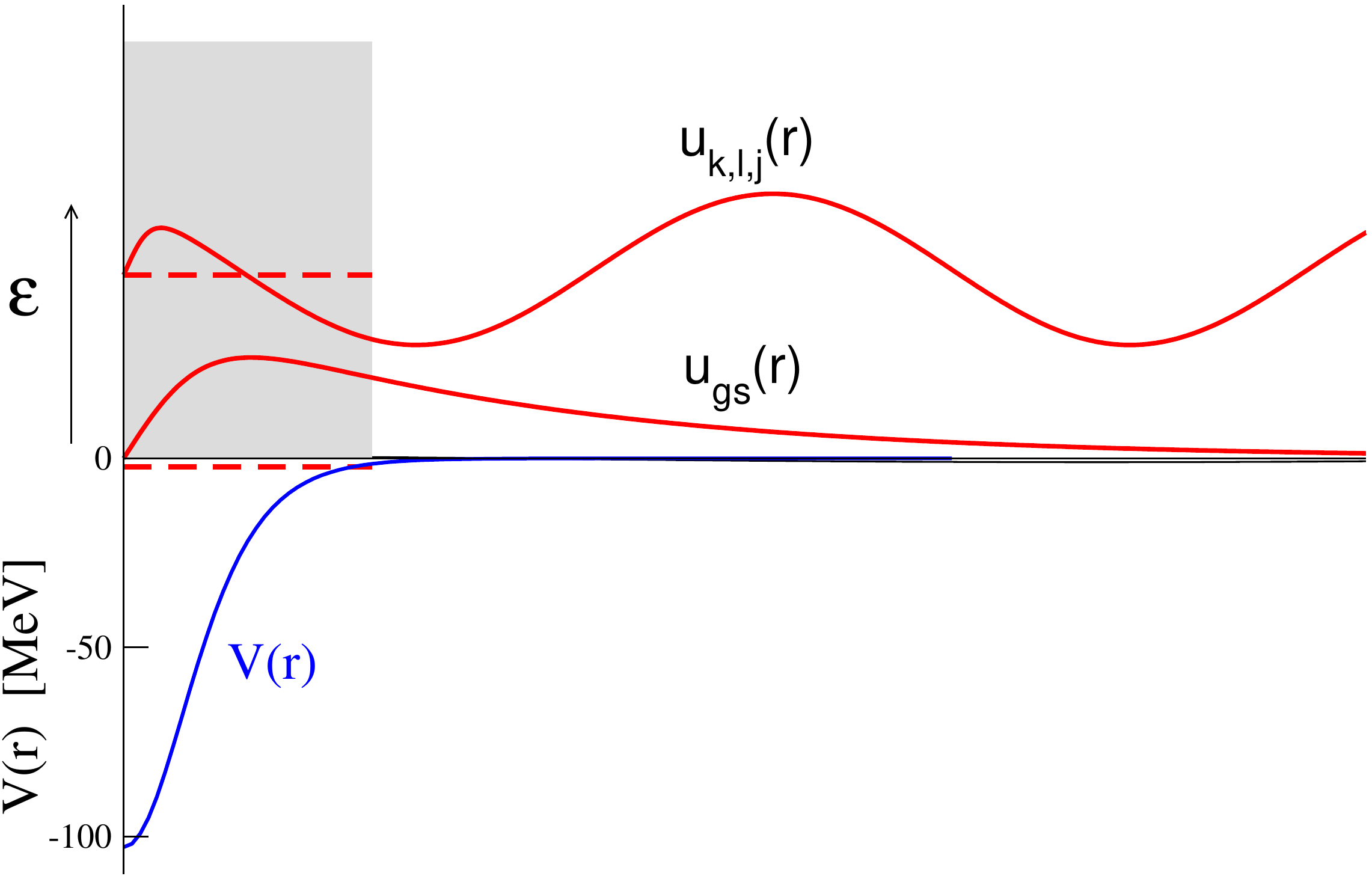} 
\end{minipage}
\begin{minipage}[t]{.48\linewidth}
\includegraphics[width=0.85\columnwidth]{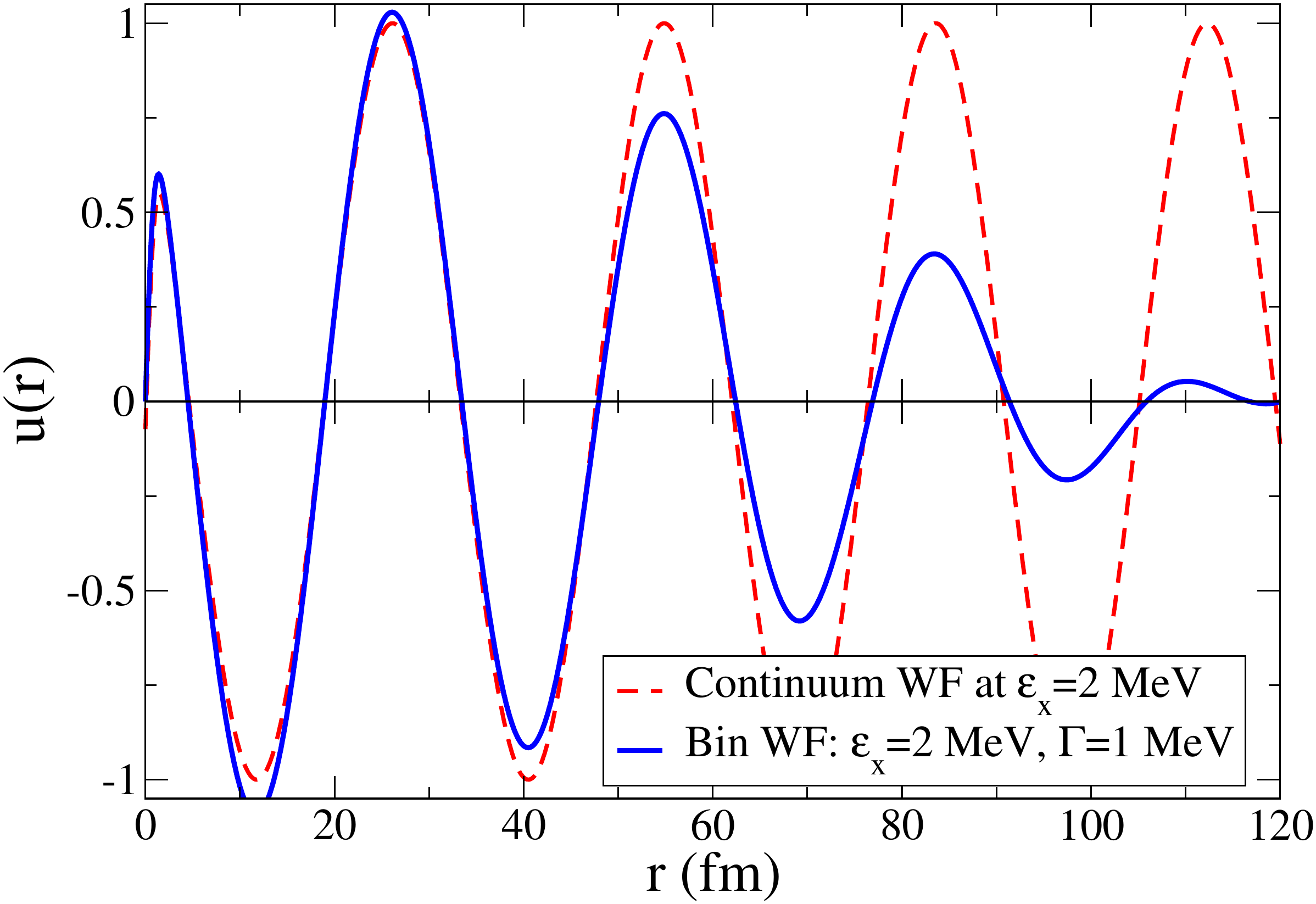}
\end{minipage}
\caption{\label{fig:wfbin}Left: The $p$-$n$ potential showing the bound state wave function (deuteron) and an unbound state wave function. Right: Comparison of a scattering state with a {\it bin} wave packet centered around the same nominal energy and with a width of 1~MeV.}
\end{center}
\end{figure}

\begin{figure}
\begin{center}
\begin{minipage}[t]{.47\columnwidth}
\includegraphics[width=0.85\columnwidth]{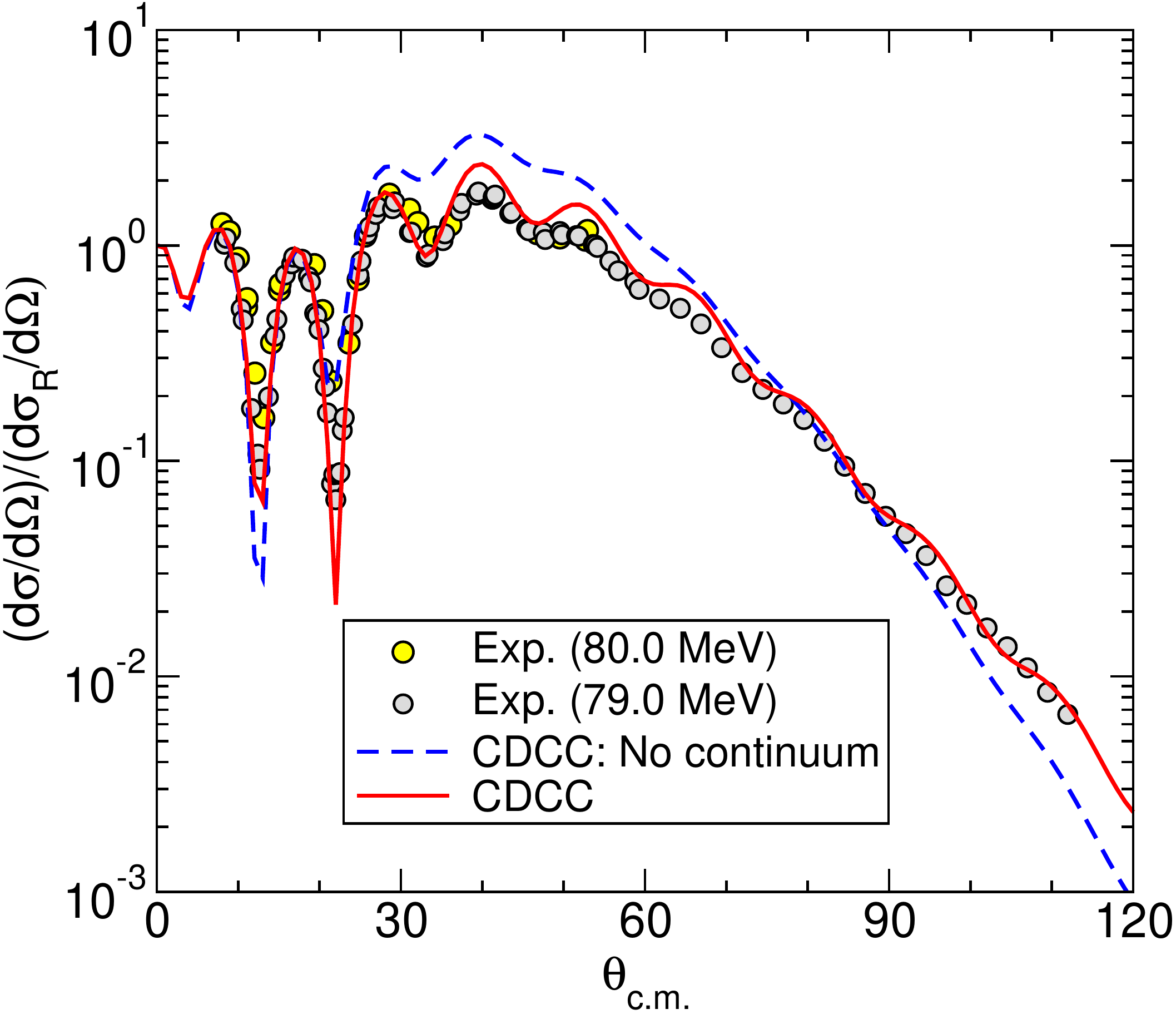} 
\end{minipage}
\begin{minipage}[t]{.47\columnwidth}
\includegraphics[width=0.85\columnwidth]{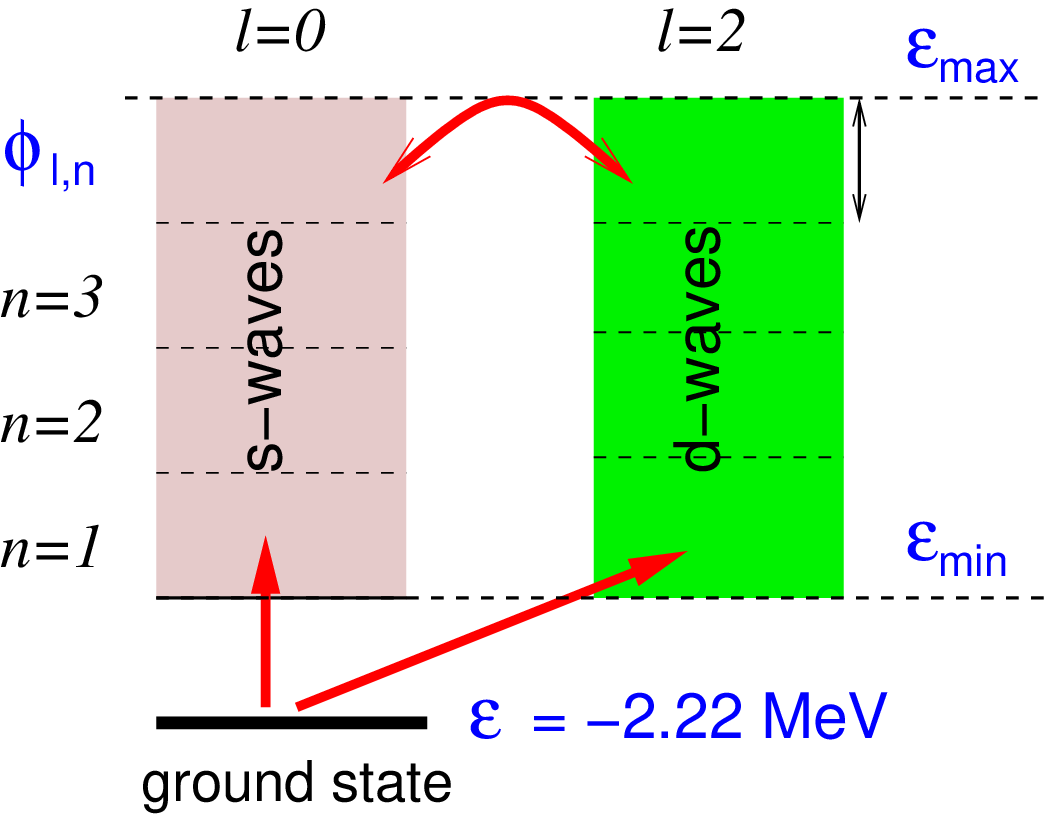}
\end{minipage}
\caption{Left: Application of the CDCC method to $d$+$^{58}$Ni elastic scattering at $E_d=80$~MeV. The solid line is the full CDCC calculation. The dashed line is the calculation omitting the breakup channels. The calculations were performed ignoring the internal spins of the proton and neutron and so $\ell=j$. Right: Illustration of continuum discretization for the same problem.}
\label{fig:dni_cdcc}
\end{center}
\end{figure}
These difficulties motivated the development of the continuum-discretized coupled-channels (CDCC) method. 
This method was originally introduced by G.\ Rawitscher \cite{Raw74} and later refined by the Pittsburgh-Kyushu collaboration \cite{Yah86,Aus87} to describe the effect of the breakup channels on the elastic scattering of deuterons. Denoting the reaction by $a+A$, with $a=b+x$ (referred hereafter as the {\it core} and {\it valence} particles, respectively), the method assumes the effective three-body Hamiltonian
\begin{equation}
\label{eq:Heff}
H= H_\mathrm{proj} + \hat{T}_{\bR}   + U_{bA}(\br_{bA}) + U_{xA}(\br_{xA}) ,
\end{equation}
with $H_\mathrm{proj}=T_{\br}+V_{bx}$ the projectile internal Hamiltonian,   $\hat{T}_{\br}$ and $\hat{T}_{\bR}$ are kinetic energy operators, $V_{bx}$ the inter-cluster interaction and $U_{bA}$ and $U_{xA}$ are the core-target and valence-target optical potentials (complex in general) describing the elastic scattering of the corresponding $b+A$ and $x+A$ sub-systems, at the same energy per nucleon of  the incident  projectile. 
In the  CDCC  method  the  three-body wave function of the system is expanded in terms of the eigenstates of the Hamiltonian $H_\mathrm{proj}$ including both bound and unbound states. Since the latter form a continuum, a procedure of discretization is applied, consisting in representing this continuum by a finite and discrete set of square-integrable functions. In actual calculations, this continuum must be truncated in excitation energy and limited to a 
finite number of partial waves $\ell$ associated to the relative co-ordinate $\vec r$. Normalizable states representing the continuum should be obtained for each $\ell,j$ values. Two main methods are used for this purpose:
\bi
\item {\it The pseudo-state method}, in which the $b+x$ Hamiltonian is diagonalized in a basis of square-integrable functions, such as Gaussians \cite{Kaw86a} or transformed harmonic oscillator functions \cite{Mor09b}. Negative eigenvalues correspond to the bound states of the systems, whereas positive eigenvalues are regarded as a finite representation of the continuum. 
\item {\it The binning method}, in which  normalizable states are obtained by constructing a wave packet ({\it bin}) by linear superposition of the actual continuum states over a certain energy interval \cite{Aus87}.
\ei


We describe this latter method in some more detail.  Assuming for simplicity a spinless core, these discretized functions are denoted as 
%
\begin{align}
\label{eq:PsiCDCC}
 \phi_i (\br)  = {u_i(r) \over r} [Y_{\ell_i}(\hat{r}) \otimes \chi_s  ]_{j_i m_i} ,
\end{align}
where $i \equiv \{[k_i,k_{i+1}] \ell_i s j_i m_i \}$ specifies the $i$-th bin, with $[k_i,k_{i+1}]$ the wave number  interval of the bin, $\ell$ the valence-core orbital angular momentum, $s$ the valence spin, and $\vec{j}=\vec{\ell}+ \vec{s}$ the total angular momentum. 
The symbol $\otimes$ denotes angular momentum coupling. The radial part of the bin is obtained as a linear combination (i.e.\ a wave packet) of  scattering states  as 
\be
u_{i} (r) = \sqrt {\frac{2 }{\pi N_i}} ~~
            \int _ {k _ i} ^ {k _ {i+1}} w_i(k) u _{k,\ell_i s j_i} (r) dk ,
\ee
where $w_i(k)$  is a weight function (for non-resonant continuum $w(k)$ is usually taken as $e^{i \delta_{\ell_i}}$, where $\delta_{\ell_i}$ are the phase shifts of the scattering states within the bin") and $N_i$ is a normalization constant.  The effect of this averaging is to damp the oscillations at large distances, making the bin wave function normalizable (see fig.~\ref{fig:wfbin}).  

Assuming a single bound state for simplicity, the CDCC wave function reads
\be
\Psi^{\mathrm{CDCC}}(\bR,\br) =\chi_{0}^{(+)}(\bR) \phi_{0}(\br) + \sum_{i=1}^{N} \chi_{i}^{(+)}(\bR) \phi_{i}(\br) ,
\label{PsiCDCC}
\ee 
where the index $i=0$ denotes the ground state of the $b+x$ system.  

This model wave function  must verify the Schr\"odinger equation: $[H-E] \Psi^{\mathrm{CDCC}}(\bR,\br)=0$. This gives rise to a set of coupled differential equations similar to that of eq.~(\ref{eq:cc}) with the coupling potentials given by 
\be
\label{eq:Vij_3bCDCC}
U_{ij}(\vecR) = \int d\vec r \; \phi^*_{i}(\br)  \left [U_{bA}+U_{xA} \right ]  \phi_{j}(\br) \, .
\ee

\begin{figure}
\begin{center}
\includegraphics[width=0.5\columnwidth]{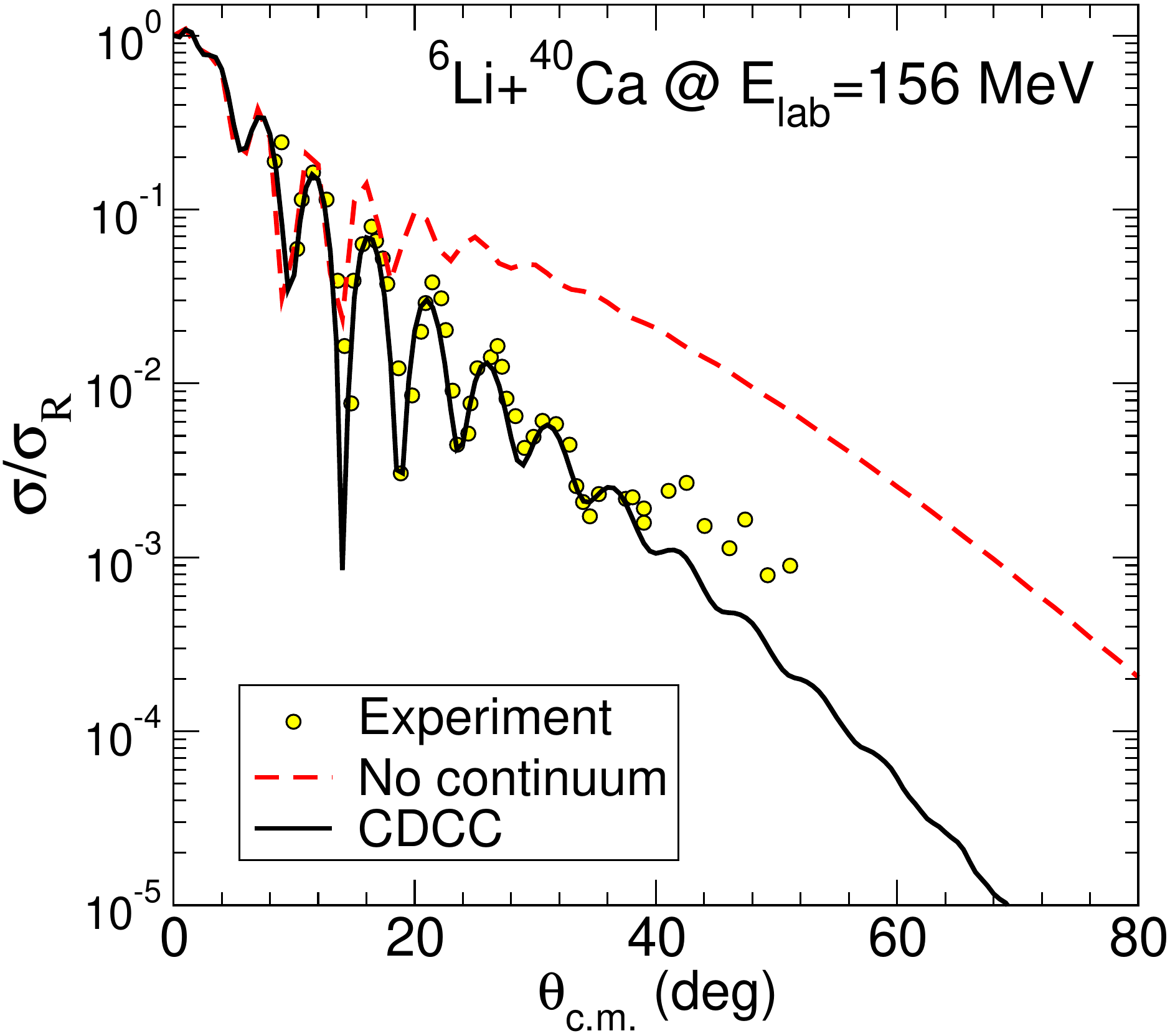} 
\caption{\label{fig:li6_cdcc}Application of the CDCC method to $^6$Li+$^{40}$Ca elastic scattering at 156 MeV. Experimental data are from ref.~\cite{Maj78}.}
\end{center}
\end{figure}

The standard CDCC method is based on  a strict three-body model of the reaction ($b+x+A$), and has proven  rather successful to describe elastic and  breakup cross sections of deuterons and other weakly bound two-body nuclei, such as $^{6,7}$Li and  $^{11}$Be (see fig.~\ref{fig:li6_cdcc}). However,  it has  limitations. The assumption of inert bodies is not always justified, since excitations of the projectile constituents ($b$ and $x$) and of the target ($A$) may take place along with the projectile dissociation. Furthermore, the two-body picture may be inadequate for some nuclei as, for example, in the case of the  Borromean systems (e.g.~$^{6}$He,  $^{11}$Li).  Some extensions of the CDCC method to deal with the these situations are outlined below.

\subsubsection{Inclusion of core and target excitations\label{sec:corex}}
Excitations of the projectile constituents ($b$ and $x$ in our case) may take place concomitantly with the projectile breakup. This mechanism is neglected in the standard formulation of the CDCC method. For example, for the scattering of halo nuclei, collective excitations of the core $b$ may be important. 
These core excitations will affect both the structure of the projectile as well as the reaction dynamics.  In the inert core picture, the projectile states will correspond to pure single-particle or cluster states but, if the core is allowed to excite, these states will contain in general admixtures of core-excited components. Additionally, the interaction of the core with the target will produce excitations and deexcitations of the former during the collision, and this will  modify the reaction observables to some extent. These two effects (structural and dynamical) have been recently investigated within extended versions of the DWBA and CDCC methods \cite{Cre11,Mor12,Sum06,Die14}. For example, considering only possible excitations of $b$, the  effective three-body Hamiltonian is generalized as follows:
\begin{equation}
\label{eq:Heff_tarx}
H= H_{\rm proj}(\br,\xi_b)   + \hat{T}_{\bR} + U_{bA}(\br_{bA},\xi_b) + U_{xA}(\br_{xA}) .
\end{equation}
The potential $U_{bA}(\br_{bA},\xi_b)$ is meant to describe both elastic and inelastic scattering  of the $b+A$ system (for example, it could be represented by a deformed potential as discussed in the context of inelastic scattering with collective models). 
Note that the core degrees of freedom ($\xi_b$) appear in the projectile Hamiltonian (structure effect) as well as in the core-target interaction (dynamic effect). 

In the weak coupling limit, the projectile Hamiltonian can be written more explicitly as
\be
H_{\rm proj}= \hat{T}_{\br} + V_{bx}(\br,\xi_b) + h_{\rm core}(\xi_b) , 
\label{Hproj}
\ee
where $h_{\rm core}(\xi_b)$ is the internal Hamiltonian of the core. The eigenstates of this Hamiltonian are of the form 
\be
\phi_{i}(\xi) 
\equiv \sum_{\alpha}  
\left[   \varphi_\alpha(\br) \otimes \Phi_{I}(\xi_b) \right]_{j_p m_p} ,
\label{wfrot}
\ee
where $i$ is an index labeling the states with angular momentum $j_p m_p$,  $\xi \equiv \{\xi_b, \br \}$,  $\alpha \equiv \{\ell,s, j,I \}$, with $I$ the core intrinsic spin, $\vec{j}=\vec{\ell} + \vec{s}$ and $\vec{j}_p=\vec{j}+\vec{I}$. 
The functions $\Phi_{I}(\xi_b)$  and  $\varphi_\alpha(\br)$ describe, respectively, the core states and the valence--core relative motion.  For continuum states, a procedure of continuum discretization is used. 

Once the projectile states (\ref{wfrot}) have been calculated, the three-body wave function is expanded in a basis of such  states, as in the standard CDCC method.  Early calculations using this extended CDCC method (XCDCC) were first performed by Summers {\it et al.}~\cite{Sum06,Sum07} for  $^{11}$Be and $^{17}$C on $^{9}$Be and  $^{11}$Be+$p$, finding a very little core excitation effect in all these cases. However, later calculations for the $^{11}$Be+$p$ reaction based on a alternative implementation of the XCDCC method using a pseudo-state representation of the projectile states \cite{Die14} suggested much larger effects.  The discrepancy was found to be due to an inconsistency in the numerical implementation of the XCDCC formalism presented in ref.~\cite{Sum06}, as clarified in \cite{Sum14}. For heavier targets, such as $^{64}$Zn or $^{208}$Pb, the calculations of \cite{Die14} suggest that the core excitation mechanism plays a minor role, although its effect on the structure of the projectile is still important.  

\begin{figure}
\begin{center}
\begin{minipage}[t]{.59\linewidth}
\includegraphics[width=0.9\columnwidth]{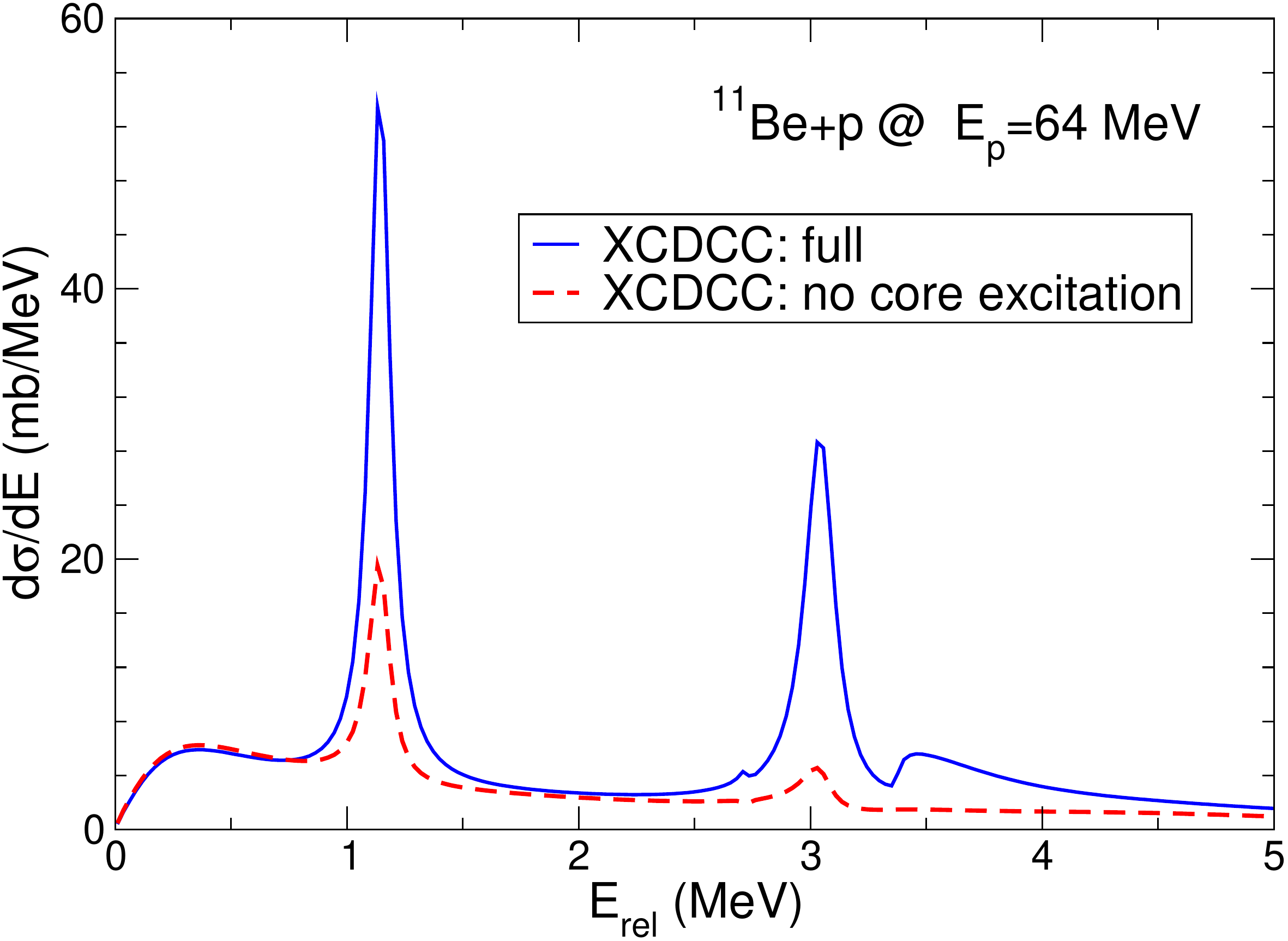} 
\end{minipage}
\begin{minipage}[t]{.4\linewidth}
\begin{center}\includegraphics[width=0.75\columnwidth]{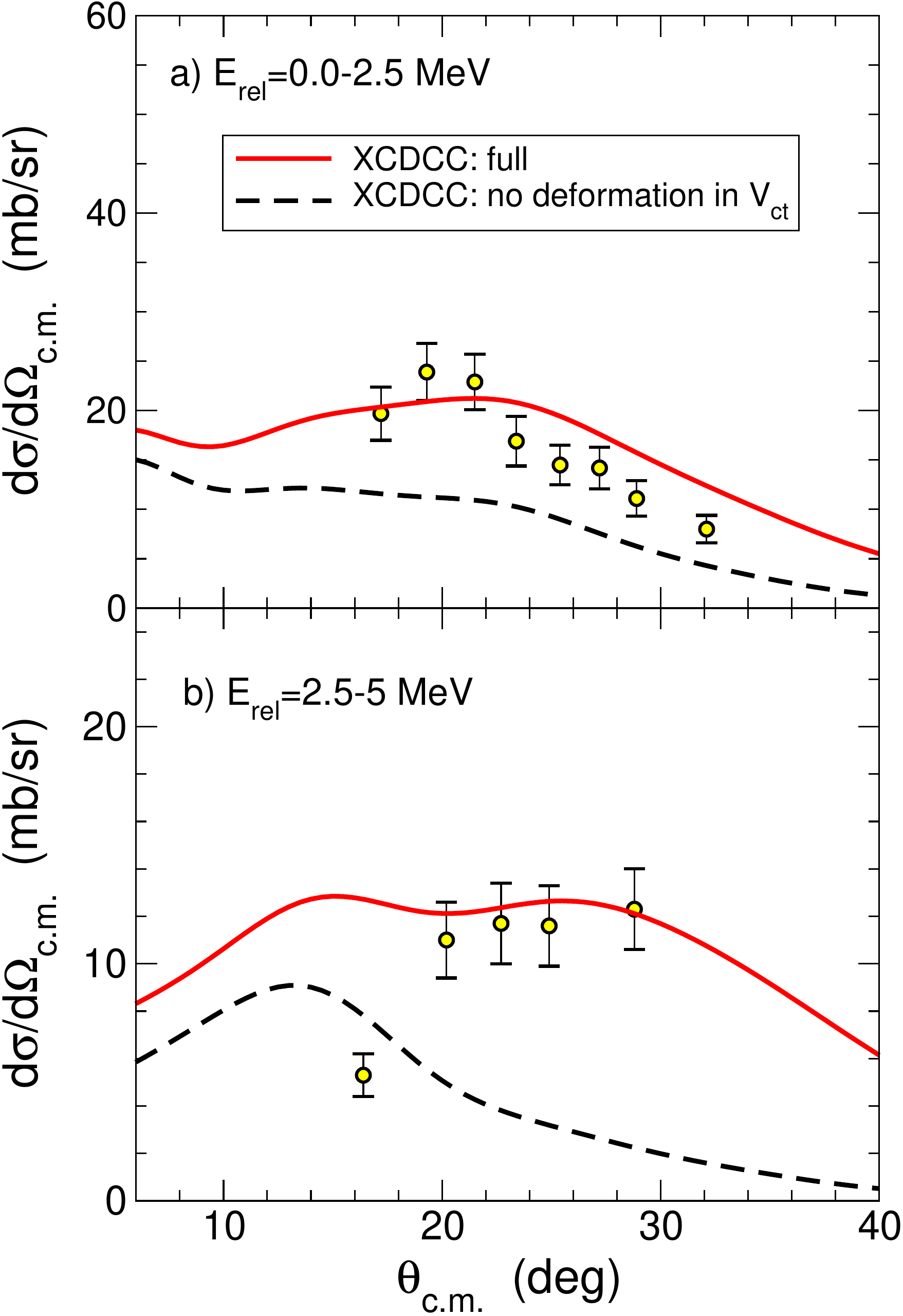}\end{center}
\end{minipage}
\caption{\label{fig:be11p_xcdcc}Left: Differential breakup cross  sections, with respect to the $n$-$^{10}$Be relative energy,  for the breakup of $^{11}$Be on protons at 63.7 MeV/nucleon. Right: breakup angular distributions for the relative-energy cuts indicated in the labels.}
\end{center}
\end{figure}





 As an example of these XCDCC calculations we show in the left panel of fig.~\ref{fig:be11p_xcdcc} the differential breakup cross section, as a function of the $n$-$^{10}$Be relative energy,  for the reaction $^{11}$Be+$p$ at 63.7~MeV/nucleon. Details of the structure model and potentials as given  in ref.~\cite{Die14}. Continuum states with angular momentum/parity $j_p=1/2^\pm$, $3/2^\pm$ and $5/2^+$ were included using a  pseudostate representation in terms of transformed harmonic oscillator (THO) functions \cite{Lay12}. To get a smooth function of the energy, the calculated differential cross sections were then convoluted with the {\it actual} scattering states of $H_\mathrm{proj}$. The two peaks at $\varepsilon=1.2$ and 3.2~MeV correspond to ${5/2}^+$ and ${3/2}^+$ resonances, respectively. The solid line is the full XCDCC calculation, including the $^{10}$Be deformation in the structure of the projectile as well as in the projectile-target dynamics. The dashed line is the XCDCC calculation omitting the effect of the core-target  excitation mechanism. It is clearly seen that the inclusion of this mechanism increases significantly the breakup  cross sections, particularly in the region of the $3/2^+$ resonance, owing to the dominant $^{10}$Be(2$^+$)$\otimes$$2s_{1/2}$ configuration of this resonance \cite{Lay12,Cre11,Mor12} The right graph shows two angular distributions, corresponding to the relative-energy intervals indicated in the labels, compared with the XCDCC calculations with (solid) and without (dashed) core excitations.


\begin{figure}
\begin{center}
\begin{minipage}[c]{.32\textwidth}
{\par \resizebox*{0.85\textwidth}{!}
{\includegraphics{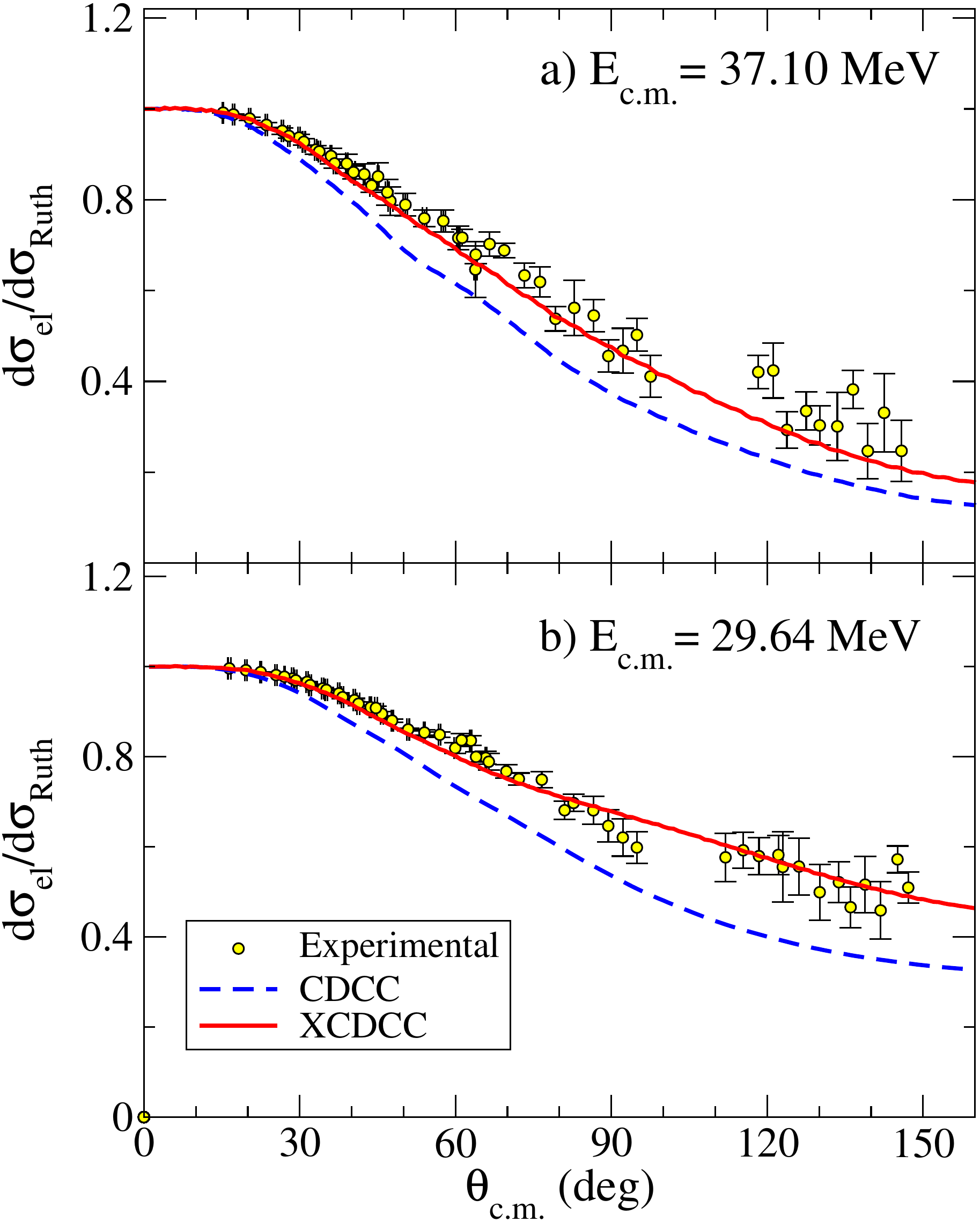}} \par}
\end{minipage}
\begin{minipage}[c]{.32\textwidth}
{\par \resizebox*{0.85\textwidth}{!}
{\includegraphics{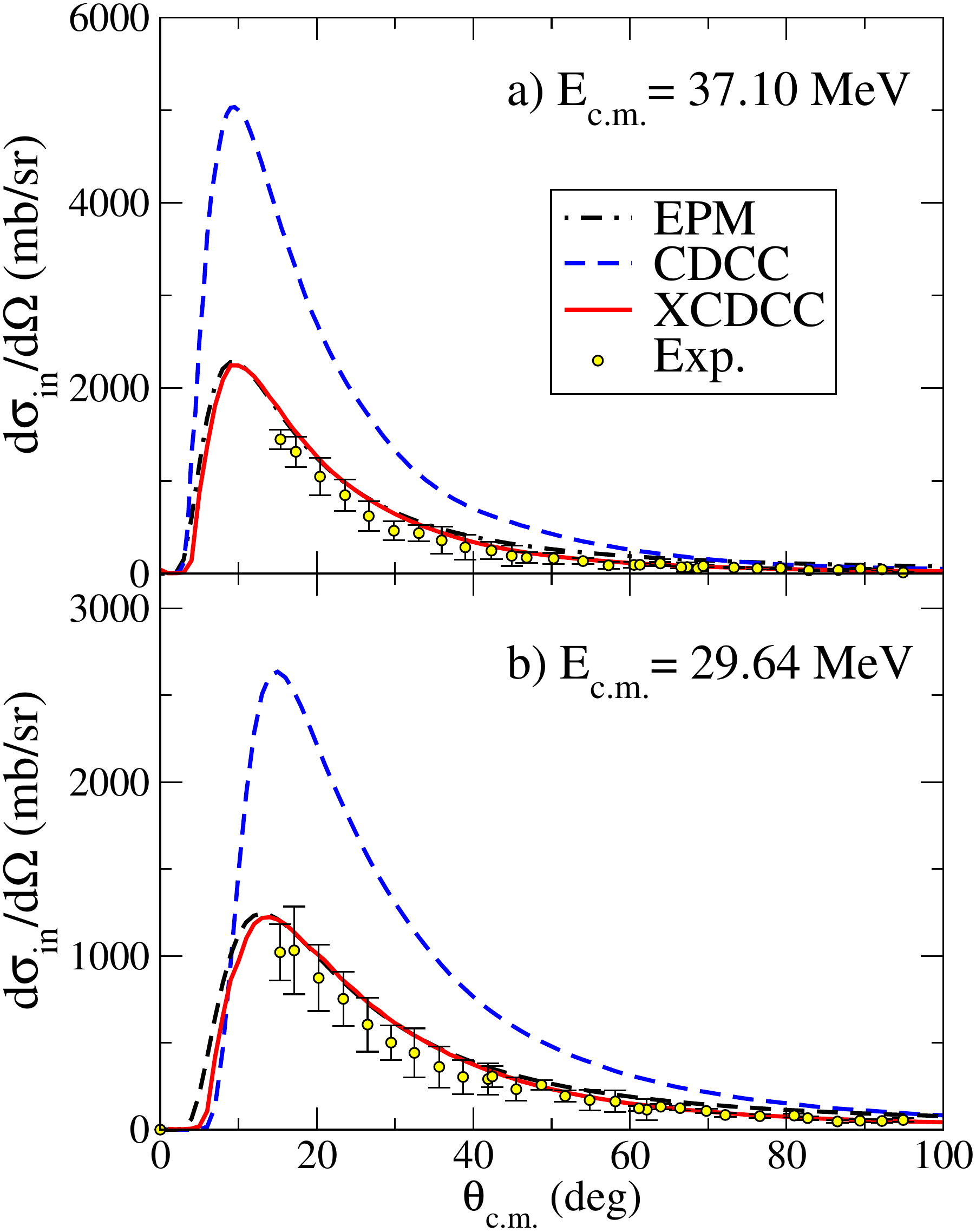}} \par}
\end{minipage}
\begin{minipage}[c]{.32\textwidth}
{\par \resizebox*{0.85\textwidth}{!}
{\includegraphics{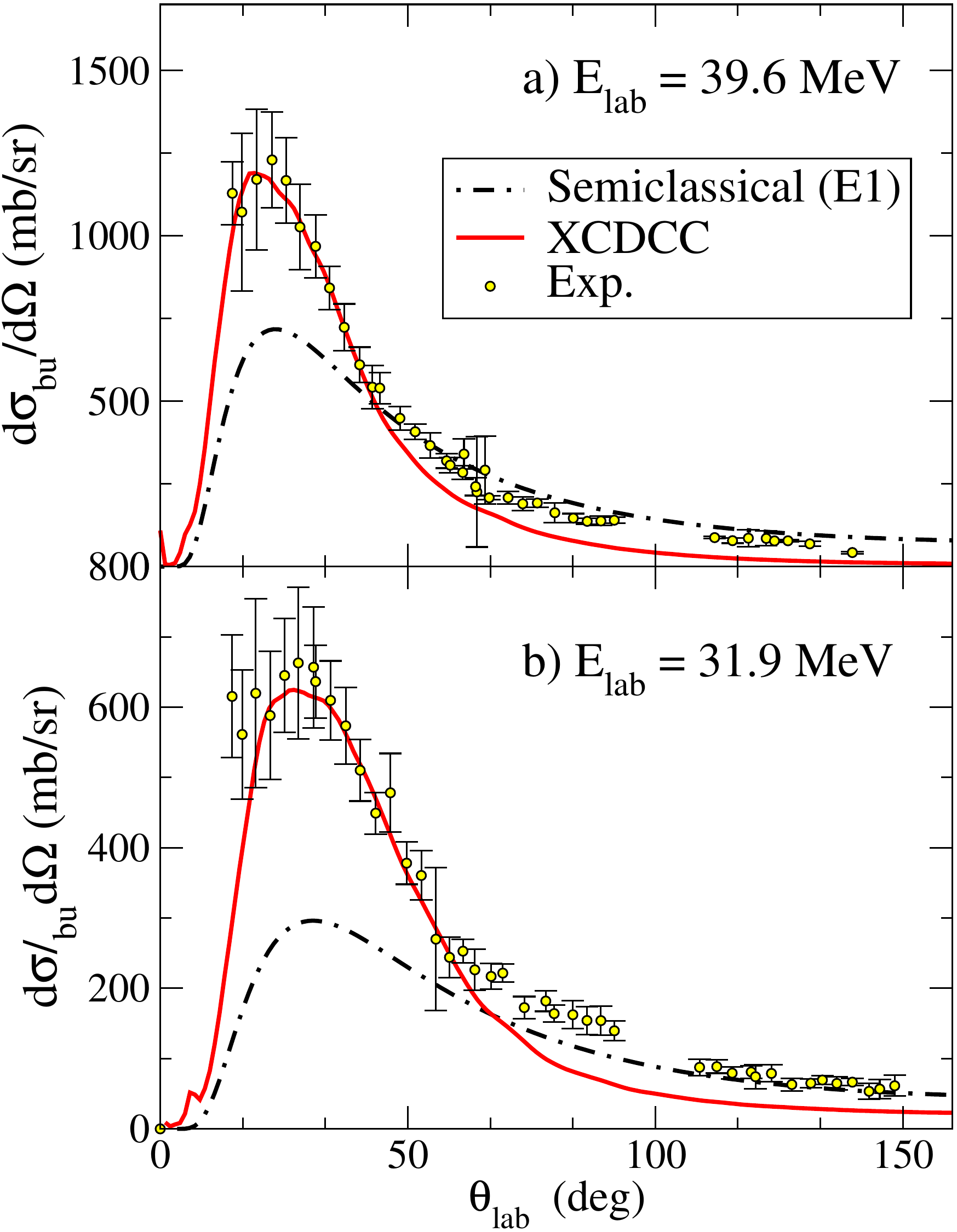}} \par}
\end{minipage}
\caption{\label{fig:be11au}Elastic (left), inelastic (middle) and breakup (right) cross sections for the reaction $^{11}$Be+$^{197}$Au at near barrier energies ($V_b \sim 40$~MeV). Experimental data are compared with CDCC, CDCC and first-order semiclassical pure $E1$ calculations (labeled EPM). Adapted from ref.~\cite{Pes17}.}
\end{center}
\end{figure}

The importance of the deformation on the structure of the projectile is clearly evidenced in the elastic and inelastic scattering of $^{11}$Be on $^{197}$Au at energies around and below the Coulomb barrier \cite{Pes17}, shown in fig.~\ref{fig:be11au}. XCDCC calculations based on the particle-plus-rotor model (solid lines) are able to reproduce simultaneously the elastic, inelastic and breakup angular distributions. By contrast, standard CDCC calculations using single-particle wave functions fail to describe the elastic and inelastic data, even describing well the breakup. This is due to the overestimation of the $B(E1)$ connecting the ground state with the bound excited state, as shown in fig.~\ref{fig:be11_dbde}.

A  simpler DWBA, no-recoil  version of the formalism (XDWBA) has been also proposed in  refs.~\cite{Cre11,Mor12}. An application of this formalism to the $^{11}$Be+$^{12}$C reaction at 69~MeV/u showed that the core excitation mechanism may interfere with the single-particle excitation mechanism, producing a conspicuous effect on the interference pattern of the resonant breakup angular distributions \cite{Mor12b}.

In addition to the excitations of the projectile constituents, excitations of the target nucleus may also take place and compete with the projectile breakup mechanism. Note that, within CDCC, the projectile breakup is treated as a inelastic excitation of the projectile to its continuum states and, thus, inclusion of target excitation amounts at including, simultaneously, projectile plus target excitations so their relative importance, and mutual influence, can be assessed. These target excitations can be treated with the collective models mentioned in sec.~\ref{sec:inelmodel}. It is worth noting that, within this three-body reaction model, target excitation arises from the non-central part of the valence-target and core-target interactions.  To incorporate this effect, the  effective Hamiltonian, eq.~(\ref{eq:Heff}), must be now generalized as:  
\begin{equation}
\label{eq:Heff_tarx}
H= \hat{T}_{\br} + \hat{T}_{\bR} + V_{bx} + U_{bA}(\br_{bA},\xi_t) + U_{xA}(\br_{xA},\xi_t)
\end{equation}
in which the $b-A$ and $x-A$ interactions depend now, in addition to the corresponding relative coordinate, on the target degrees of freedom (denoted as $\xi_t$).  
 Ideally, these $U_{xA}$ and $U_{bA}$ potentials should reproduce simultaneously  the elastic and inelastic scattering for the $x+A$ and $b+A$ reactions, respectively.

The explicit inclusion of target excitation was first done by the Kyushu group in the 1980s \cite{Yah86}, which considered the case of deuteron scattering. The motivation was to compare the roles of target-excitation and  deuteron breakup  in the elastic and inelastic scattering of deuterons. They applied the formalism to the $d$+$^{58}$Ni reaction at $E_d=22$ and 80~MeV, including the ground state and the first excited state of $^{58}$Ni ($2^+$) and finding that, in this case, the deuteron breakup process is more important than the target-excitation. 

\begin{figure}
{\par\centering \resizebox*{0.55\textwidth}{!}{\includegraphics{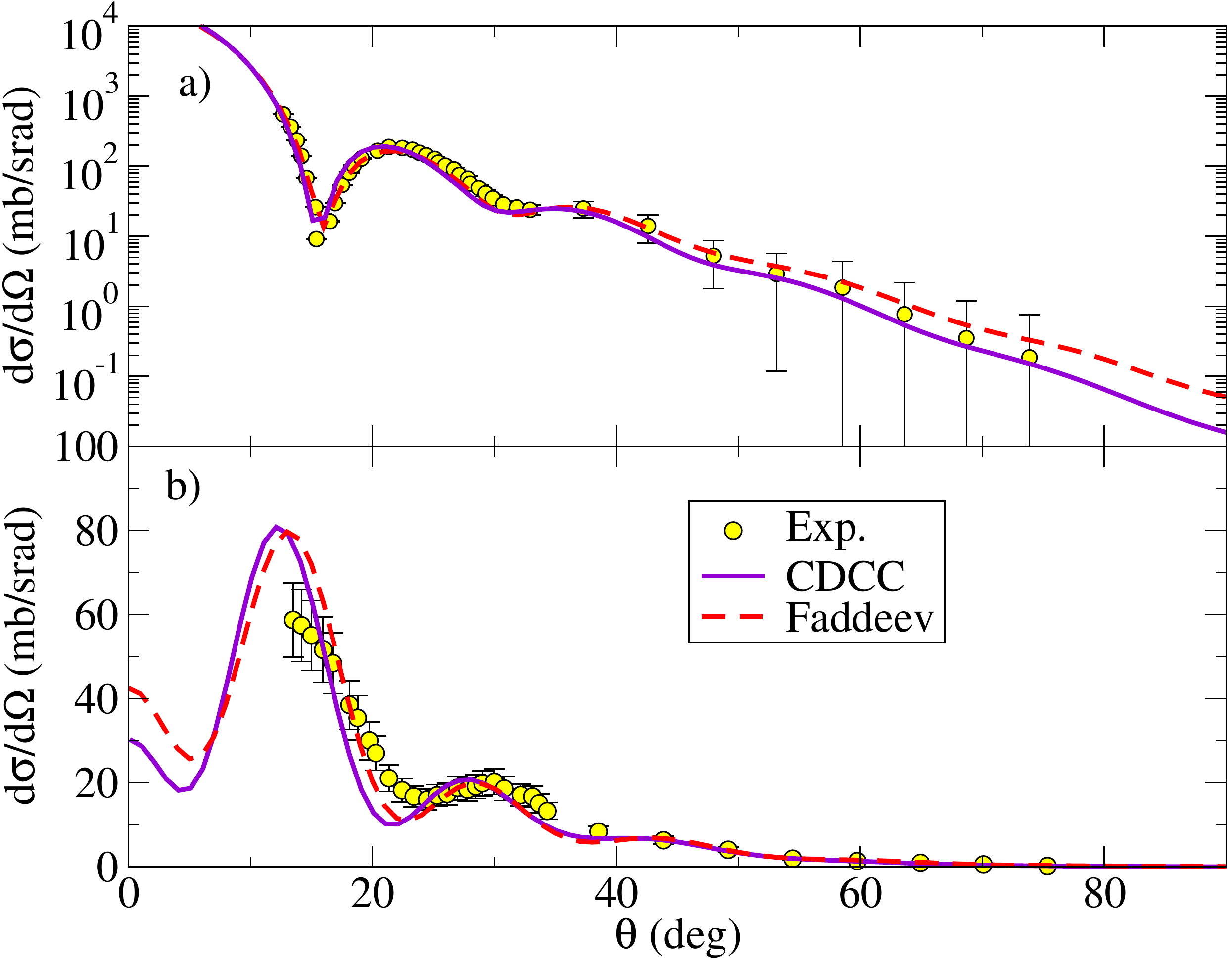}}\par}
 \caption{\label{fig:24Mg}Elastic (upper) and  $^{24}$Mg$(2^+)$ excitation (lower) differential angular cross sections for $d$+$^{24}$Mg $E_d=70$ ${\rm MeV}$. CDCC calculations are compared with Faddeev calculations from ref.~\cite{Del16} and with the data of ref.~\cite{Kis76}. The calculations  employ CH89 \cite{CH89} parameterization for the $p$+$^{24}$Mg and $n$+$^{24}$Mg potentials and treat the target excitation with a collective model, assuming a deformation of $\beta=0.5$  for the $^{24}$Mg nucleus.  The plot is adapted from ref.~\cite{Gom17a}.}
\end{figure}

Recently, the problem has been also addressed by some authors \cite{Chau11,Gom17a}, also in the context of deuteron elastic and inelastic scattering. 
A recent application of the formalism is shown in fig.~\ref{fig:24Mg}, which corresponds to the reaction $^{24}{\rm Mg}(d,d)^{24}{\rm Mg}^*$ at  $E_d=70$~MeV, including the ground and first excited states of $^{24}{\rm Mg}$, in addition to the deuteron breakup. The data are from ref.~\cite{Kis76}. The target excitation was treated within the collective model, using a quadrupole deformation parameter of $\beta_2=0.5$. Also included are Faddeev calculations performed by A.~Deltuva \cite{Del16}. Both calculations reproduce equally well the elastic differential cross section. The calculated  inelastic angular distributions are slightly out of phase with the data, but they agree well with each other, pointing to some inadequacy of the structure or potential inputs. Although these inelastic cross section can be also well reproduced with standard DWBA calculations based on a deformed deuteron-target potential, it was shown in \cite{Del16} that the extracted deformation parameter obtained with the three-body approach is more consistent with that derived from nucleon-nucleus inelastic scattering.

\subsubsection{Extension to three-body projectiles}

\begin{figure}
\begin{minipage}[c]{.45\linewidth}
\begin{center}\includegraphics[width=0.95\columnwidth]{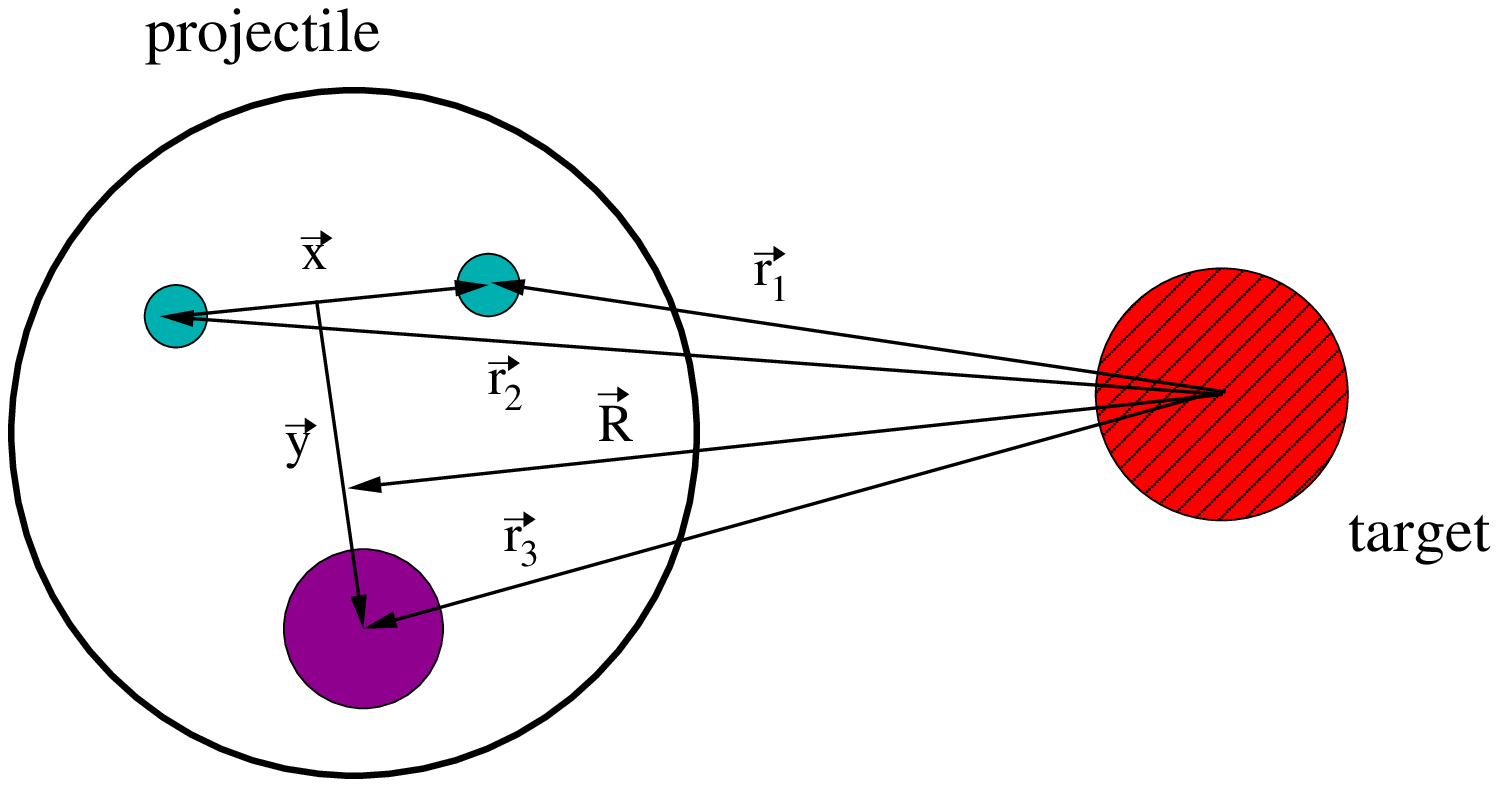} \end{center}
\end{minipage}
\begin{minipage}[c]{.45\linewidth}
\begin{center}\includegraphics[width=0.95\columnwidth]{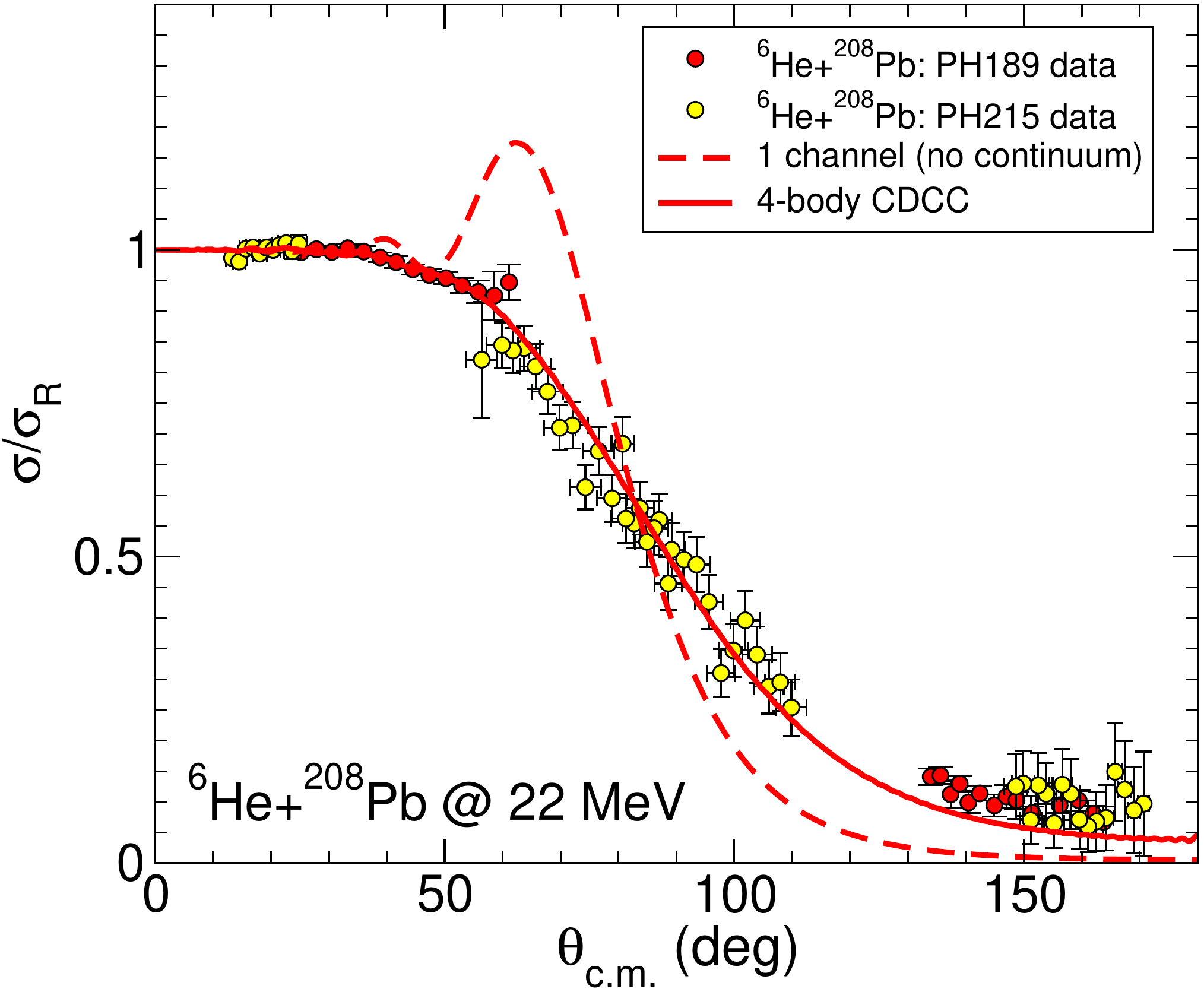} \end{center}
\end{minipage}
\caption{Left: Relevant coordinates of the four-body scattering problem.  Right: Differential elastic cross section, relative to Rutherford, for the $^{6}$He+$^{208}$Pb reaction at $E_\mathrm{lab}=22$~MeV. The solid (dashed) line is the four-body CDCC calculation including (excluding) $^{6}$He continuum states.  Experimental data are from refs.~\cite{San08,Aco11} and the CDCC calculations from ref.~\cite{Rod09}.}
\label{fig:he6pb_cdcc}
\end{figure}

To study the scattering of three-body projectiles,  such as Borromean nuclei, the  Hamiltonian  (\ref{eq:Heff}) must be generalized in order to take into account the three-body structure of the projectile. For example, for a two-neutron Borromean system with a structure of the form $a=b+n + n$ one may use the Hamiltonian
\begin{equation}
\label{eq:Heff_4b}
H= H_{\rm proj}(\vec{x},\vec{y})   + \hat{T}_{\bR} + U_{n A}(\br_{1}) + U_{n A}(\br_{2})+  U_{bA}(\br_{3})  ,
\end{equation}
where $H_{\rm proj}(\vec{x},\vec{y})$ is the  projectile (three-body) Hamiltonian, depending now on two relative coordinates (for example, the Jacobi coordinates shown in the cartoon of
 fig.~\ref{fig:he6pb_cdcc}), and  $U_{nA}(\br_{i})$ the valence-target effective interactions.  Clearly, the calculation of the projectile states will be much more involved than in the two-body case. In general, a given projectile state with angular momentum $j_p,m_p$ will consist of a superposition of many configurations involving the internal orbital angular momenta and spins of the constituents, 
\be
\phi_{i}(\xi) 
=\sum_{\alpha} \left[   \varphi_\alpha(\vec{x},\vec{y}) \right]_{j_p m_p} ,
\label{wf3b}
\ee
where $\xi \equiv \{\vec{x},\vec{y} \}$  are the internal coordinates of the three-body system. The label $\alpha$  denotes the set of quantum numbers $\{\ell_1, \ell_2, s_1, s_2,\ldots \}$, required to characterize the three-body state, which may vary depending on the three-body approach used. Note that $\phi_{i}(\xi)$ is an (approximate) eigenstate of the projectile three-body Hamiltonian, with a given energy and angular momentum. 

Once the internal states are obtained, coupling potentials are computed using a generalized form of eq.~(\ref{eq:Vij_3bCDCC})
\be
V_{i;i^\prime}(\vecR) = 
\int d \vecr  \, \phi_i^{*}(\vec{x},\vec{y})
	\Big [ U_{nA} (\br_{1}) + U_{nA} (\br_{2})  + U_{\alpha A} (\br_{3}) \Big ]
 \phi_{i^\prime}(\vec{x},\vec{y}) 
\ee
The formulation and first applications of this four-body CDCC method can be found in refs.~\cite{Mat04a} and  \cite{manoli08}.   As an illustrative example, we show in fig.~\ref{fig:he6pb_cdcc} the elastic scattering of $^{6}$He on $^{208}$Pb at 22 MeV \cite{Rod09}. The solid line represents the four-body CDCC calculation, whereas the dashed line is the calculation omitting the breakup channels. The full calculation shows a very good agreement with the data from refs.~\cite{San08,Aco11}. We finally see that the disappearance of the Fresnel peak, already discussed in the  context of the phenomenological OM, can be understood as a consequence of the strong coupling to the breakup channels.

\begin{figure}
\begin{minipage}[c]{.45\linewidth}
\begin{center}\includegraphics[width=0.9\columnwidth]{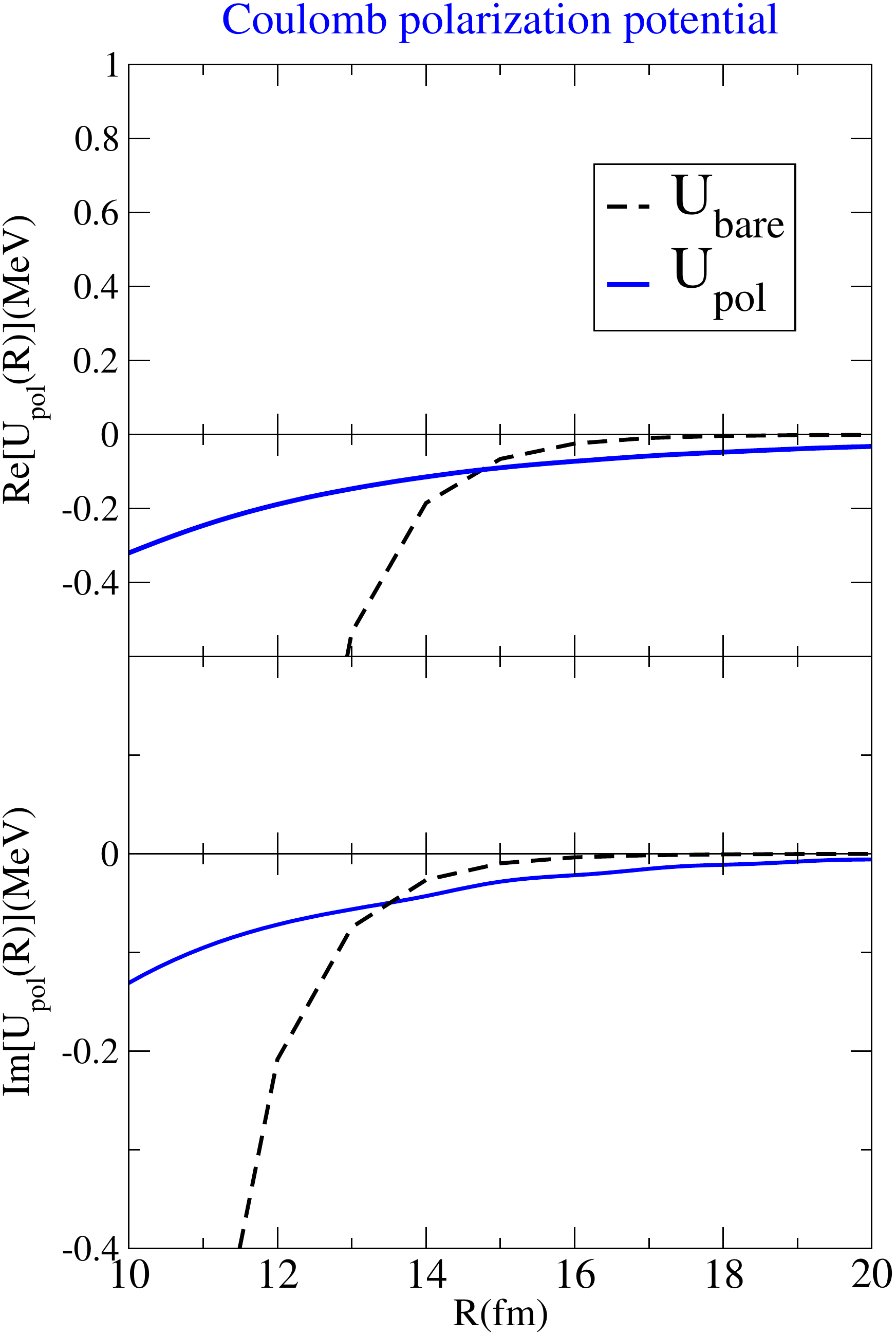} \end{center}
\end{minipage}
\begin{minipage}[c]{.45\linewidth}
\begin{center}\includegraphics[width=0.9\columnwidth]{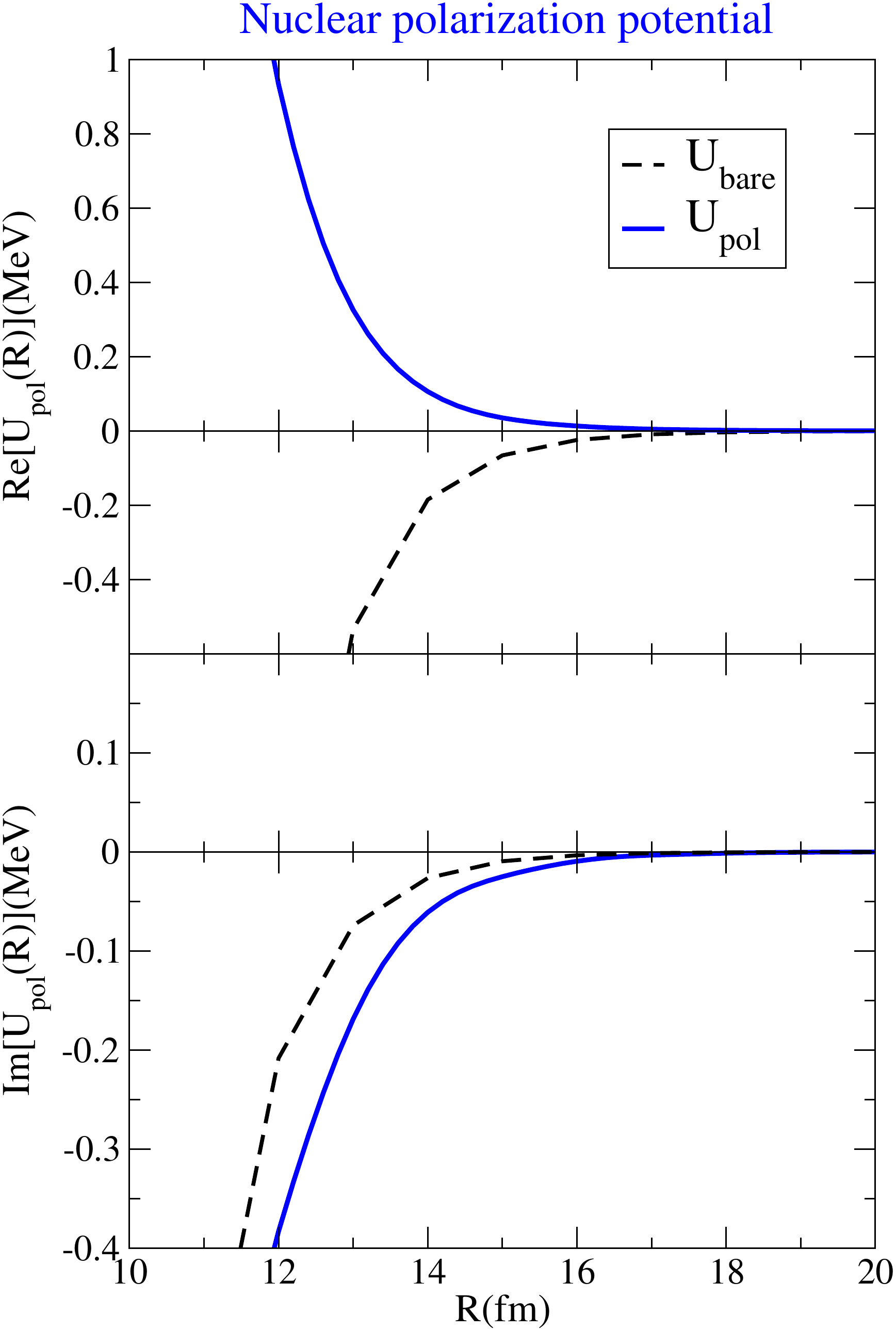} \end{center}
\end{minipage}
\caption{Coulomb (left) and nuclear (right) polarization potentials extracted from the $^{6}$He+$^{208}$Pb CDCC calculations of fig.~\ref{fig:he6pb_cdcc}. The {\it bare} potential corresponds to the ground-state diagonal potential. Adapted from ref.~\cite{Fer10}}
\label{fig:he6pb_telp} 
\end{figure}

From a coupled-channel calculation, one may infer an effective polarization potential, also called {\it trivial equivalent local polarization potential} (TELP). This is done rewriting the elastic channel equation as
\begin{equation}
\label{eq:cc_telp}
\left[E-\varepsilon_{0}-\hat{T}_\bR -V_{0,0}(\bR) \right] \chi_{0}(\bR)  = 
\sum_{i \neq 0} V_{i,0}(\bR) \chi_{i}(\bR) \equiv U_\mathrm{TELP}(R) \chi_{0}(\bR) .
\end{equation}
In an angular momentum representation, in which actual calculations are standardly performed, one has an independent equation of each total angular momentum of the system ($J$). Each of these equations defines a angular-momentum dependent TELP, $U^J_\mathrm{TELP}(R)$. An approximate TELP can be  obtained by averaging the $U^J_\mathrm{TELP}(R)$ potentials using as weights the cross sections for each $J$ \cite{Tho89}. The result of such angular-momentum averaged TELP extracted from the aforementioned CDCC calculation for $^{6}$He+$^{208}$Pb at 22~MeV is displayed in 
fig.~\ref{fig:he6pb_telp}. To isolate the Coulomb and nuclear effects two different calculations were performed, one including only Coulomb breakup couplings and the other including only nuclear breakup. The corresponding TELPs are shown in the left and right panels of this figure, respectively. The bare potential (dashed line) is represented in this case by the   
ground-state diagonal potential $V_{0,0}$. The most noticeable feature of the Coulomb polarization potential is its long range, with respect to the bare potential. This is consistent with the behaviour of the phenomenological optical potential discussed in sec.~\ref{sec:elas_weakly}.  Note also the different character of the Coulomb and nuclear real parts: the former is attractive (recall the adiabatic limit, eq.~(\ref{eq:adpol})), whereas the latter is strongly repulsive.   The nuclear polarization potential is also found to be of long range.


\subsubsection{Connection with the Faddeev formalism}
\begin{figure}
\begin{minipage}[t]{.49\linewidth}
\begin{center}\includegraphics[width=0.8\columnwidth]{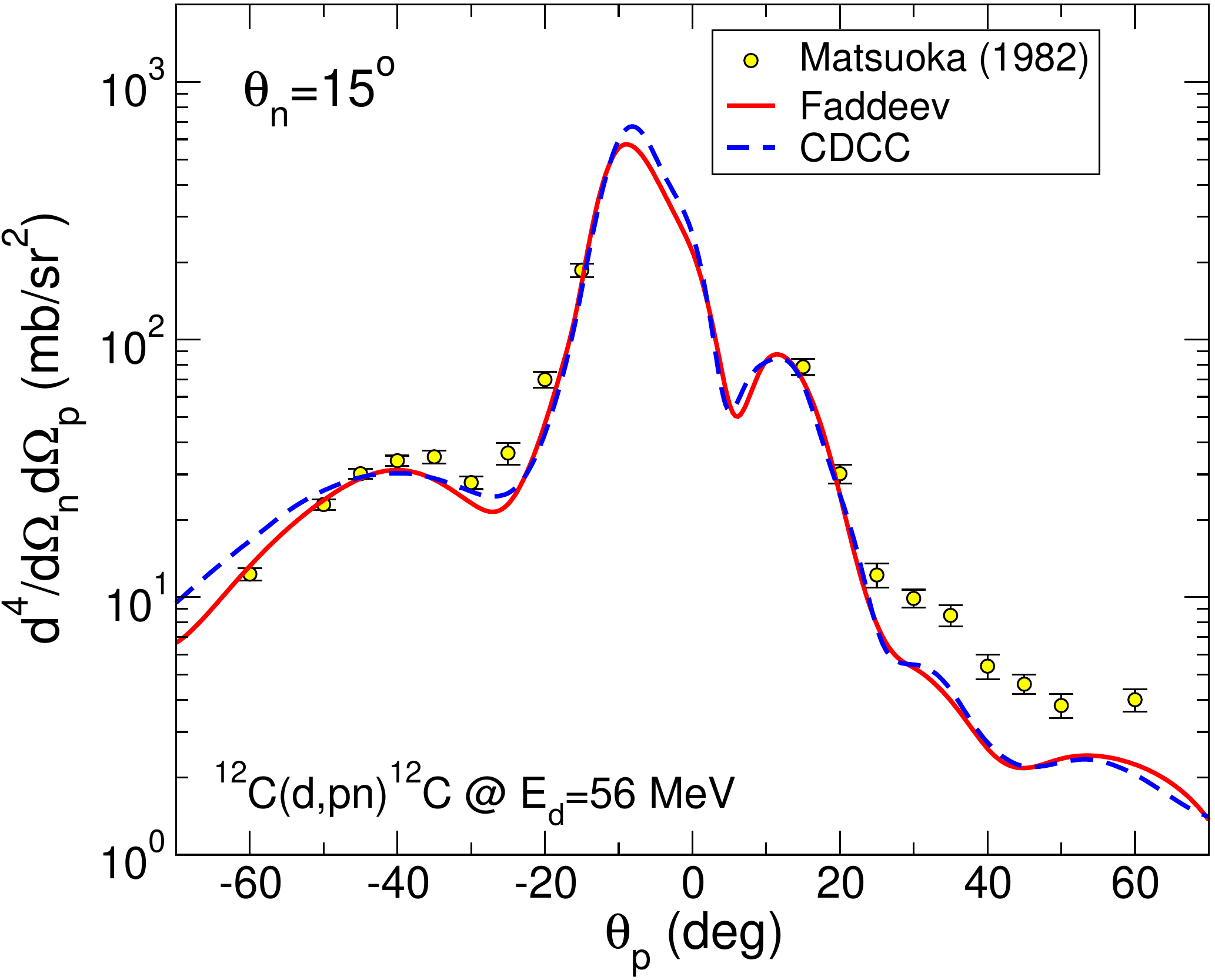} \end{center}
\end{minipage}\hfill
\begin{minipage}[t]{.49\linewidth}
\begin{center}\includegraphics[width=0.75\columnwidth]{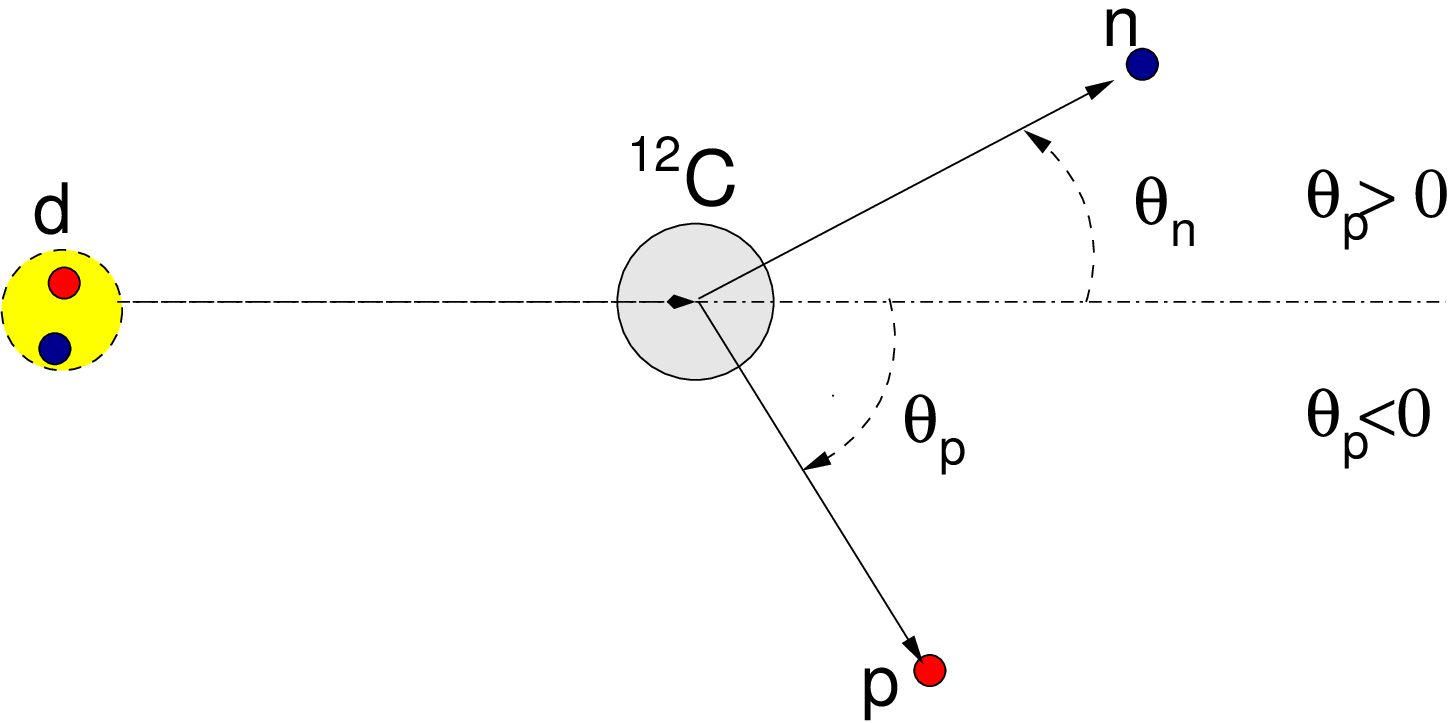} \end{center}
\end{minipage}
\caption{Breakup differential cross section for $^{12}$C($p$,$pn$)$^{12}$C at $E_d=56$~MeV, as a function of the proton scattering angle and for a fixed neutron detection angle of 15$^\circ$. The Faddeev and CDCC calculations are given by the solid and dashed lines, respectively. The circles are the data from ref.~\cite{Mat82}. From ref.~\cite{Del07}.}
\label{fig:fad_cdcc} 
\end{figure}

The CDCC method was originally devised as a  physically sound and numerically appealing  {\it ansatz} for the three-body wave function, rather than as a formally rigorous solution of the three-body problem. This rigorous solution  was provided by Faddeev in the 60s \cite{Fad60}, who showed that this solution  can be obtained from a system of coupled-differential equations, named the Faddeev equations after him. The numerical solution of these equations is very involved  but, recently, it has been possible to solve them for a number of situations \cite{Del07,Upa12}, thus providing a valuable benchmark for more approximate models. These comparative studies have shown that the elastic and breakup observables calculated with the CDCC method agree in general very well with the Faddeev solution (see fig.~\ref{fig:fad_cdcc}). However,  there are also kinematical situations and observables  \cite{Upa12} for which differences appear. This result calls for additional studies and, possibly, for extensions and improvements of the CDCC formalism. 

\subsubsection{Microscopic CDCC}
\begin{figure}
\begin{minipage}[c]{.5\linewidth}
\begin{center}\includegraphics[width=0.6\textwidth]{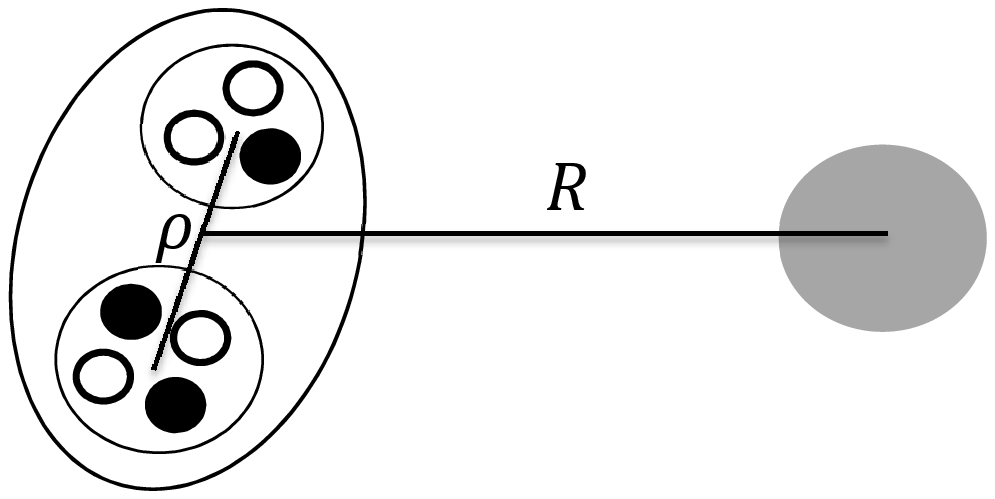} \end{center}
\end{minipage}
\begin{minipage}[c]{.5\linewidth}
\begin{center}\includegraphics[width=0.85\textwidth]{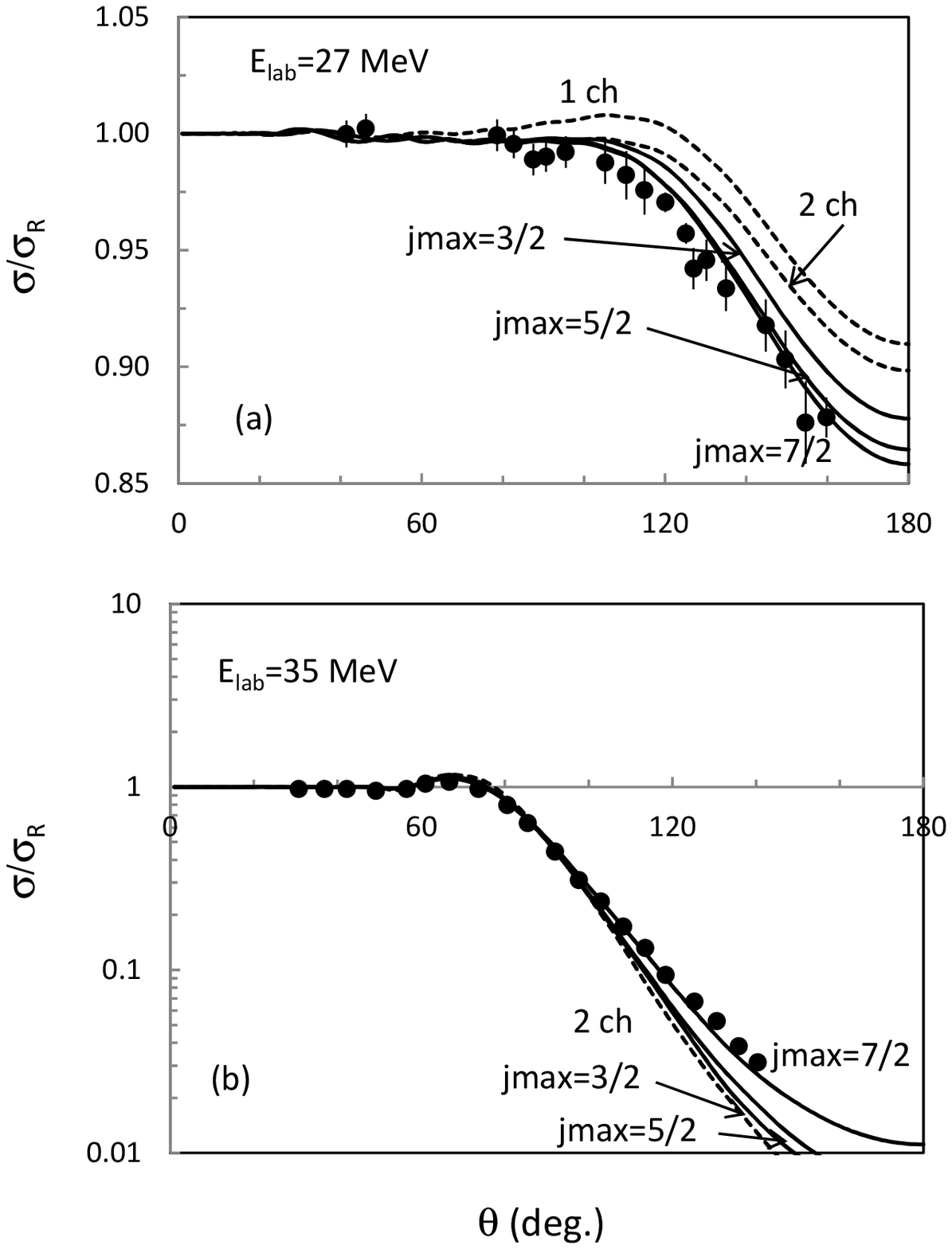} \end{center}
\end{minipage}
\caption{\label{fig:li7pb_mcdcc}Left:  Schematic picture of the projectile-target system, with
a microscopic cluster structure of the projectile. Right: $^{7}$Li+$^{208}$Pb elastic cross section \cite{Mar96}, relative to Rutherford, compared with microscopic CDCC calculations. Dotted lines represent the calculations without breakup channels  and the solid lines are the full calculations with increasing  $\alpha+t$ maximum angular momentum $j_p$. Taken from \cite{Des13}.}
\label{fig:be11pb_xcdcc} 
\end{figure}

The standard CDCC method assumes a cluster (two-body or three-body) description of the projectile nucleus. This simplification has of course limitations and drawbacks. For example: (i) it requires cluster-target optical potentials, which are are not always well determined; (ii) the extension to more than three bodies is very challenging and currently not available; (iii)  excitations of the fragments are ignored altogether or, at most, approximately included with some collective model. To overcome these problems, a microscopic version of the CDCC method (MCDCC) has been proposed by Descouvemont and co-workers \cite{Des13,Des18}. The method uses a many-body description of the projectile states, based on a cluster approximation, known as resonating group method (RGM). In the RGM, an eigenstate of the projectile Hamiltonian is written
as an antisymmetric product of cluster wave functions. For example, for a $^{7}$Li projectile, described as $\alpha+t$, the RGM wave function is expressed as:
\be
\phi_i(\xi_p)={\mathcal A}\bigl[ [\phi_{\alpha} \otimes \phi_t]^{1/2} \otimes 
Y_{\ell}(\Omega_{\rho})\bigr]^{jm} g^{\ell j}_i(\rho),
\label{eq:li7rgm}
\ee
where $\phi_{\alpha}$ and $\phi_{\alpha}$ are  shell model wave functions of the $\alpha$ and $t$ clusters, $\ell$ their relative orbital angular momentum and $j$ the total spin.  In eq.~(\ref{eq:li7rgm}), $\vec{\rho}$ is the relative coordinate (see fig.~\ref{fig:li7pb_mcdcc}), and  ${\mathcal A}$ is the 7-body antisymmetrization operator which takes into account the Pauli principle among the 7 nucleons of the projectile.  The function $g^{\ell j}_i(\rho)$ is determined from a Schr\"odinger equation associated with the projectile Hamiltonian.  Continuum states are included using a pseudo-state basis.  
The projectile-target interaction is given by the sum of nucleon-nucleus interactions (instead of cluster-target interactions), for which reliable parametrizations are available. 
 
Figure \ref{fig:li7pb_mcdcc} shows an application of the method to the reaction $^{7}$Li+$^{208}$Pb at near-barrier energies. Experimental data are compared with a one-channel calculation (only $^{7}$Li g.s.),  two-channels calculations (ground plus first excited state) and several CDCC calculations including continuum states up to a certain projectile angular momentum. It is seen that a good description of the data is achieved when a sufficiently large number of continuum states is included. 

\subsection{Exploring the continuum with breakup reactions}
%
\begin{figure}
\begin{minipage}[c]{.5\linewidth}
\begin{center}\includegraphics[height=0.75\columnwidth]{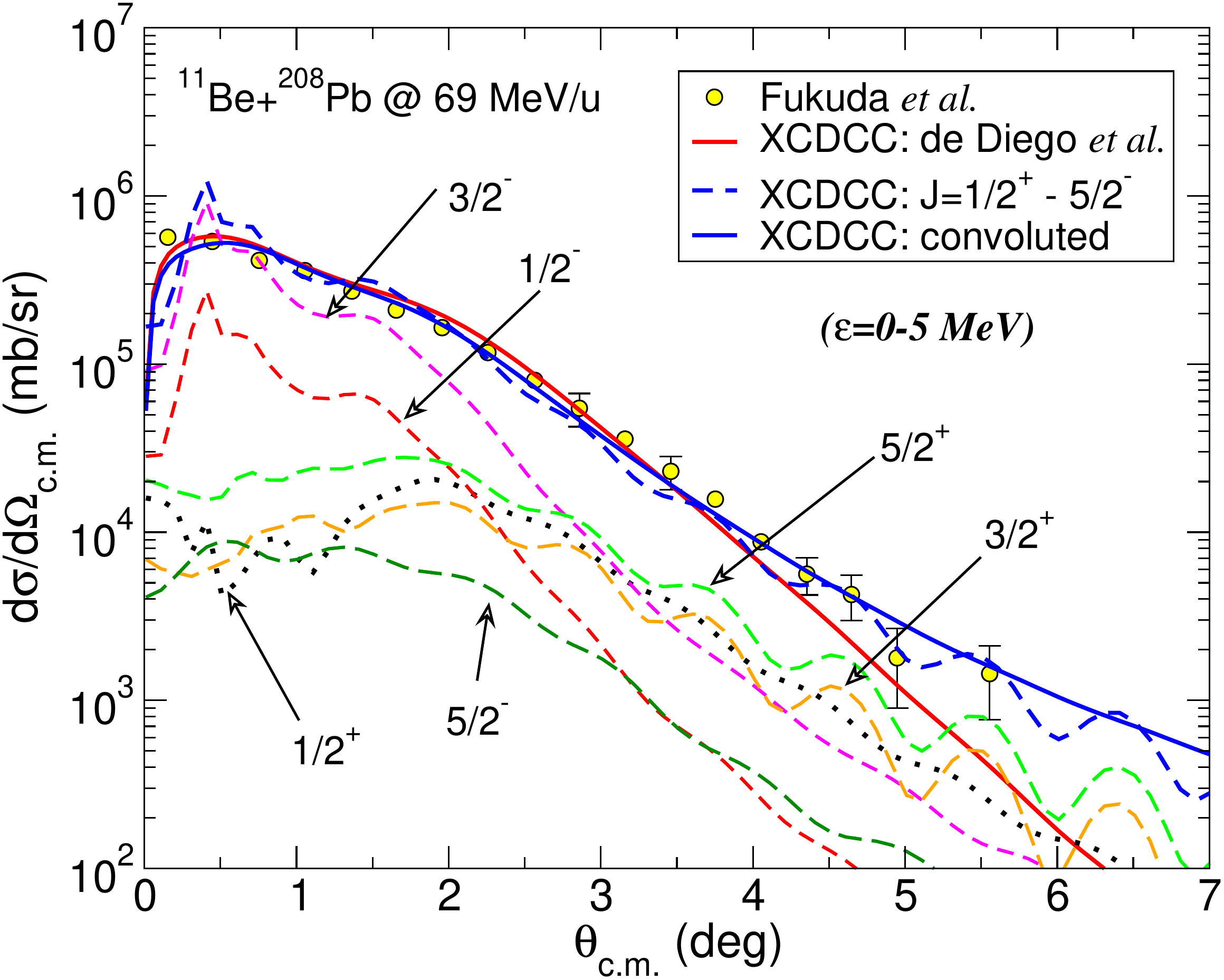} \end{center}
\end{minipage}
\begin{minipage}[c]{.5\linewidth}
\begin{center}\includegraphics[height=0.75\columnwidth]{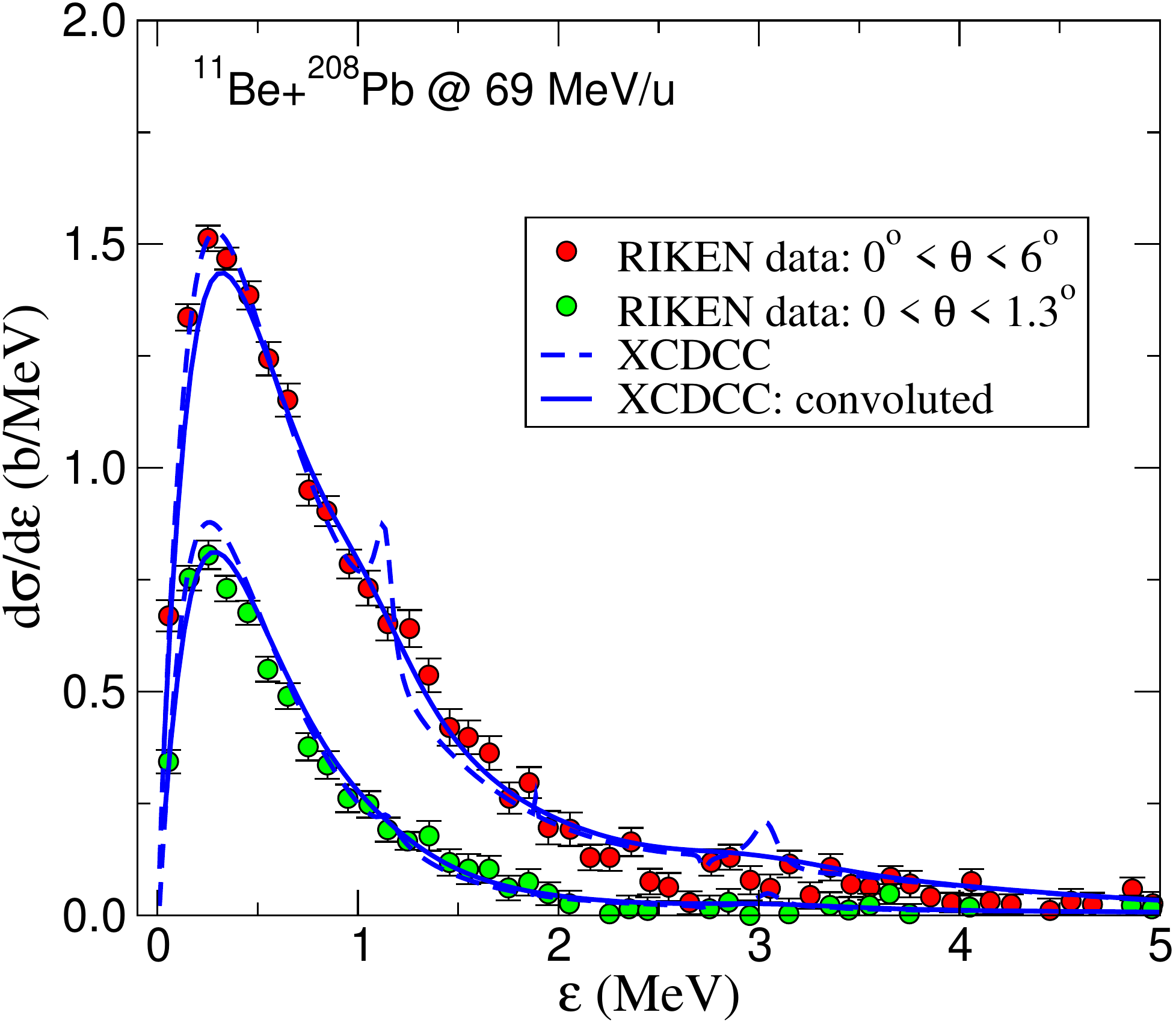} \end{center}
\end{minipage}
\caption{Left: Experimental \cite{Fuk04} and calculated (XCDCC) breakup angular distributions for the $^{11}$Be+$^{208}$Pb reaction at 69~MeV/u. The separate contributions of the differrent continuum angular momenta are shown, as indicated by the labels. Right: Measured and calculated angle-integrated relative-energy distribution for the outgoing $n$-$^{10}$Be system.}
\label{fig:be11pb_xcdcc} 
\end{figure}

\subsubsection{Coulomb breakup}
Breakup observables, which provide valuable information about the dissociated nucleus, can be computed within the CDCC formalism. An example if shown in fig.~\ref{fig:be11pb_xcdcc} for the reaction $^{11}$Be+$^{208}$Pb $\rightarrow$ $n$+$^{10}$Be+$^{208}$Pb at $E_\mathrm{lab}$=69~MeV/u, measured at RIKEN \cite{Fuk04}. The left panel is the angular distribution, with respect to the center of mass of the outgoing system $n$+$^{10}$Be. The different lines are the  contributions coming from different continuum states of $^{11}$Be ($j^\pi_p$=$1/2^\pm$, $3/2^\pm$ and $5/2^\pm$), obtained from a XCDCC calculation \cite{Die14}. The thick dashed line is the total contribution which, after convoluting with the experimental angular resolution, yields the thick solid line. This is found to reproduce very well the data. Moreover, it is seen that the main contributions come from the $3/2^-$ and, to a lesser extent, the $1/2^-$ waves. Since the g.s.\ has $j_p=1/2^+$ these correspond to dipole ($\lambda=1$) transitions. At these very small scattering angles, the nuclear contribution is very small so the breakup is mostly due to the Coulomb interaction. The right panel shows the breakup cross section as a function of the relative energy of the $n$ and $^{10}$Be fragments, and integrated up to $\theta_\mathrm{c.m.}=1.3^\circ$ and $\theta_\mathrm{c.m.}=6^\circ$. The most notable feature is the enhanced breakup cross section near the breakup threshold which, in turn, is a consequence of the large $B(E1)$ strength at these energies. The dipole strength distribution is given by 
\be
\frac{dB}{d \varepsilon} = \frac{3}{4\pi}(Z_\mathrm{eff} e^2) \langle \ell 0 1 0 | \ell' 0 \rangle^2 
\left | \int dr ~ r^3 ~u^*_{k,\ell',s,j'}(r)  u_{\ell}(r) \right |^2
\ee
where $Z_\mathrm{eff}$ is the effective charge ($Z_\mathrm{eff}=-Z/A$ for a neutron halo nucleus), and $u_\ell(r)$ and $u_{k,\ell',s,j'}(r)$ are the  radial parts of the ground and scattering states with orbital angular momenta $\ell$ and $\ell'$, respectively. For weakly bound nuclei, the integral is dominated by the asymptotic region, which allows to replace the bound state by its asymptotic form and the scattering state by a plane wave. For a $s$ to $p$ transition, as it is the case of $^{11}$Be, this yields \cite{Nag05}
\be
\frac{dB}{d \varepsilon}(s \rightarrow p) = \frac{3 \hbar^2}{\pi^2 \mu} (Z_\mathrm{eff} e)^2
 \frac{\sqrt{\varepsilon_b} \varepsilon^{3/2}} {(\varepsilon + \varepsilon_b)^4} ,
\ee
where $\varepsilon_b$ is the separation energy. This distribution has a maximum at $\varepsilon=3/5\varepsilon_b$. Hence, the smaller the binding energy, the closer is this peak to the breakup threshold. 

\begin{figure}[!ht]
\begin{center}
\includegraphics[width=0.65\linewidth]{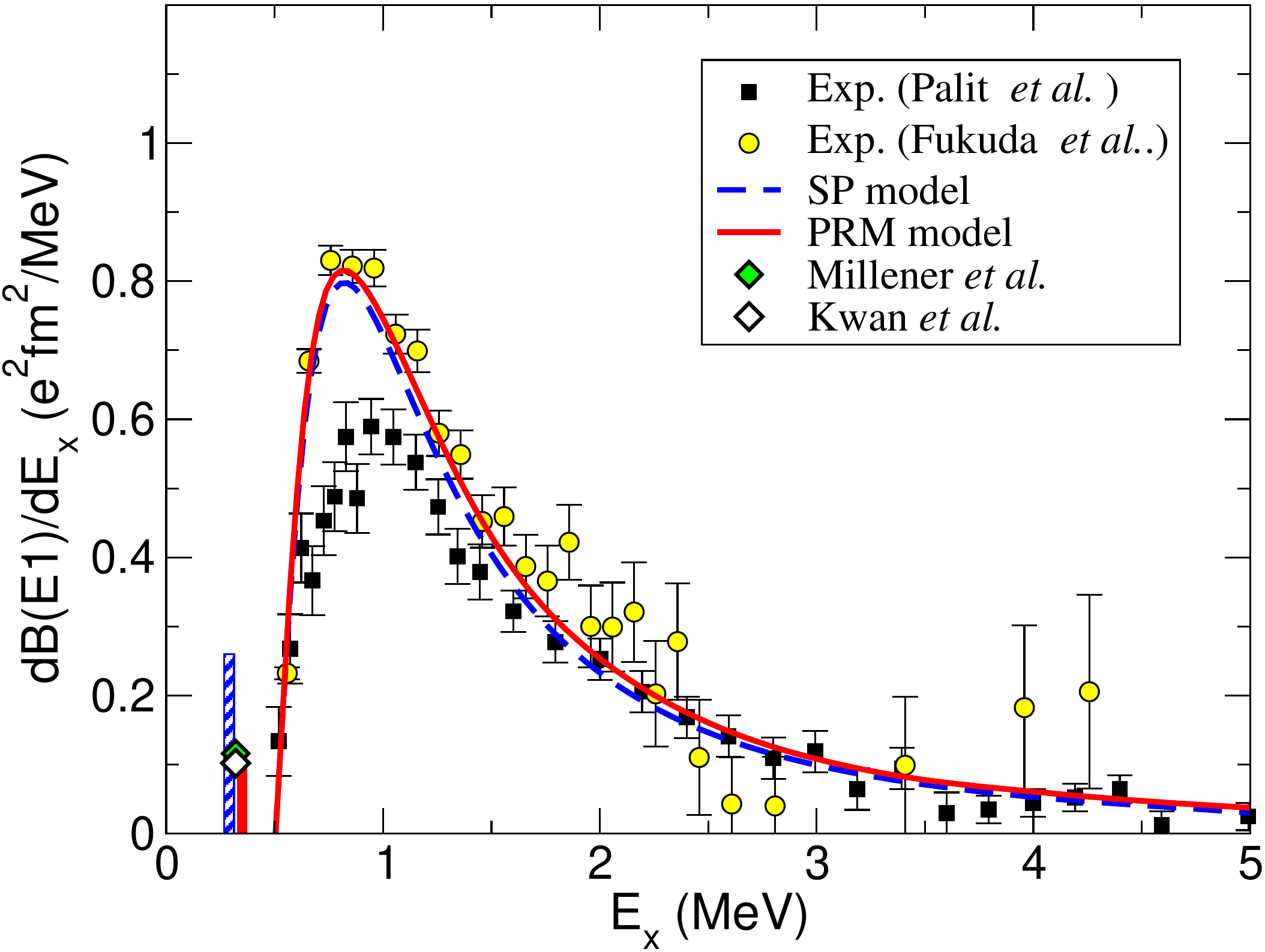}
\caption{\label{fig:be11_dbde}Experimental and calculated $B(E1)$ strength distributions for $^{11}$Be as a function of the excitation energy $E_x= \varepsilon_b + \varepsilon$. The experimental distributions
 are from the Coulomb dissociation experiments of Palit {\it et al.} \cite{Pal03} and Fukuda {\it et al.} \cite{Fuk04}. 
 The filled diamond and the vertical bars represent the experimental and calculated $B(E1; 1/2^+ \rightarrow 1/2^-)$ value (in units of e$^2$fm$^2$)   between the bound states of $^{11}$Be \cite{Mil83, Kwa14}. The theoretical distributions correspond to the single-particle model of Ref.~\cite{Cap04} and the particle-plus-rotor model of ref.~\cite{Tar03}.}
\end{center}
\end{figure}

Figure \ref{fig:be11_dbde} shows some $dB/dE_x$ distributions for the $^{11}$Be nucleus extracted from Coulomb dissociation experiments \cite{Pal03,Fuk04}, compared with two theoretical predictions, one based on a single-particle (SP) model of $^{11}$Be and the other on a particle-plus-deformed-core (PRM).   The large concentration of strength near threshold can be attributed to the low separation energy ($\varepsilon_b=0.5$~MeV), according to the simple model just presented.  

\subsubsection{Resonant nuclear breakup}
\begin{figure}
\begin{center}
\begin{minipage}[c]{.46\linewidth}
\includegraphics[width=0.75\columnwidth]{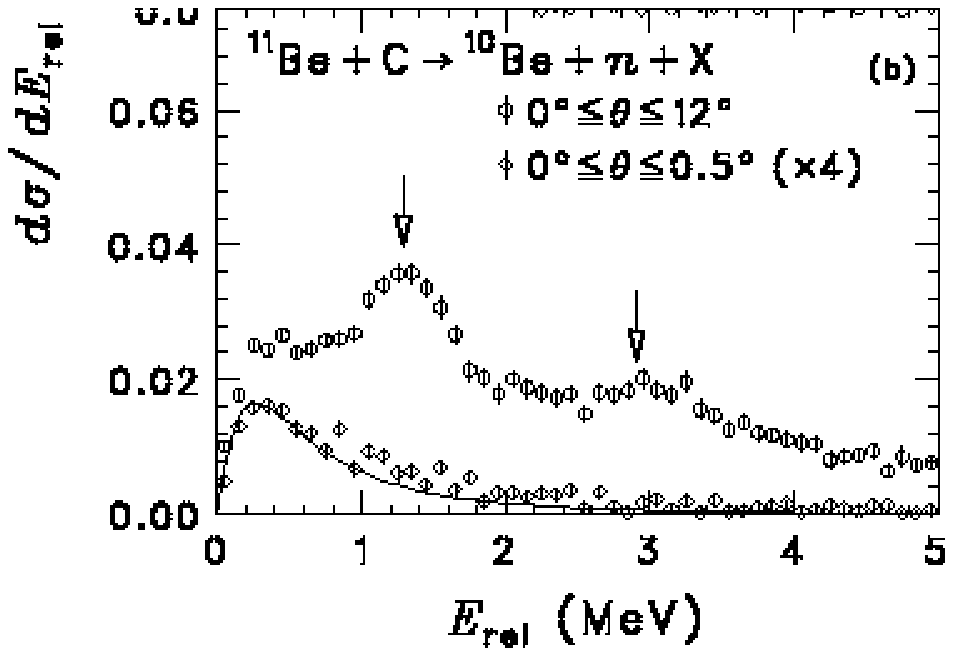} \\
\includegraphics[angle=-90,width=0.75\columnwidth]{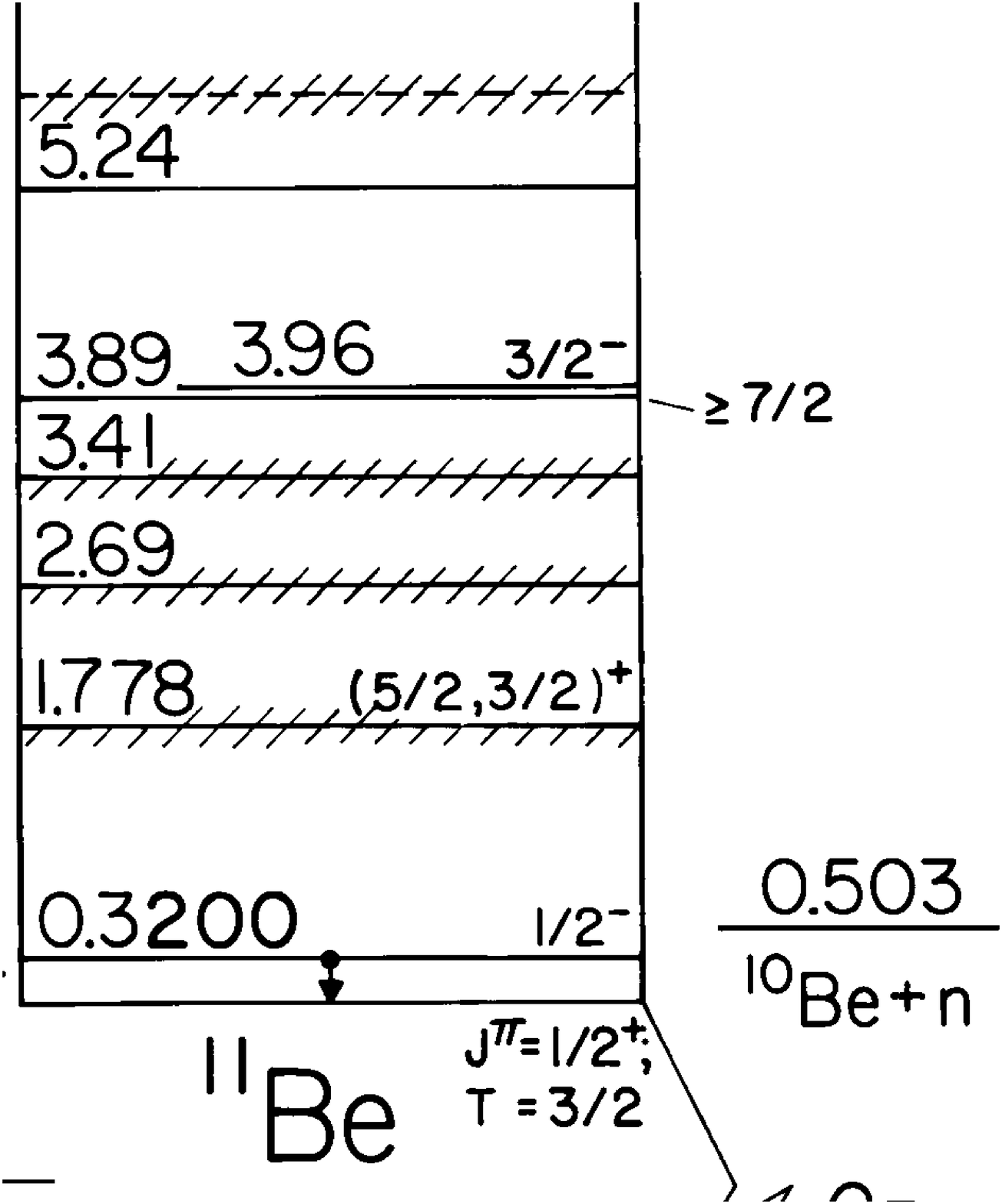} 
\end{minipage}
\begin{minipage}[c]{.46\linewidth}
\begin{center}\includegraphics[width=0.9\columnwidth]{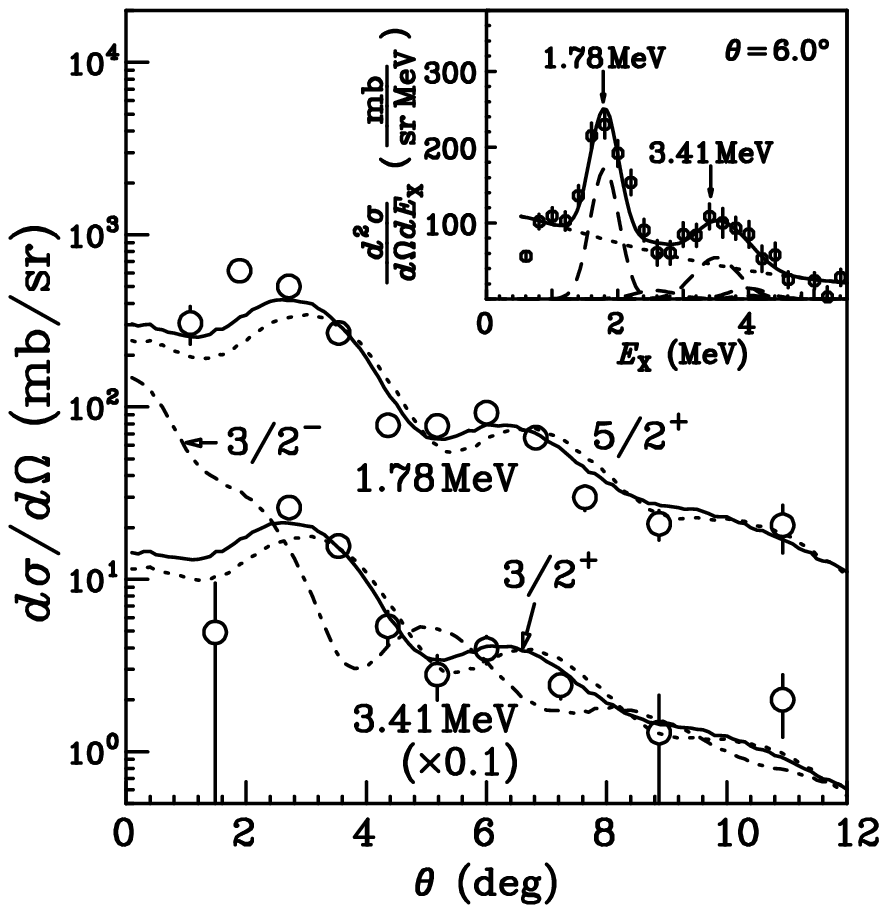} \end{center}
\end{minipage}
\end{center}
\caption{\label{fig:be11c_exp}Left: Experimental breakup cross section for $^{11}$Be+$^{12}$C at 69~MeV/u from ref.~\cite{Fuk04} (top) and $^{11}$Be low-lying spectrum (bottom). Right: Comparison of the angular distribution of the resonant peaks with DWBA calculations based on a collective model of $^{11}$Be.  Adapted from ref.~\cite{Fuk04}.}
\end{figure}

The scattering of a weakly-bound nucleus with a heavy target emphasizes the electric response of this nucleus, which can be quantified in terms of the electric reduced probability $dB/dE$. Conversely, when the nucleus scatters by a light target, such as $^{9}$Be or $^{12}$C, Coulomb breakup will be less important and nuclear breakup will gain importance. This kind of measurements are helpful to probe low-lying resonances  which would be otherwise hindered by the dipole strength. An example of this is depicted in fig.~\ref{fig:be11c_exp}, corresponding to the data for the $^{11}$Be+$^{12}$C breakup reaction at 69~MeV/u. The upper panel shows the experimental data from ref.~\cite{Fuk04}. Superimposed on a smooth structure, one can see two peaks which, according to the energies shown in the spectrum of the bottom panel, correspond to  low-lying resonances of $^{11}$Be with spin-parity assignment  $5/2^+$ and $3/2^+$.   

\begin{figure}
\begin{minipage}[c]{.5\linewidth}
\begin{center}\includegraphics[width=0.8\columnwidth]{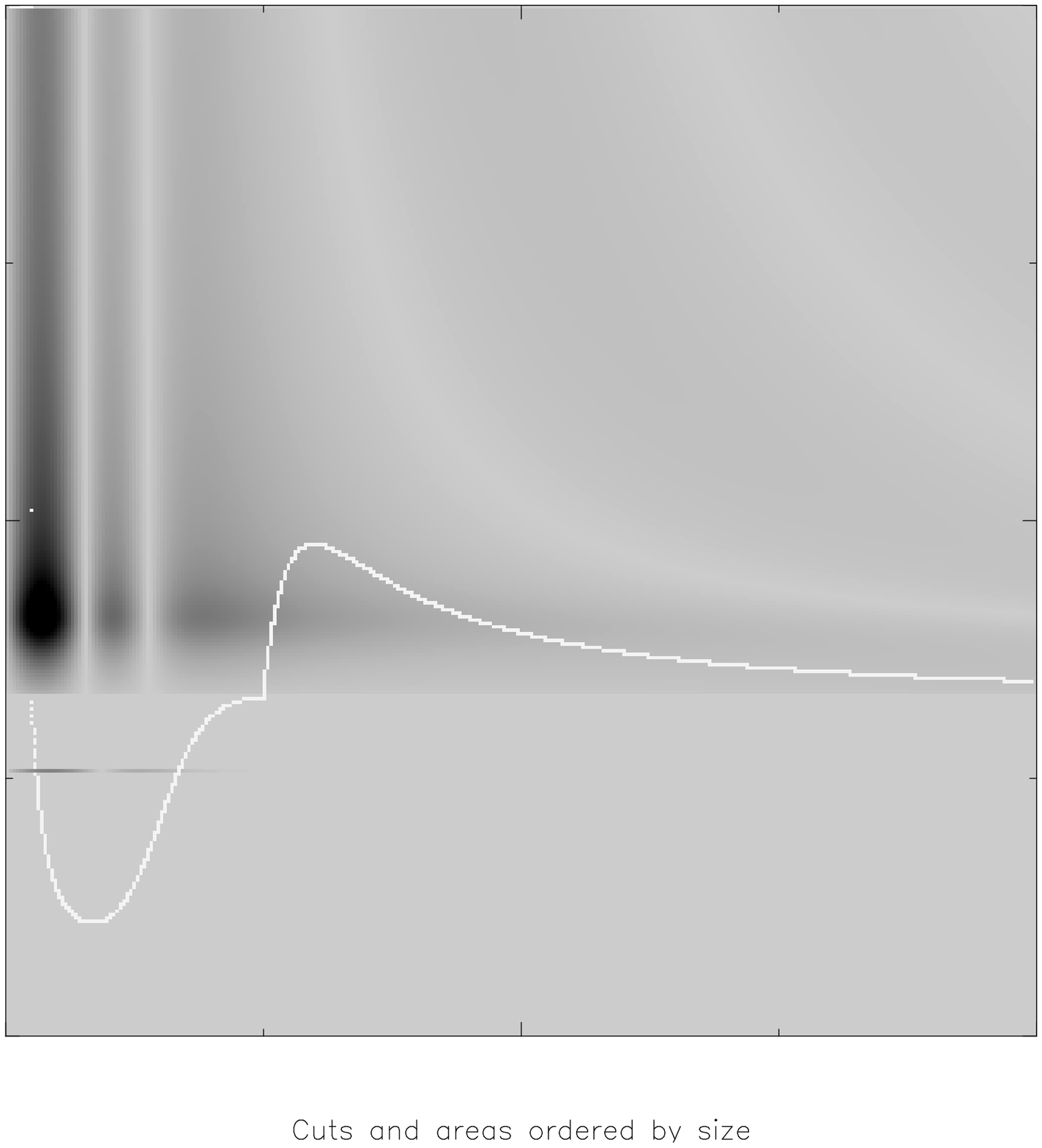} \end{center}
\end{minipage}
\begin{minipage}[c]{.5\linewidth}
\begin{center}\includegraphics[width=0.9\columnwidth]{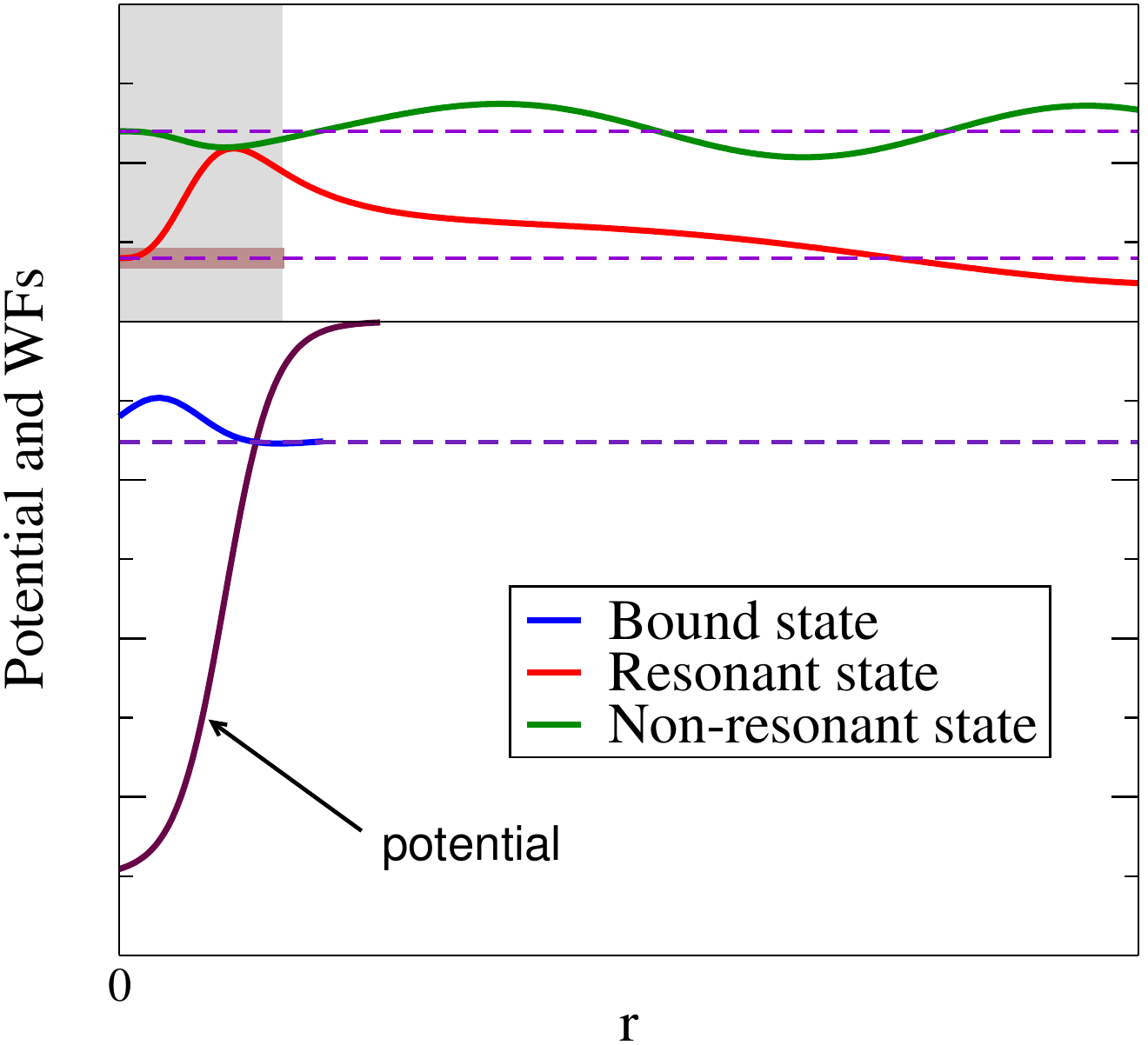} \end{center}
\end{minipage}
\caption{Left panel: potential well and density probabilities (square of wave functions). Courtesy of C.H.~Dasso. Right panel: wave functions for a bound state, a non-resonant state and a resonant state.}
\label{fig:res} 
\end{figure}

Before addressing the problem of the theoretical analysis of these reactions, let us briefly recall the origin and meaning of continuum resonances. The simplest case is that of a potential resonance, illustrated in fig.~\ref{fig:res}. The left panel shows a potential well consisting on an attractive potential and a repulsive one (the latter can be due to the Coulomb interaction, the centrifugal barrier, $\hbar^2 \ell (\ell+1)/2 \mu r^2$, or the combination of both). In general, the potential will contain a finite number of bound states at negative energies and a infinite, and continuous, number of unbound states at positive energies.  A particle moving in this well would remain confined in the case of bound states,  but it would scape for the case of unbound states. However, due to the presence of the potential barrier, the system  exhibits also quasi-bound structures (or resonances) at positive energies. For an infinitely high barrier, these states would be also permanently confined but, for a finite barrier, they will eventually decay by means of barrier penetration. If the resonant region extends over an energy range $\Gamma$, the system will ``survive'' in this state over a period $\tau \approx \hbar /\Gamma$ (the lifetime of the resonance). 
In terms of their wave functions, these quasi-bound regions, or resonances, have therefore a higher presence probability. This is depicted in fig.~\ref{fig:res} using  a gray scale, with the darker regions corresponding to larger probabilities (given by the modulus squared of the wave function). One can observe a bound state and a resonant continuum region, with the node structure of the wave functions.
 The right panel of this figure shows three wave functions:  a bound state wave function,  a non-resonant continuum wave function, and a resonant wave function. It is seen that the resonant wave function presents a much larger probability inside the range of the potential as compared to the non-resonant one. 

\begin{figure}
\begin{center}
\includegraphics[width=0.6\columnwidth]{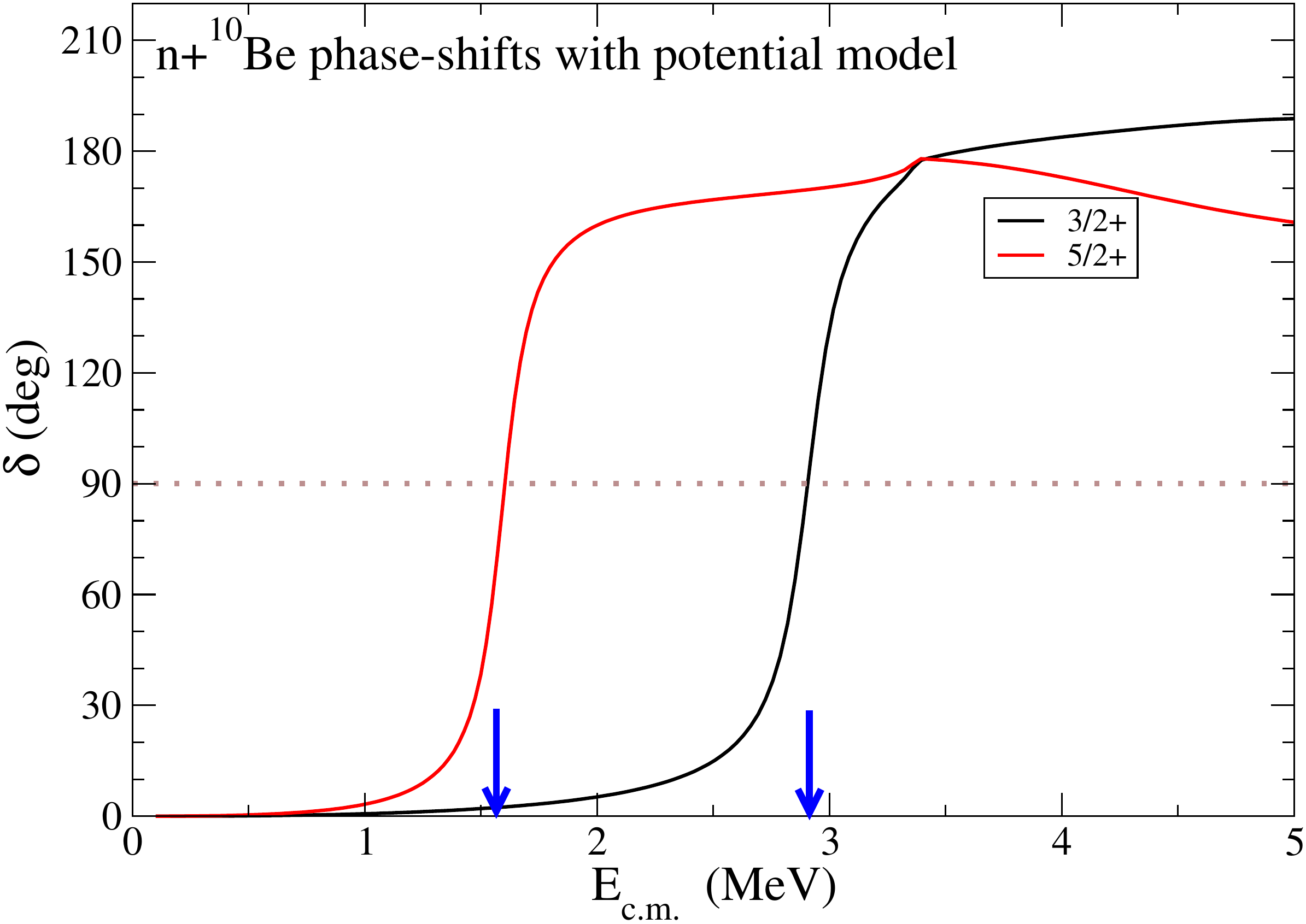}
\caption{\label{fig:be10n_psh}Elastic phase shifts for the $5/2^+$ and $3/2^+$ waves in $n$+$^{10}$Be elastic scattering using a deformed potential, which includes the coupling to the $^{10}$Be(2$^+$) excited state. The vertical arrows indicate the nominal position of the resonances, defined as the energies for which the phase-shift jumps over $\pi/2$.}
 \end{center}
\end{figure}

 The resonant character is also apparent in the S-matrix and phase-shifts (recall sec.~\ref{chap:elas}). For a potential resonance, the phase-shift displays a rapid increase as a function of the continuum energy, when crossing the nominal energy of the resonance. This is shown in fig.~\ref{fig:be10n_psh} for the $n$+$^{10}$Be phase shifts, where the jumps at the position of the $5/2^+$ and $3/2^+$ resonances are clearly seen. 


Whereas the energy differential cross section gives information on the position of resonances, the angular momentum assignment can be inferred from the shape of the angular distribution. 
 These calculations have been performed using different formalisms: DWBA with collective form factors \cite{Fuk04}, semiclassical models \cite{Cap04}, CDCC \cite{How05} and, more recently, also the generalized CDCC method with inclusion of $^{10}$Be excitations, XCDCC \cite{Mor12}. In the right panel of fig.~\ref{fig:be11c_exp} we show the experimental angular distributions of the resonances (after subtracting the non-resonant background) compared with DWBA calculations based on a collective model of the $^{11}$Be nucleus. From this analysis, the authors of ref.~\cite{Fuk04} concluded that these resonances were populated by $\lambda=2$ transitions, which led the authors to propose the spin-parity assignments $5/2^+$ and $3/2^+$, respectively.

\subsection{The problem of  inclusive breakup}
\begin{figure}
\begin{center}\includegraphics[width=0.75\columnwidth]{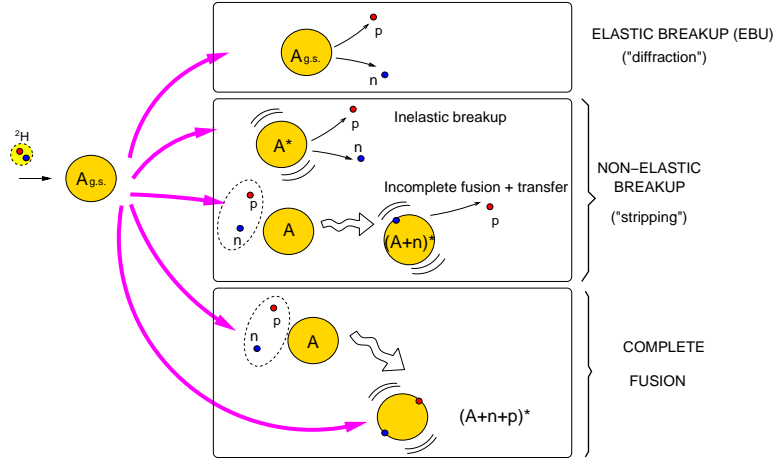} \end{center}
\caption{\label{fig:dA_neb}Pictorial representation of elastic and non-elastic breakup channels for a $d+A$ reaction.}  
\end{figure}

In previous sections, we have considered only exclusive breakup processes in which the three fragments ($b$, $x$ and $A$) survive after the collision and are observed in a definite internal state. In particular, when all fragments end up in their ground state, the process is called {\it elastic breakup}  (EBU).  In many experiments, however,  the final state of one or more fragments is not determined in the final state. This is the case, for example, of reactions of the form $A(a,b)X$, in which only one of the projectile constituents ($b$ in this case) is observed. The angular and energy distributions of the $b$ fragments will contain contribution from all possible final states of the $x+A$ system.    
This includes the EBU channel, in which $x$ and $A$ remain in their ground state, but also $x$ transfer, breakup accompanied by excitations of $A$, and  $x$+$A$ fusion [named {\it incomplete fusion} (ICF)].  These processes are schematically depicted in fig.~\ref{fig:dA_neb} for a deuteron-induced reaction.   These non-elastic breakup components (NEB) must be added to the  EBU component to give the total inclusive breakup.   Whereas the  EBU contribution can be accurately calculated within the CDCC method and other approaches, the evaluation of NEB is more involved due to the large number of accessible states.

\subsubsection{The IAV model for inclusive breakup}
 The problem of inclusive breakup  was addressed in the 1980s by several groups, which developed formal techniques to reduce the sum over final states to a closed form. We cite here the theory developed by Ichimura, Austern and Vincent (IAV) \cite{Aus81,Aus87,Ich85}, in which the double differential cross section for  NEB  with respect to the angle and energy of the $b$ fragments is given by
%
%
%
%
\begin{equation}
\label{eq:iav_3b}
\left . \frac{d^2\sigma}{dE_b d\Omega_b} \right |_\mathrm{NEB} = -\frac{2}{\hbar v_{i}} \rho_b(E_b)  \langle \varphi_x (\vec{k}_b) | \mathrm{Im}[U_{xA}] | \varphi_x (\vec{k}_b) \rangle   ,
\end{equation}
where $\rho_b(E_b)=k_b \mu_{b} /[(2\pi)^3\hbar^2]$,  $U_{xA}$ is the optical potential describing $x+A$ elastic scattering, and  $\varphi_x(\vec{k}_b,\br_{xA})$ is the wave function  describing the evolution of the $x$ particle after dissociating from the projectile, when the core is scattered with momentum $\vec{k}_b$ and  the target remains in its ground state. This function is obtained by solving the  inhomogeneous equation
\begin{equation}
\label{eq:pz_3b}
(E^+_x - \hat{T}_x - {U}_{xA})  \varphi_x (\vec{k}_b,\br_{xA}) =  (\chi_b^{(-)}(\vec{k}_b)| V_\mathrm{post}|\Psi^{3b} \rangle ,
\end{equation}
where $E_x=E-E_b$,  $\chi_b^{(-)}(\vec{k}_b,\br_{bB})$ is the distorted-wave describing the scattering of the outgoing $b$ fragment  with respect to the $B\equiv x+A$ system, obtained with some optical potential $U_{bB}$, and $V_\mathrm{post} \equiv V_{xb}+U_{bA}-U_{bB}$ is the post-form transition operator.  This equation is to be solved with outgoing boundary conditions. 

Although IAV suggested approximating the three-body wave function appearing in the source term of eq.~(\ref{eq:iav_3b}), $\Psi^{3b}$, by the CDCC one, a simpler choice is to use the DWBA approximation $\psi^{3b} \approx \chi^{(+)}_{a}(\bR) \phi_{a}(\br)$, where $\chi^{(+)}_{a}$ is a distorted wave describing $a+A$ elastic scattering, obtained with some optical potential, and 
$\phi_{a}$ is the projectile ground state wave function. 

This DWBA version of the IAV model has been recently revisited by some groups, and applied to several deuteron \cite{Pot15,Car16,Jin15} and $^6$Li induced reactions \cite{Jin15,Jin17}, showing in most cases a very promising agreement with the existing  data. Further applications and developments are in progress. 

\begin{figure}
\begin{center}\includegraphics[width=0.78\columnwidth]{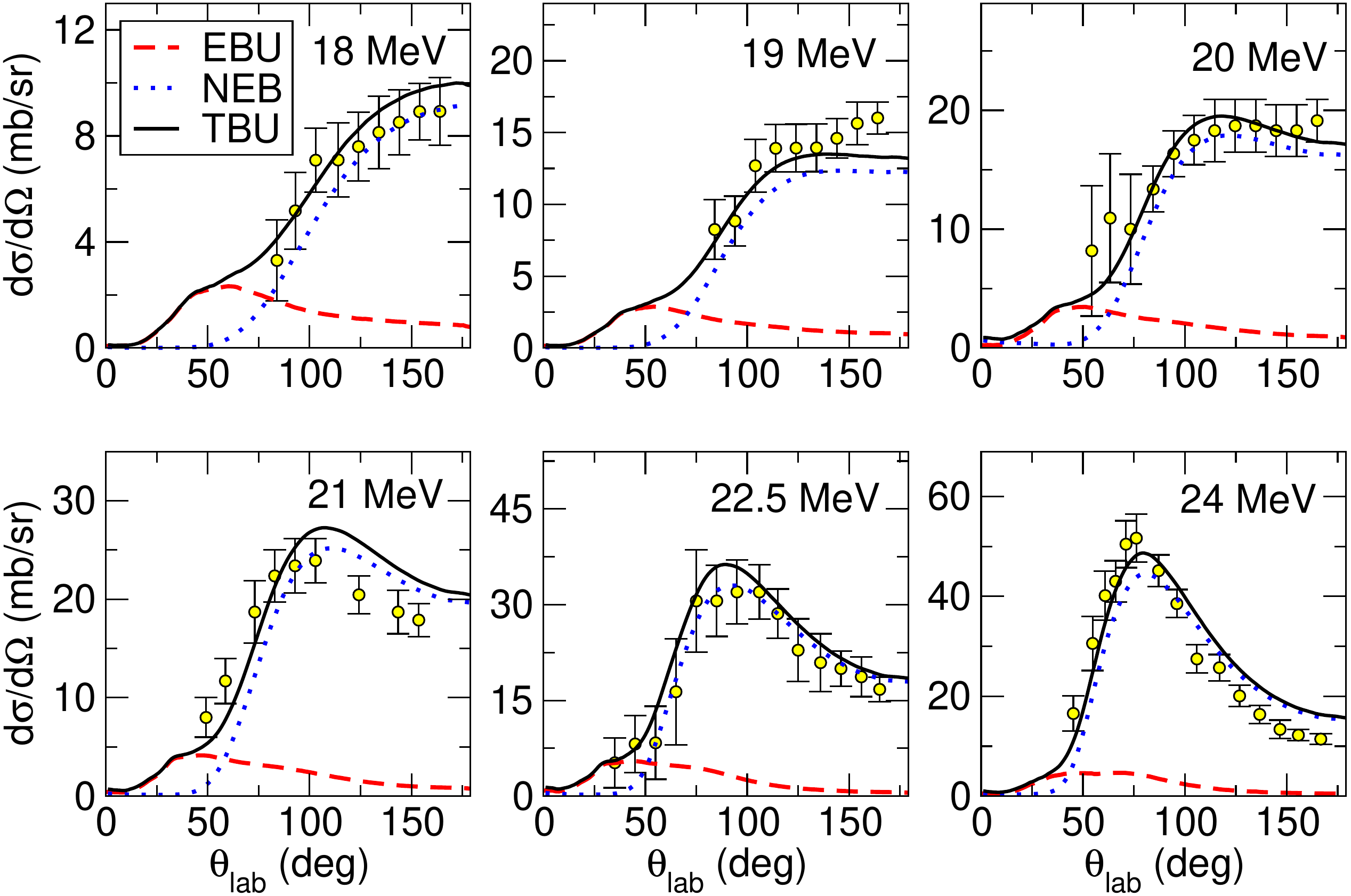} \end{center}
\caption{Energy distribution of $\alpha$ particles produced in the reaction $^{118}$Sn($^{6}$Li,$\alpha$)$X$ at the incident energies indicated by the labels. The dashed and dotted lines are the EBU and NEB contributions, and the solid line is their incoherent sum (TBU).}  
\label{fig:neb} 
\end{figure}

 As an example, in fig.~\ref{fig:neb} we show the experimental and calculated energy distribution of $\alpha$ particles emitted in the reaction induced by $^6$Li on $^{118}$Sn at the incident energies indicated by the labels (adapted from ref.~\cite{Jin17}). The EBU contribution (dashed line) was evaluated with the CDCC method whereas the NEB part (dotted line) was obtained with the IAV method, in its DWBA form.  Interestingly, one can see that the inclusive $\alpha$-yield is largely dominated by the NEB mechanism. The EBU is only important at small scattering angles (distant collissions, in a classical picture).  

\subsubsection{Eikonal approximation to inclusive breakup}\label{sec:KO}
At sufficiently high energies, inclusive breakup is usually assumed to proceed via a one-step mechanism in which a portion of the projectile (a nucleon, for instance) is suddenly removed, without disturbing the rest of the system. So, for a reaction of the form $a+A \rightarrow b + X$, one assumes that the observed states of $b$ are already contained in the ground-state of the $a$ system.  The probability for the production of a given $b$ state is therefore proportional to the parentage  with the initial projectile. This parentage, that  will be addressed also in sec.~\ref{sec:transfer}, is directly linked to  the normalization of the overlap function between the  projectile wave function ($a$) and that of the core ($b$). For that, one can use the so-called   {\it fractional parentage decomposition} (see e.g.\ Chap.\ 7 of \cite{Glen83}), which is a technique to expand a given antisymmetrized many-body wave function in terms of antisymmetrized wave functions of two of its constituents. For the present purposes, it refers to the expansion of the $a$ wave function in terms of wave functions of the $b$ subsystem (the {\it core}). This expansion will contain, in general, one  term (or more than one) involving exactly the state  $\phi_b(\xi)$ observed in the experiment, plus additional terms containing other $b$ states.  Assuming for simplicity a single contribution we may write
\be
\label{eq:a_exp}
\sqrt{a} \, \phi^{*}_a (\xi,\br) = \phi_b(\xi) \varphi_{bx}(\br) ~ + ~\textrm{\Big \{ other  b states \Big \}}
\ee 
where the factor $\sqrt{a}$ accounts for the fact that there are ``$a$'' identical nucleons that can be ``isolated'' from the rest ``$a-1$'' nucleons.  Note that, for simplicity, we have omitted Clebsch-Gordan coefficients (needed to ensure conservation of angular momentum). This expansion defines the overlap function $\varphi_{bx}(\br)$. From the previous equation: 
\be
\label{eq:over_ba}
\sqrt{a} \int d\xi \, \phi^{*}_a (\xi,\br) \phi_b(\xi) \equiv   \varphi_{bx}(\br) .
\ee
The normalization of this overlap is known as {\it spectroscopic factor } (denoted hereafter as $S$),
\be
\label{eq:c2s}
S^{a}_{bx} = \int d\br \, |\varphi_{bx}(\br) |^2  .
\ee
The evaluation of the overlap integrals (\ref{eq:over_ba}) represents a difficult problem, because they require the knowledge of the many-body wave functions of $a$ and $b$. For that reason, in many practical applications, it is customary to approximate  $\varphi_{bx}(\br)$ by a single-particle wave function of $x$ relative to the core $b$:
\be
\label{eq:Camp}
\varphi_{bx}(\br)   \rightarrow C^a_{bx}  \tilde{\varphi}^{n \ell j I}_{bx}(\br) ,
\ee
where $I$ is the spin of the core $b$, $\ell$ the orbital angular momentum and $\vec{j}=\vec{\ell} + \vec{s}$, with $\vec{s}$ the intrinsic spin of $x$. The single-particle wave function is obtained as a solution of a one-body equation of the form:
\be
\label{Hbx}
[\hat{T}_{bx} + V_{bx} -\varepsilon_{\ell j I}] \tilde{\varphi}^{n \ell j I}_{bx}(\br) =0
\ee
where $\varepsilon_{\ell j I}$ is the effective separation energy of the particle $x$ in the system $a$, when the core is left in the state $I$, and $V_{bx}$ some mean-field potential, for example, of Woods-Saxon type. By definition, the solution $\tilde{\varphi}^{n \ell j I}_{bx}(\br)$ is normalized to unity. 

The coefficient $C^a_{bx}$ is called spectroscopic amplitude  and, according to (\ref{eq:c2s}),  $|C^a_{bx}|^2 = S^a_{bx}$. The spectroscopic factor can be understood as the occupation number of the orbital  ${\ell sj}$ with the core in a given state $I$ (of course, other quantum numbers may be required, but we use a loosely notation here assuming that the core states are fully  characterized by $I$). Clearly, the solution of eq.~(\ref{Hbx}) will depend on the choice of the mean-field potential $V_{bx}$. Consequently, the spectroscopic amplitudes (and the corresponding spectroscopic factors), defined according to eq.~(\ref{eq:Camp}), will depend on this choice. This model dependence leads to some ambiguity  in the determination of spectroscopic factors.%
\footnote{Strictly speaking, the occupation probability of a single-particle orbital and its associated energies are not  physical observables. This is because these single-particle states are solutions of a given mean-field potential, but the latter is model-dependent since there is an infinite number of ways of splitting the full many-body Hamiltonian into  mean-field plus residual Hamiltonians. Still, for some standard mean-field potential (eg. Hartree-Fock, Woods-Saxon with some ``standard'' parameters) they provide useful quantities for the interpretation of experimental data. A more formal discussion about the non-observability of spectroscopic factors can be found in ref.~\cite{Dug15}.}

Returning to the evaluation of the inclusive breakup cross section,  the nucleon-removal inclusive cross section is then expressed as \cite{Tos01}:
\be
\sigma_b = \sum_{n \ell j} S^a_{bx}(I;n\ell j)  \sigma_\mathrm{sp}(I;n \ell j) ,
\ee
where $\sigma_\mathrm{sp}(I; n \ell j)$ is the single-particle cross section for the removal of a nucleon from the $n \ell j$ configuration, leaving the core in the state $I$. The sum extends over all allowed configurations $n \ell j$. Owing to the inclusive character of this reaction,  $\sigma_\text{sp}(I; n \ell j)$ contains contributions from elastic and non-elastic  breakup mechanisms which, in the context of this intermediate-energy reactions, are referred to as {\it diffraction} and {\it stripping contributions}, respectively. Thus, we may write $ \sigma_\text{sp}=  \sigma^\text{str}_\mathrm{sp} +  \sigma^\text{diff}_\mathrm{sp}$. 

Hussein and McVoy \cite{HM85} showed that a simple expression for the stripping contribution can be obtained using the so-called eikonal limit, in which one assumes that the projectile  constituents move along  straight-line trajectories, characterized by their asymptotic velocity and impact parameter ($\vec b$). In this limit, known as Glauber optical limit approximation, the distorted waves describing the scattering by an optical potential $U(\bR)$ can be expressed as
\be
\chi(\bK,\bR) = e^{ i \bK \cdot \bR } \exp \left [  -\frac{i}{2 \hbar v} \int_{-\infty}^{z} dz' U(b,z') \right] ,
\ee
where $z$ is the component  of the vector $\bR$ along the $z$-axis. The quantity
\be
S(b)=  \exp \left [ - \frac{i}{\hbar v}  \int_{-\infty}^{\infty} dz' U(b,z') \right] 
\ee
is the so-called profile function, which is nothing but the elastic S-matrix as a function of the continuous parameter $b$. 

In terms of these S-matrices, the stripping single-particle cross section can be written as:
\be
\sigma_\text{sp}^\text{str}= 2 \pi 
  \int b db \int d \vecr \,  |\varphi_{bx}(\vecr)|^2 (1- |S_{xA}(b_x)|)^2   |S_{bA} (b_b) | ^2 \, ,
\ee
where $b_x$ and $b_b$ are the impact parameters of the valence and core, respectively. This equation has an appealing and intuitive form: the integrand contains the product of the probabilities for the core being elastically scattered by the target, $|S_{bA} (b_b) | ^2$, times the probability of the valence particle being  absorbed, $(1- |S_{xA}(b_x)|)^2$. These probabilities are weighted by the projectile wave function squared, and integrated over all possible impact parameters.

Likewise, for the diffraction contribution, one obtains
\begin{equation}
\sigma_\text{sp}^\text{diff}=2 \pi \int b d b \,
\Big[ \langle \varphi_{bx} |\;| S_{b} S_{x} |^2| \varphi_{bx} \rangle-  
|\langle \varphi_{bx} | S_b S_x | \varphi_{bx} \rangle|^2 \Big].
\label{two}
\end{equation}

\begin{figure}
\begin{minipage}[c]{.48\linewidth}
\begin{center}\includegraphics[width=0.8\columnwidth]{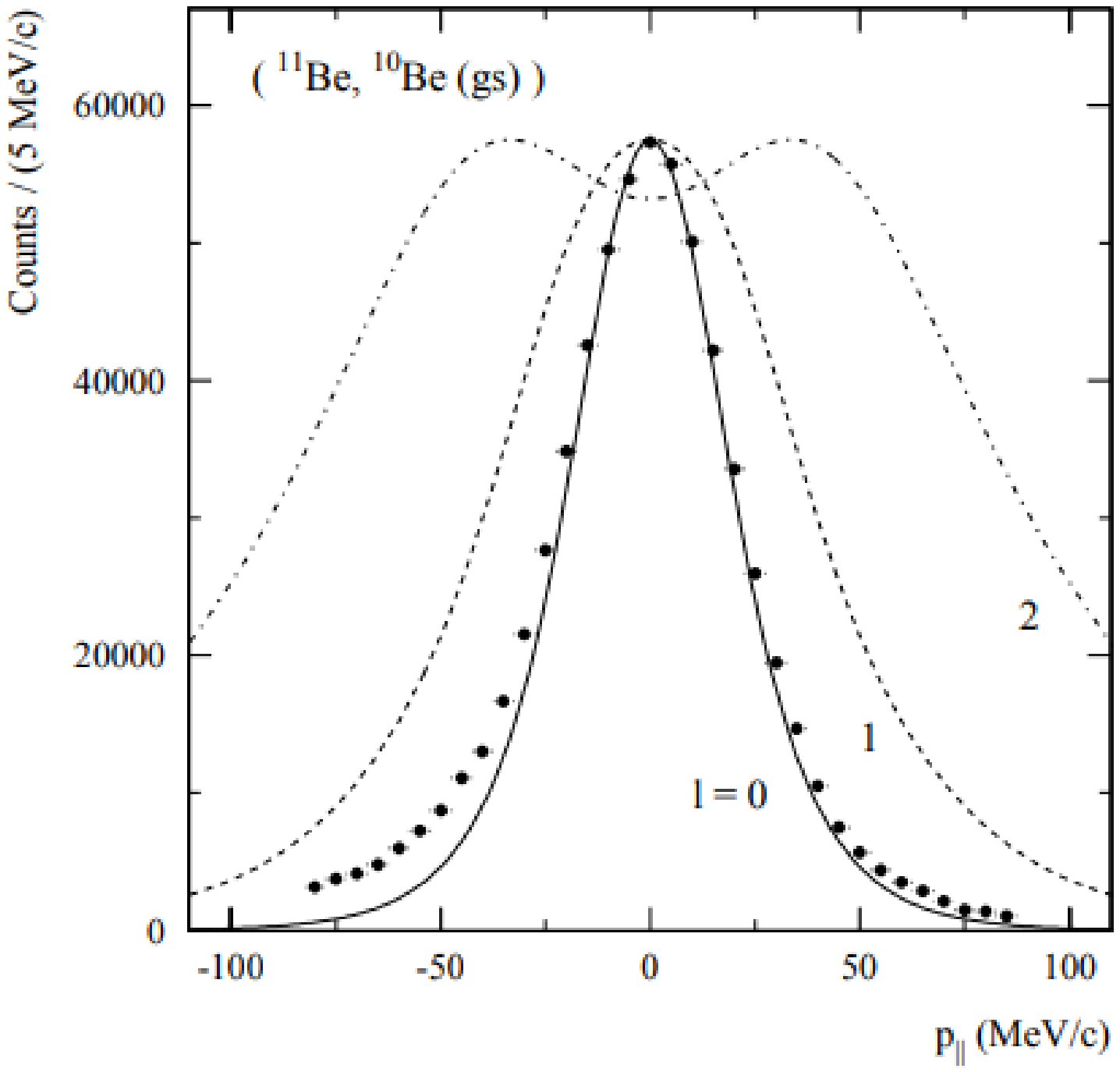} \end{center}
\end{minipage}
\begin{minipage}[c]{.51\linewidth}
\begin{center}\includegraphics[width=0.9\columnwidth]{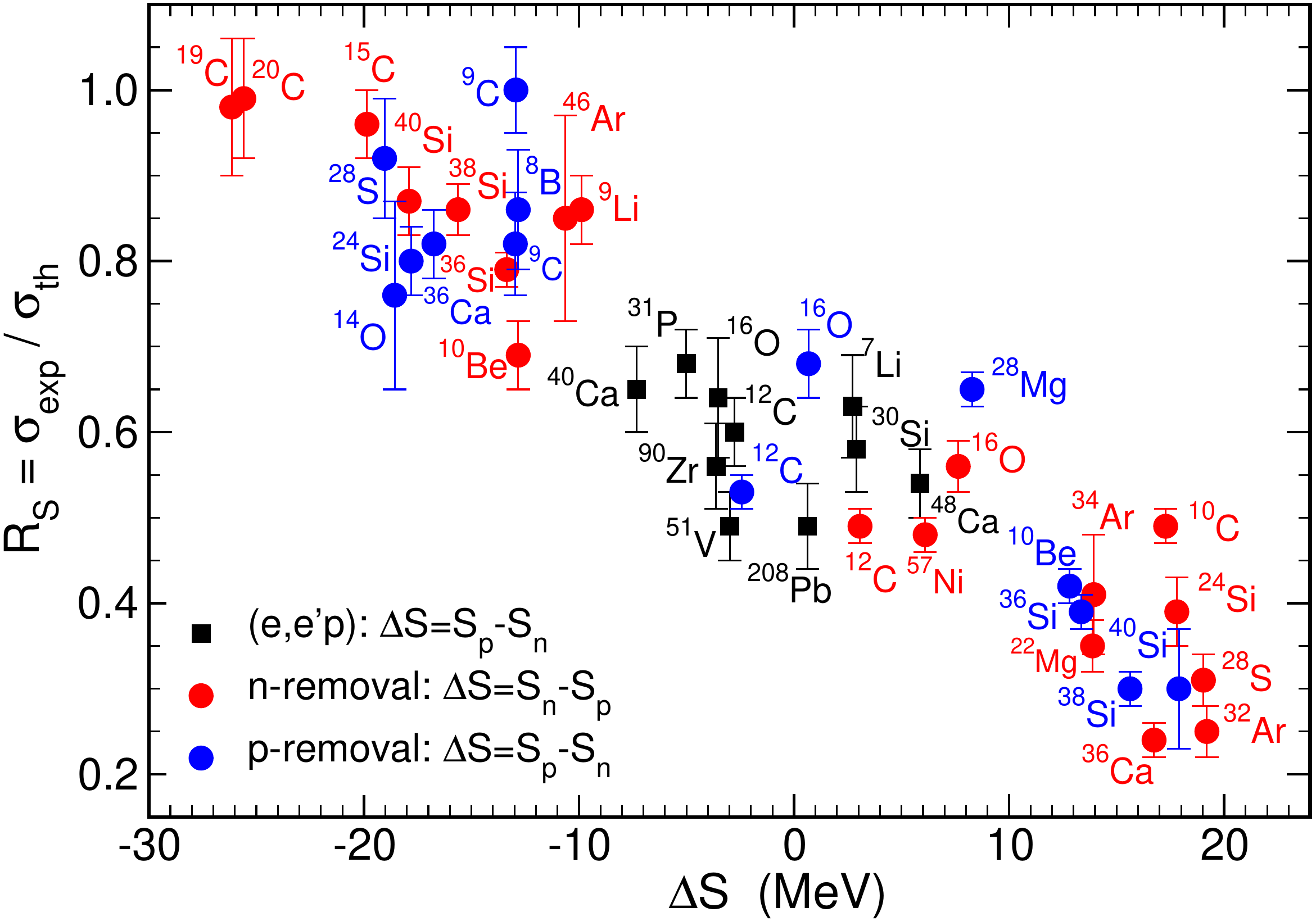} \end{center}
\end{minipage}
\caption{\label{fig:rs}Left panel: longitudinal momentum distribution of the $^{10}$Be fragments, in the rest frame of the projectile, from the knockout reaction $^{9}$Be($^{11}$Be,$^{10}$Be) at 60~MeV/u. The curves are eikonal calculations assuming a nucleon knockout from $s$, $p$, and $d$ states. Quoted from ref.~\cite{Aum00}. Right:  Compilation of the computed ratios  $R_s$
 of the experimental and theoretical inclusive one-nucleon-removal cross sections for each of the projectile nuclei indicated, as a function of the parameter $\Delta S = \pm (S_p - S_n)$,  used as a measure of the asymmetry of the neutron and proton Fermi surfaces. The red points are for neutron-removal cases and the blue points those for proton removal. Taken from ref.~\cite{Tos14}.}
\end{figure}

Comparison of knockout absolute cross sections with the eikonal model  permits in principle the extraction of information about the spectroscopic factors. A recent compilation of these results can be found in ref.~\cite{Tos14}. Usually, these analyses start from some theoretical spectroscopic factors, such as those obtained from truncated-space shell-model calculations and compute the ratio $R_s = \sigma^\mathrm{exp}_{b} / \sigma^\mathrm{theo}_{b}$. Typically, one obtains $R_s<1$, which has been interpreted as an effect of additional correlations not present in small-scale shell-model calculations, and which lead to a larger fragmentation of single-particle strengths (and the subsequent reduction of spectroscopic factors). Moreover, these studies have found a systematic dependence of this ratio on the separation energy of the removed nucleon, with $R_s$ becoming smaller and smaller as the separation energy becomes larger (see right panel of fig.~\ref{fig:rs}). Some authors have interpreted this result as an indication of additional correlations (coming from tensor and short-range portion of the nucleon-nucleon interaction) \cite{Tos14}. However, this interpretation has been recently questioned by other authors, because this trend is apparently not observed in other reactions, such as transfer and $(p,pN)$ reactions, to be discussed below. These results are still under debate.
 
Knockout reactions provide also valuable information about the angular momentum content of the struck nucleon. Assuming a sudden removal of a nucleon, momentum conservation in the projectile rest frame gives
\be
\vec{k}_x = \frac{A-1}{A} \vec{k}_a - \vec{k}_{b} ,
\ee
where $\vec{k}_x$ is the momentum of the struck particle. This expression shows that the latter  can be directly inferred from the  momentum of the observed residue $b$. Normally, either the component of the momentum along the beam direction (longitudinal momentum) or that perpendicular to it (transverse momentum) is used for this purpose. An example is shown in the left panel of fig.~\ref{fig:rs}, corresponding to the longitudinal momentum of the $^{10}$Be(g.s.) fragments produced in the $^{9}$Be($^{11}$Be,$^{10}$Be) reaction at 60~MeV/u, taken from ref.~\cite{Aum00}. The calculations, based on the eikonal model, clearly support a dominance of the $\ell=0$ component, as confirmed indeed by other experiments.

\subsection{Quasi-free $(p,pN)$ reactions}
Breakup experiments of the form $A(p,pn)C$  and $A(p,2p)C$ were used extensively in the 1970s as a tool to extract spectroscopic information on proton-hole and neutron-hole states in nuclei, such as  separation energies, spin-parity assignments, and occupation probabilities. In these reactions, an energetic proton beam ($E > 100$~MeV) collides with a stable target nucleus, removing one or more nucleons, and leaving a residual nucleus ($C$),  either in its ground state, or in an excited state.   

Theoretical analyses of  ($p$, $pN$) reactions with stable nuclei have  traditionally relied on  the distorted-wave impulse approximation (DWIA) \cite{Jac66a,Jac66b},  which  assumes that the binding potential of the removed particle can be neglected in comparison with the projectile--target kinetic energy. At sufficiently high energies (several hundreds of MeV per nucleon) this approximation is expected to be well justified.

 In the DWIA formalism, the transition amplitude for a $A$($p$,$pN$)$C$ reaction (with $A=C+N$ and $N=\{n, p\}$) is given by
\be
{\cal T}_{if}  =  \langle  \chi^{(-)}_{p}(\br_p) \chi^{(-)}_{N}(\br_N)  |  
t_{pN}  | \varphi_a \chi^{(+)}_{pA} \rangle ,  
\label{T3b_Vpn}
\ee
where $\varphi_a$ is the wave function of the removed nucleon relative to the ``core'' nucleus $C$ and 
\begin{align}
\label{tpn_3b}
t_{pN}(E) & = V_{pN} 
           + V_{pN} \frac{1}{E^+ - \hat{K}_r - \hat{K}_R - U_{pC} - U_{NC}-V_{pN}}  V_{pN} , 
\end{align}
which is the T-matrix describing the scattering of the incident proton with the struck nucleon in the presence of the interactions with the core. This is to be compared with the {\it free} $p$-$N$ transition amplitude, i.e., 
\begin{equation}
\label{eq:tfree}
t^f_{pN}(E_{pN})= V_{pN} + V_{pN} \frac{1}{E^{+}_{pN} -\hat{K}_r -V_{pN}}  V_{pN} .
\end{equation}
They are formally related by
\begin{equation}
t_{pN}(E) = t^f_{pN}(E-K_R - U_{pC} - U_{NC}) .
\end{equation}
At sufficiently high incident energies, one may neglect  $U_{pC}$ and $U_{NC}$ in the propagator of eq.~(\ref{tpn_3b}), thus resulting
\begin{align}
{\cal T}_{if} & =  \langle  \chi^{(-)}_{p}(\br_p) \chi^{(-)}_{N}(\br_N)  |  
t^f_{pN}(E-\hat{K}_R)  | \varphi_a  \chi^{(+)}_{pA} \rangle .  
\label{T3b_free}
\end{align}
This corresponds to the distorted-wave impulse approximation (DWIA). The amplitude $t^f_{pN}$ is difficult to evaluate because it contains the operator $\hat{K}_R$. A simple approach to overcome this difficulty is by making he approximation $\hat{K}_R \approx \frac{1}{2}E$, so that $t_{pN}(E) \approx t^f_{pN}(\frac{1}{2}E)$. This approximation has been used, for example, in the context of proton inelastic scattering \cite{Ama80}.
Another common approximation in $(p,pN)$ analyses is the assumption that the T-matrix entering (\ref{T3b_free}) varies sufficiently slowly with momenta, so that its arguments may be replaced by their asymptotic values. In this case, the matrix elements of this T-matrix between these asymptotic momenta can be singled out from the integral, giving rise to a factorized expression for the scattering amplitude, and to  cross sections which are proportional to the free NN cross section. Furthermore, if one describes the motion of the initial and outgoing nuclei by plane waves, one gets the plane wave impulse approximation (PWIA) 
\be
\label{eq:pwia}
{\cal T}^\text{PWIA}_{if} = t_{pN}(\veckp_{pN}, \veck_{pN}; E) F(\vec{Q})
\ee
with the structure formfactor 
$
F(\vec{Q})= \int d\vec{r} e^{-i \vec{Q} \vecr} \varphi_{a}(\vecr) ,
$
where $\vec{Q}= \veckp_p - \veck_N-\frac{A}{A-1} \veck_p$ is the so-called missing momentum.  Expression (\ref{eq:pwia}) shows very clearly the connection between the $(p,pN)$ cross section and the transferred momentum. 
Although formally appealing, the PWIA approximation  is not expected to describe correctly the magnitude of the experimental cross sections because the distortion and absorption effects, arising from the interaction of the core with the incoming proton and with the outgoing nucleons, can be rarely neglected. In DWIA, these distortion effects are taken into account by the (complex) optical potentials used to generate the distorted waves. 


In recent years,  $(p,pN)$ reactions have experienced a renewed interest because of the possibility of performing these measurements with unstable nuclei, using inverse kinematics, i.e., bombarding a hydrogen target with a energetic radioactive beam. This technique is analogous to the knockout experiments with composite targets 
discussed in sec.~\ref{sec:KO}, but it is expected to be sensitive to deeper regions of the nuclear density because of the larger penetrability of the proton probe. In addition to the DWIA method (eg.~\cite{Oga15}), some other approaches are being developed and applied to these reactions, such as the so-called transfer to the continuum method, which is based on the CDCC method \cite{Mor15}, and the Faddeev formalism \cite{Cre14,Cre16}.

While {\it normal} kinematics experiments $A(p,pN)C$  provide typically exclusive observables corresponding to specific outgoing angles of the emitted nucleons, in inverse kinematics experiments  measured observables usually correspond to parallel or longitudinal momentum distributions of the residual nucleus $C$. As we saw in the case of knockout reactions between composite nuclei, the shape of these momentum distributions carries information on the orbital angular momentum of the struck nucleon. Moreover, their magnitude is proportional to the occupation probability of the orbital from which this nucleon has been removed (spectroscopic factor). Therefore, the comparison of the measured distributions with a suitable reaction framework provides useful spectroscopic information.

\section{Breakup reactions (II): semiclassical methods  \label{sec:AW}} 
When the de Broglie wavelength of the projectile is small compared to some characteristic distance of the collision process  one may describe its motion in terms of classical trajectories. This provides a more intuitive and, normally, mathematically simpler description of the reaction. This approximation cannot be done to the internal motion of the nucleons inside the nucleus because their typical wavelength is of the same order as the size of the nucleus and hence quantum effects are important (for example, for typical energy of 30~MeV, $v\sim c/5$ and so $\lambdabar=\hbar/p \approx$ 1~fm). The methods in which the internal excitations are treated quantum-mechanically, while the projectile--target relative motion is treated classically, are called semiclassical methods. There exists in the literature a large variety of such models \cite{Typ94,Esb96,Kid94,Typ01,Cap04,Gar06}.  As an example, we discuss here the one developed  by Alder and Winther and its application to Coulomb excitation experiments. 

\subsection{The semiclassical formalism of Alder and Winther}

In its simplest form, the theory of Alder and Winther \cite{Ald75} assumes that the projectile moves along a classical trajectory, which is not much affected by the internal excitations of the colliding nuclei. This means that: 
$$
\frac{\Delta \ell}{\ell} \ll 1
~ ~
\textrm{and}
~~
 \frac{\Delta \varepsilon_n}{E} \ll 1 .
 $$
The projectile--target interaction is assumed to consists of  two terms: $V(\bR,\xi)= V_0(\bR) + V_\mathrm{coup}(\bR,\xi)$, where $V_0(\bR)$ is independent of the internal coordinates and determines the classical trajectory $\vec{R}(t)$.  The time evolution of the total wave function of the system verifies  the 
 Schr\"odinger equation 
\be
i \hbar {d \Psi(\xi,\theta,t) \over dt} = \left[ V(\bR (\theta,t), \xi) + H_p(\xi) \right] \Psi(\xi,\theta, t) 
\label{eq:TDSE}
\ee
subject to the initial condition $|\Psi (- \infty)\rangle= |0 \rangle$. 

In the spirit of the coupled-channels method, the total wave function is expanded in the internal states of the projectile Hamiltonian $[H_p(\xi) - \varepsilon_n]\phi_n(\xi) $=0   :
\be
\Psi(\xi,\theta,t) = \sum_{n=0} c_n(\theta, t) e^{-i \varepsilon_n t/\hbar} \phi_n(\xi)
\ee
which, when inserted in (\ref{eq:TDSE}), gives rise to the following set of coupled equations for the expansion coefficients: 
\be
\label{eq:cc_clas}
i \hbar {d c_n(\theta, t) \over dt} = \sum_m e^{-i (\varepsilon_m-\varepsilon_n) t/\hbar} 
V_{nm}(\theta,t) c_m(\theta, t)
\ee
with the initial condition $c_n(\theta,- \infty)= \delta_{n0}$. 
The time-dependent coupling potentials $V_{nm}(\theta,t)$ are given by:
\be
V_{nm}(\theta,t) = \int d \xi \phi_n^*(\xi) V(\bR(\theta, t), \xi)
\phi_m(\xi)
\ee

Once the coefficients are obtained, the excitation probability for a $0\rightarrow n$ transition is given by:
$$
  P_n(\theta)=|c_n(\theta, \infty)|^2 ,
$$
and the differential cross section 
 $$
 \left({d \sigma \over d \Omega}\right)_{0\to n} = 
 \left({d \sigma \over d \Omega}\right)_\mathrm{clas} P_n(\theta) .
 $$
where $(d \sigma / d \Omega)_\mathrm{clas}$ is the classical differential elastic cross section which, for a pure Coulomb case, coincides with the Rutherford cross section.

Due to the conservation of the total probability (flux), one has
$$
\sum_n P_n(t) = \sum_{n} |c_n(t)|^2 =1 .
$$

When the couplings are weak, one may solve (\ref{eq:cc_clas}) perturbatively, assuming that $c_0 \approx 1$ and $c_n \ll 1$ for $n>0$. This gives the first order solution
\be
c_n(\theta) \equiv c_n(\theta, \infty) \simeq 
 {1 \over i \hbar}\int_{-\infty}^{\infty}
 e^{-i (\varepsilon_0-\varepsilon_n) t/\hbar}
 V_{n0}(\theta,t) dt .
\ee

In the important case of pure Coulomb scattering, which was the case studied in detail by Alder and Winther \cite{Ald75}, one finds analytical expressions for the excitation probabilities. In particular, the first-order excitation probability for a $0\rightarrow n $ transition, due to the electric Coulomb operator E$\lambda$ , results
%
%
\be
\left( {d \sigma \over d \Omega}\right)_{0\rightarrow n}= \left({ Z_t  e^2 \over   \hbar v}\right)^2
{B(E \lambda, 0 \to n) \over  e^2 a_0^{2 \lambda-2}} f_\lambda(\theta,\xi)
\quad \quad \quad 
(\theta < \theta_\text{gr})
\ee
which is valid only for angles smaller than the grazing one ($\theta_\text{gr}$) and 
where 
$a_0$ is half of the distance of closest approach in a head-on collision, $\xi_{0 \rightarrow n} =  \frac{(\varepsilon_n-\varepsilon_0)}{\hbar} \frac{a_0}{v}$ is the adiabaticity parameter and $f_\lambda(\theta,\xi)$ is an analytic function, depending on the kinematical conditions, but independent of the structure of the projectile. 

For weakly-bound nuclei, the excitation occurs to unbound (continuum) states. The previous formula can be generalized to:
\be
\label{eq:dsdwde}
\frac{d \sigma(E\lambda)}{d \Omega d \varepsilon} = \left({ Z_t  e^2 \over   \hbar v}\right)^2
{1 \over  e^2 a_0^{2 \lambda-2}}
\frac{dB(E \lambda)}{d \varepsilon} 
{df_\lambda(\theta,\xi) \over d\Omega}
\quad \quad \quad 
(\theta < \theta_\text{gr})
\ee
where  $df_\lambda(\theta,\xi) / d\Omega$ is also a well-defined analytic function and   $dB(E \lambda)/d\varepsilon$ is the  electric reduced probability to the continuum states. 

An appealing feature of these formulas is that they provide a neat separation between the structure of the projectile (through $B(E\lambda)$) and the reaction dynamics. This separation allows in principle to extract the ${dB(E \lambda)}/{d\varepsilon}$ distribution from the analysis of small-angle Coulomb dissociation data. For that, one can integrate the double differential cross section (\ref{eq:dsdwde}) up to a maximum angle ($\theta_\mathrm{max}$) such that for $\theta < \theta_\mathrm{max}$ the breakup can be assumed to be purely Coulomb and hence  (\ref{eq:dsdwde}) is valid ($ \theta_\mathrm{max} <\theta_\text{gr}$). This gives:
\be
\frac{d\sigma}{d\varepsilon} (\theta < \theta_\mathrm{max}) = 
\int_{0}^{\theta_\mathrm{max}}  \frac{d \sigma(E\lambda)}{d \Omega d \varepsilon} d\Omega
\propto \frac{dB(E \lambda)}{d \varepsilon}  \, .
\ee
\begin{figure}
\begin{minipage}[c]{.5\linewidth}
\begin{center}\includegraphics[width=0.9\columnwidth]{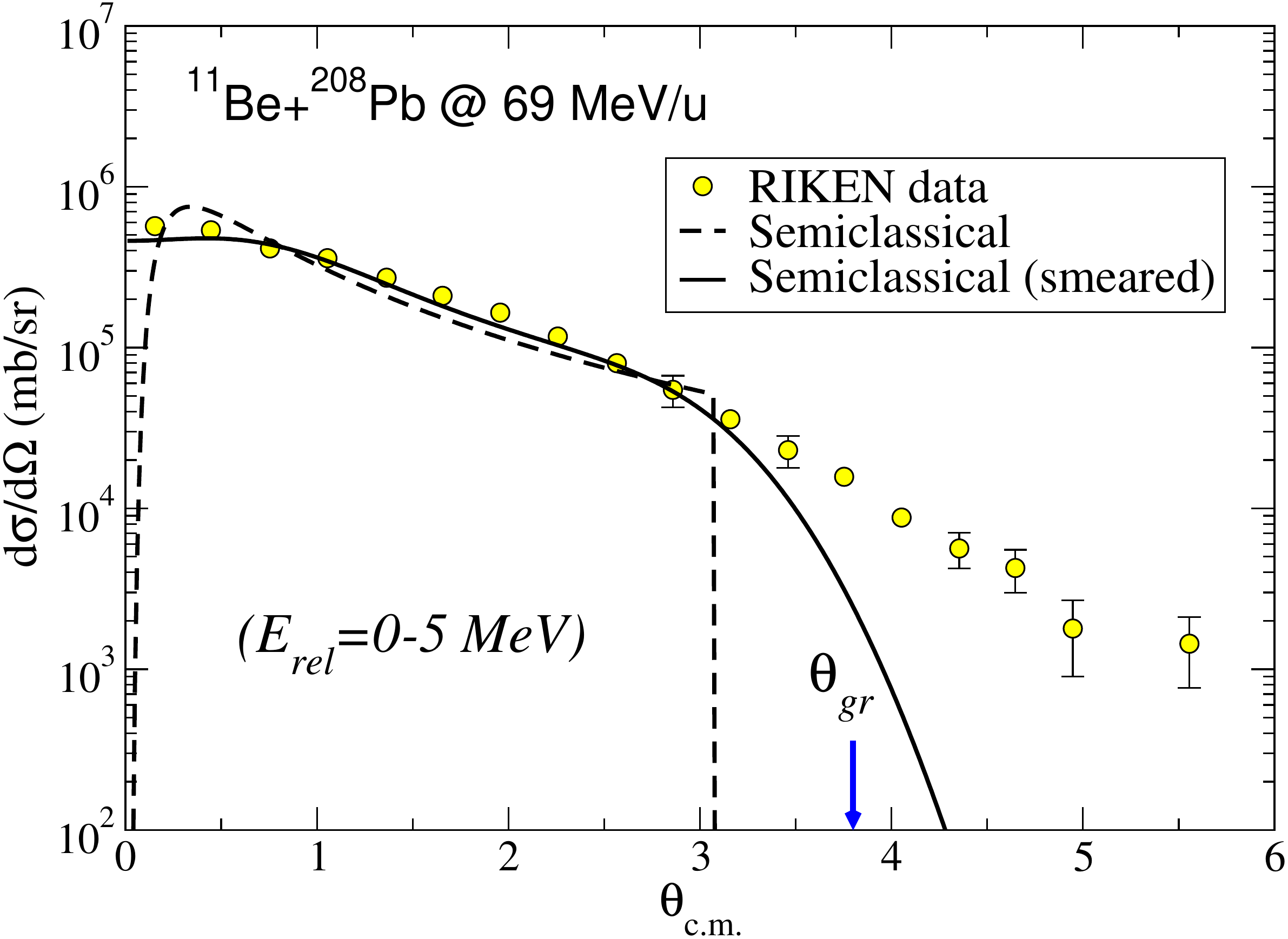} \end{center}
\end{minipage}
\begin{minipage}[c]{.5\linewidth}
\begin{center}\includegraphics[width=0.9\columnwidth]{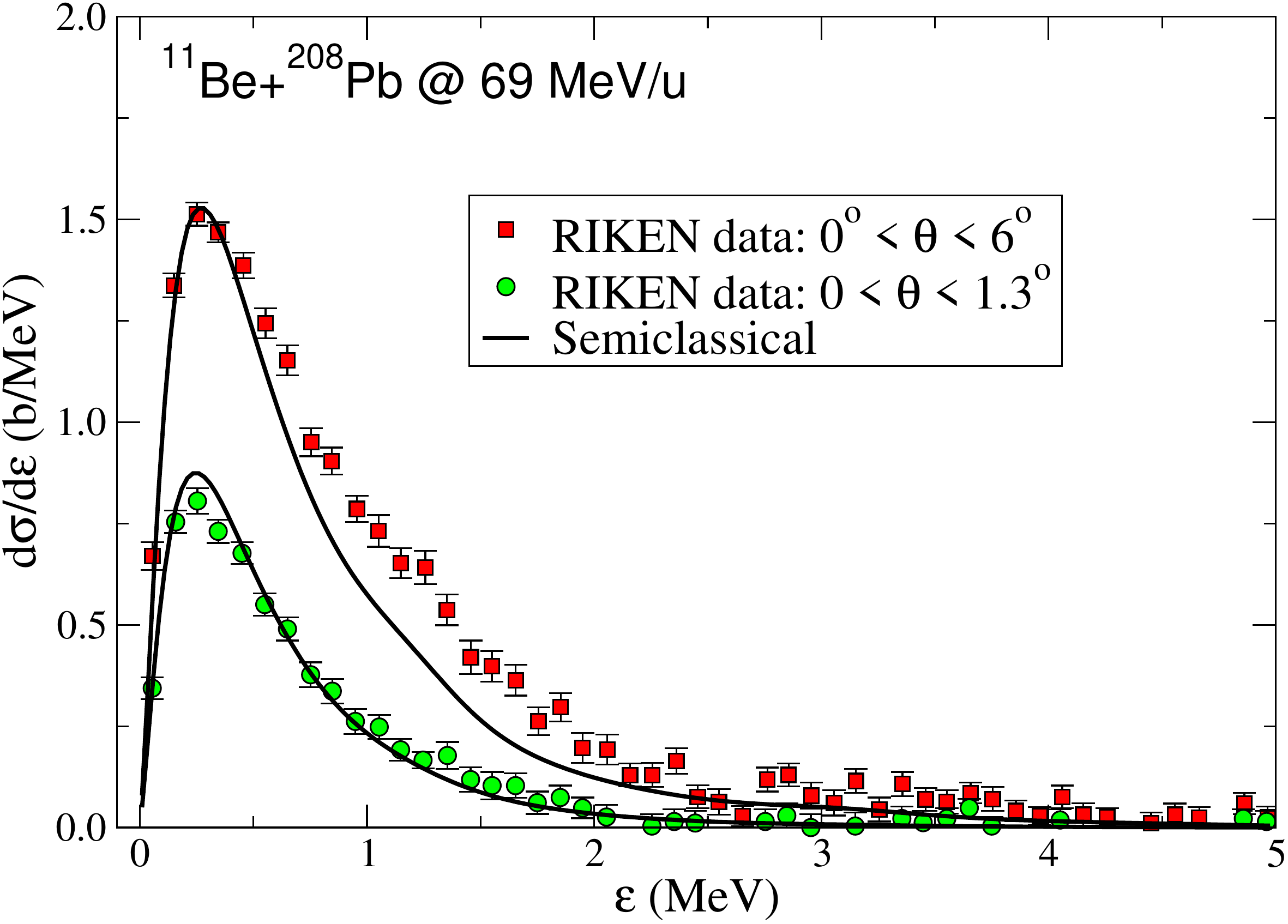} \end{center}
\end{minipage}
\caption{Left: angular distribution of the c.m.\ of the outgoing $n$+$^{10}$Be system, following the breakup of  $^{11}$Be on $^{208}$Pb at 69~MeV/u. Right: relative energy distribution between the $n$+$^{10}$Be  fragments, integrated up to two different maximum angles, as indicated by the labels. The curves are semiclassical calculations based on the theory of Alder and Winther and smeared with the experimental resolution. }
\label{fig:be11pb_epm} 
\end{figure}

An application of this method to the  $^{11}$Be+$^{208}$Pb  reaction discussed above is shown in fig.~\ref{fig:be11pb_epm}. The curves are the result of semiclassical calculations, convoluted with the experimental energy and angular resolutions. On the left panel, it is apparent that the semiclassical calculation reproduces well the data up to
 $\theta_\text{c.m.} \sim 3^\circ$. For larger angles, the assumption of pure Coulomb scattering is no longer accurate (the grazing angle is estimated to be $\theta_\text{gr}=3.8^\circ$). In the right panel, the corresponding relative-energy distributions are compared with the data, for two angular cuts. It is seen that, for the wider angular range ($\theta_\mathrm{c.m.} < 6^\circ$), there is some underestimation of the data, which may be attributed to the omission of nuclear effects and higher Coulomb multipoles. 

\begin{figure}
\begin{center}\includegraphics[width=0.5\columnwidth]{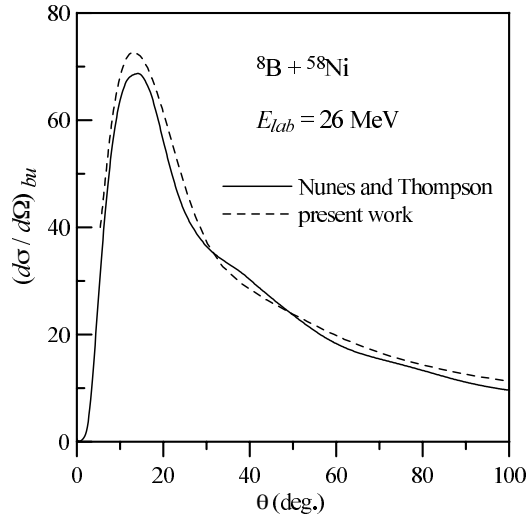} \end{center}
\caption{\label{fig:b8ni_dsdw}Breakup of $^{8}$B$\rightarrow$ $^{7}$Be+p on a $^{58}$Ni target at 26~MeV. Coupled-channel semiclassical calculations, using Coulomb+nuclear trajectories, are compared with CDCC calculations, which include also nuclear and Coulomb couplings. Quoted from \cite{Mar02}.}  
\end{figure}
The assumption of  pure Coulomb trajectories can be relaxed, at the expense of loosing some of the simplicity of the method. A compelling application is shown in fig.~\ref{fig:b8ni_dsdw} (taken from \cite{Mar02}) where semiclassical coupled-channels calculations, using trajectories modified by the nuclear interaction, are compared with CDCC calculations for the breakup angular distribution of the $^{8}$B+$^{58}$Ni reaction at the near-barrier energy of 26 MeV.

\subsection{Dynamic Coulomb polarization potential  from the AW theory}
We saw in sec.~\ref{sec:cdp} that the effect of the coupling to the breakup channels due to the dipole Coulomb interaction is given in the  adiabatic limit  by a simple analytical form.  The resultant expression [eq.~(\ref{eq:adpol})] is purely real, and depends on the structure of the dissociated nucleus through the so-called polarizability parameter. 

This  adiabatic expression assumes that the excitation energies are large and, therefore,  is not applicable to the Coulomb breakup of very weakly-bound nuclei, for which the average excitation energies are typically small (of the order of 1~MeV or less). 

\begin{figure}
\begin{center}\includegraphics[%
  width=0.65\textwidth]{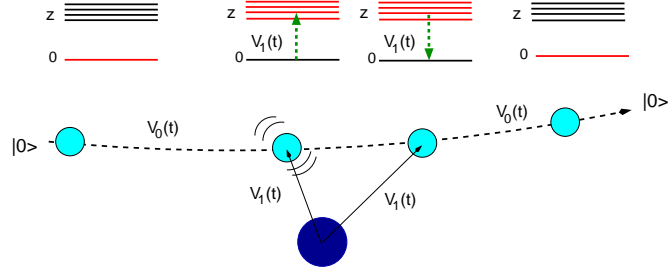}\end{center}
\caption{\label{fig:a1_traj} Second-order process describing the excitation and de-excitation of a nucleus moving along a classical trajectory determined by the potential $V_0(t)$, due to the action of the potential $V_1(t)$.}
\end{figure}

A non-adiabatic (also called {\it dynamic}) polarization potential can be derived from the Alder and Winther theory \cite{And94,And95}. For that, we consider the second order process in which the projectile is excited to a continuum state due to the dipole Coulomb interaction ($V_1(t)$) and then de-excites, returning to the ground state. The situation is schematically depicted in fig.~\ref{fig:a1_traj}.
The amplitude probability for this process is given by:
\begin{align}
c^{(2)}_n = & \sum_{z} \left({-i \over \hbar}\right)^2 
  \int_{-\infty}^{+\infty} dt  \langle  n | V_1(t) | z \rangle
  \exp \left\{ {i \over \hbar}(\varepsilon_n - \varepsilon_z) t \right\} 
  \int_{-\infty}^{t} dt'  \langle  z | V_1(t') | 0 \rangle
  \exp \left\{ {i \over \hbar}(\varepsilon_z - \varepsilon_0) t' \right\} ,
\end{align}
where $|z\rangle$ denotes the intermediate states populated during the reaction. 
Then, one requires that this second-order amplitude coincides with the first-order amplitude associated with the polarization potential  for all classical trajectories corresponding to a given scattering energy. 
This gives rise to:
\begin{align}
\label{eq:Upol}
U_\mathrm{pol}(R)&=-\frac{4\pi }{9}\frac{Z_t^{2}e^{2}}{\hbar v}\frac{1}{(R-a_{o})^{2}R} 
\int ^{\infty }_{0}d\varepsilon \frac{dB(E1,\varepsilon )}{d\varepsilon }
\left[ g\left(\frac{R}{a_{o}}-1,\xi \right)+if\left(\frac{R}{a_{o}}-1,\xi \right) \right] , 
\end{align}
where \emph{g} and \emph{f} are analytic functions defined as
\begin{eqnarray}
\label{eq:ucdp}
f(z,\xi ) &=& 4\xi ^{2}z^{2}\exp{(-\pi \xi )}K_{2i\xi }''\left(2\xi z\right), \\
g(z,\xi ) &=& \frac{P}{\pi }\int _{-\infty }^{\infty }\frac{f(z,\xi ')}{\xi -\xi '}d\xi ',
\end{eqnarray}
and $\xi$   is the Coulomb adiabaticity 
parameter corresponding to the excitation to the continuum energy $\varepsilon$ of the nucleus.
An important feature of this potential is that when the breakup energy \( \varepsilon _{b} \)
is large enough, the purely real adiabatic dipole potential is re-obtained. In
the opposite limit, for small breakup energies, \( f\left(\frac{R}{a_{o}}-1,\xi \right)\rightarrow 1 \)
and \( g\left(\frac{R}{a_{o}}-1,\xi \right)\rightarrow 0 \), and the polarization potential
becomes purely imaginary, depending on $R$ as \( {1}/[(R-a_{o})^{2}R]. \)

\begin{figure}
\begin{center}
\begin{minipage}[t]{.49\textwidth}
{\par \resizebox*{0.8\textwidth}{!}
{\includegraphics{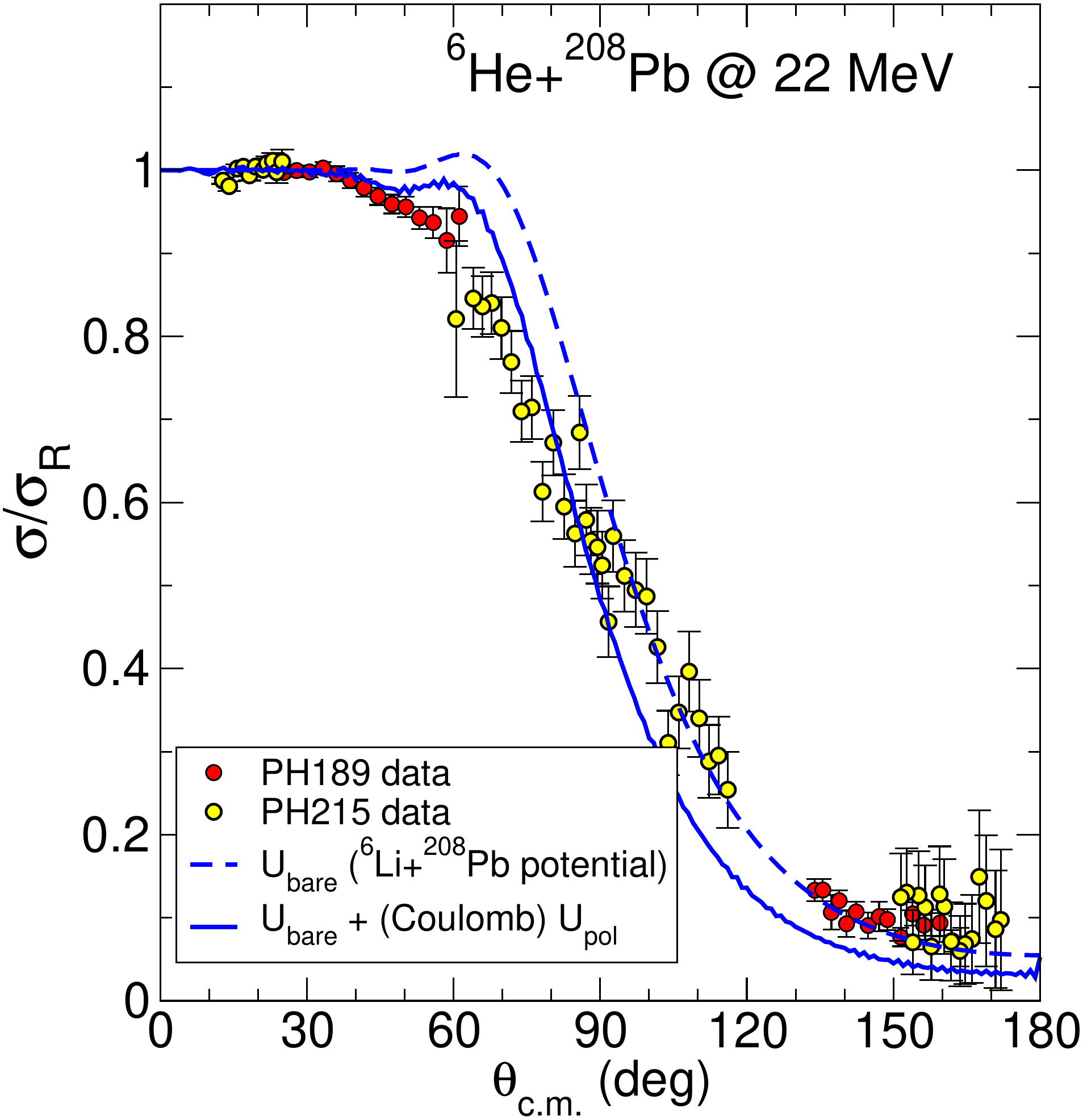}} \par}
\end{minipage}
\begin{minipage}[t]{.49\textwidth}
{\par \resizebox*{0.8\textwidth}{!}
{\includegraphics{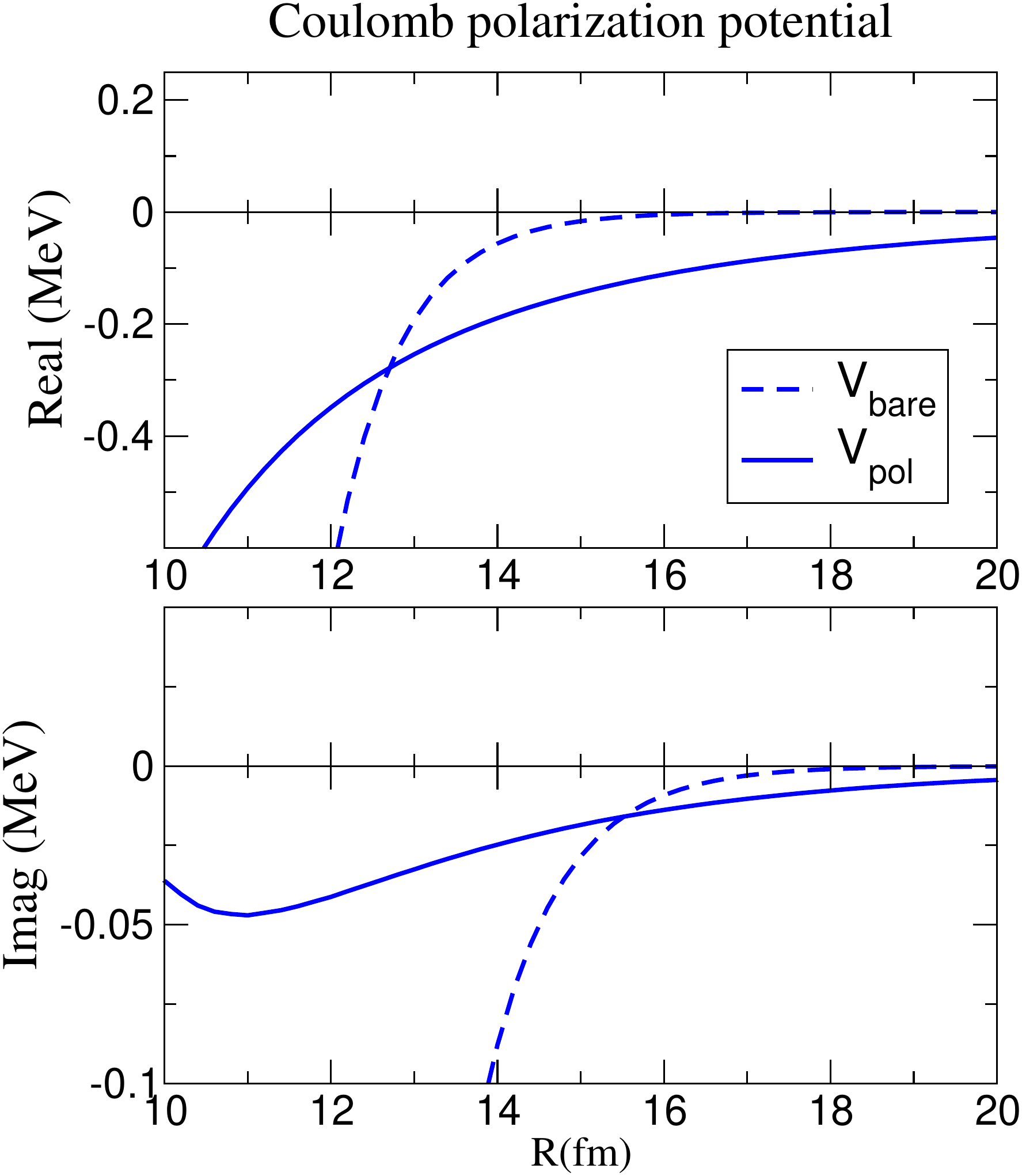}} \par}
\end{minipage}
\caption{\label{fig:he6pb_cdp} Left: Elastic scattering differential cross section for  $^{6}$He+$^{208}$Pb at $E_\mathrm{lab}=22$~MeV. The dashed line is a single-channel calculation performed with a {\it bare} potential, given by a $^{6}$Li+$^{208}$Pb optical potential. The solid line is the optical model calculation obtained with the bare plus dynamic Coulomb polarization potential of eq.~(\ref{eq:Upol}). Right: Real and imaginary parts of the bare and CDP potentials.}
\end{center}
\end{figure}

As an example, we show in the LHS of fig.~\ref{fig:he6pb_cdp} the elastic scattering data for $^{6}$He+$^{208}$Pb at 22~MeV compared with the optical model calculations obtained with a bare potential alone (dashed) and with the bare plus the Coulomb dipole polarization (CDP) potential (solid line). For the bare interaction we have used an optical potential for the nearby projectile $^{6}$Li, who has a similar structure to that of $^{6}$He but which does not exhibit the strong dipole excitation mechanism. The elastic cross section predicted by the bare potential presents a Fresnel-like diffraction pattern, not observed in the data. The inclusion of the CDP reduces significantly the elastic cross section and suppresses the Fresnel peak. On the RHS, the real and imaginary parts of the bare and CDP potentials are shown. Their most remarkable feature is their long-range, which was anticipated from the optical model analysis of fig.~\ref{fig:hepb_om}, and is associated to the long-range of the dipole Coulomb couplings, as we have already discussed in the context of the CDCC method. 


\section{Transfer reactions \label{sec:transfer}}
Another example of direct reaction is that of transfer reactions. In this kind of processes, the projectile and target exchange one or more nucleons. Compared to inelastic scattering, the modeling of transfer reactions has the added difficulty that the initial and final nuclei are different so one has to deal with two different mass partitions  (see fig.~\ref{fig:post-prior}). The analysis of transfer reactions has been traditionally performed using the DWBA method. In the next section, we derive the DWBA formula from the exact transition amplitude. 

\subsection{An exact expression for the transfer amplitude}  
\begin{figure}
\begin{center}
{\par \resizebox*{0.6\textwidth}{!}
{\includegraphics{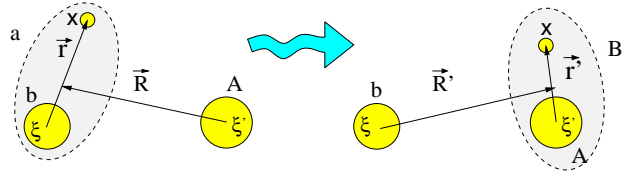}} \par}
\end{center}
\caption{Schematic representation of a transfer reaction, showing the relevant coordinates for the initial and final states.}
\label{fig:post-prior} 
\end{figure}

Following the notation introduced in the introductory section, the initial and final partitions will be denoted as $\alpha$ and $\beta$, respectively.
Accordingly, the full Hamiltonian can be written in two different forms, depending on whether we choose the coordinates of the initial (LHS picture of fig.~\ref{fig:post-prior}) or the final (RHS picture) configuration. For example, if we choose the final ({\it post}) representation, it reads
\be
  H = \hat{T}_{\bR'} + H_\beta(\xi_\beta) + V_\beta(\bR',\brp)
      = \hat{T}_{\bR'} + \underbrace{ H_{B}(\xi',\brp) +
       H_b(\xi)}_{H_\beta(\xi_\beta)} + \underbrace{V_{bx}+U_{bA}}_{V_\beta(\bR',\brp)} 
\ee 
where we have used the notation $\xi_\beta=\{ \xi, \xi', \brp \}$ to denote the set of internal coordinates.

The total wave function, which must contain at least the incident (elastic) and the transfer channel of interest, must behave asymptotically according to eq.~(\ref{eq:Psi-asym}).
To obtain the transfer scattering amplitude ($f_{\beta,\alpha}(\theta)$ in that equation), we could proceed as in the coupled-channels method, by writing an {\it ansatz} for $\Psi^{(+)}_{\bK}(\bR,\xi_\alpha)$ which includes the desired transfer channels $\beta$. This procedure gives rise to a set of coupled integro-differential equations involving non-local couplings. This is the so-called Coupled Reaction Channels (CRC) method \cite{Sat83,Tho09}. This is in fact the procedure followed by some computer codes, such as the popular code {\tt FRESCO} \cite{fresco}. From the formal point of view, a more straightforward derivation of the scattering amplitude can be obtained making use of the general expression (\ref{eq:Tdist}) for the transition amplitude, which we rewrite as
\be
\label{eq:T-trans_exact}
{\cal T}^\mathrm{post}_{\beta,\alpha} = 
\int \int \chi_{\beta}^{(-)*}(\bK', \bR') 
  \Phi^{*}_\beta (\xi_\beta) (V_\beta-U_\beta)  \Psi^\mathrm{(+)}_{\bK}(\bR,\xi_\alpha) d\xi_\beta d\bR' \, ,
\ee
where $\Psi^\mathrm{(+)}_{\bK}(\bR,\xi_\alpha)$ is the  exact wave function, 
$\Phi_\beta (\xi_\beta)$ represents the internal state of the final nuclei, which are eigensolutions of the internal Hamiltonian $H_\beta$: 
$$
 H_\beta \Phi_\beta (\xi_\beta)= \varepsilon_{\beta}  \Phi_\beta (\xi_\beta) \quad \text{with} \quad
 \Phi_\beta (\xi_\beta)=\phi_b(\xi)\phi_B(\xi',\brp),
$$ 
where $\varepsilon_{\beta}$ represents the sum of the internal energies of the outgoing nuclei. Finally, $\chi_{\beta}^{(-)}$ is the time-reversal of $\chi_{\beta}^{(+)}$, which is the distorted wave generated by the auxiliary potential $U_\beta$
\be
\left[E-\varepsilon_{\beta}-\hat{T}_{\bR'}-U_{\beta}(\bR') \right] \chi^{(+)}_{\beta}(\bK',\bR')  =  0 \, .
\ee 

\subsection{The DWBA approximation}
The transition amplitude (\ref{eq:T-trans_exact}) can not be directly evaluated because it contains the exact scattering of the many-body problem ($\Psi^\mathrm{(+)}_{\bK}$). A solvable formula can be obtained making the approximation 
\be
\Psi^\mathrm{(+)}_{\bK} (\bR,\xi_\alpha) \approx \chi^{(+)}_{\alpha}(\bK,\bR) \Phi_\alpha(\xi_\alpha),
\ee
where $\Phi_\alpha (\xi_\alpha)=\phi_a(\xi,\br)\phi_A(\xi')$ and 
where the distorted wave $\chi^{(+)}_{\alpha}$, which describes the projectile--target relative motion in the entrance channel, is generated with some potential $U_\alpha$, that is:
\begin{equation}
\left[E-\varepsilon_{\alpha}-\hat{T}_{\bR}-U_{\alpha}(\bR) \right] \chi^{(+)}_{\alpha}(\bK,\bR)  =  0 
\end{equation}
 Typically, $U_\alpha$ is chosen so as to  reproduce  the elastic scattering differential cross section. This approximation gives rise to the DWBA scattering amplitude:
\be
\label{eq:T-dwba}
{\cal T}^\mathrm{DWBA}_{\beta,\alpha} = 
\int \int \chi_{\beta}^{(-)*}(\bK', \bR') 
  \Phi^{*}_\beta (\xi_\beta) (V_\beta-U_\beta)    \chi^{(+)}_{\alpha}(\bK,\bR) \Phi_\alpha(\xi_\alpha) d\xi_\beta d\bR' \, .
\ee

Let us consider for simplicity the important $(d,p)$ case. The post-form interaction $V_\beta$ is given by 
$V_\beta= V_{pn}+ U_{pA}$. Moreover, the internal states and internal coordinates are given in this case by:
\begin{align}
\Phi_\alpha(\xi_\alpha) & = \varphi_d(\br) \phi_A(\xi')  &    \xi_\alpha= \{ \xi', \br \} \\
\Phi_\beta(\xi_\beta) & =   \Phi_B(\xi',\brp)            &   \xi_\beta= \{ \xi', \brp \}
\end{align}
where $\br$ is the proton-neutron relative coordinate and $\brp$ that of the transferred neutron relative to the target nucleus (c.f.\ fig.~\ref{fig:post-prior}). Moreover, for not very light targets, we can further approximate: $U_{pA} \approx U_{pB}$ $\Rightarrow$ $V_{pn}+ U_{pA}-U_{pB} \approx V_{pn}(\br)$. The amplitude is then dominated by the $V_{pn}$ interaction and hence by small $p$-$n$ separations.   Then, (\ref{eq:T-dwba}) becomes in the $(d,p)$ case:
\be
\label{eq:Tdwba-dp}
{\cal T}^\mathrm{DWBA}_{d,p} = 
\int \int \chi_{p}^{(-)*}(\bK_p, \bR') 
  \Phi^{*}_B(\xi',\brp)   V_{pn}(\vecr) \chi^{(+)}_{d}(\bK_d,\bR)  \varphi_d(\br) \phi_A(\xi') d\xi_\beta d\bR' \, .
\ee

\begin{figure}
\begin{minipage}[t]{.35\linewidth}
\begin{center}\includegraphics[width=0.7\columnwidth]{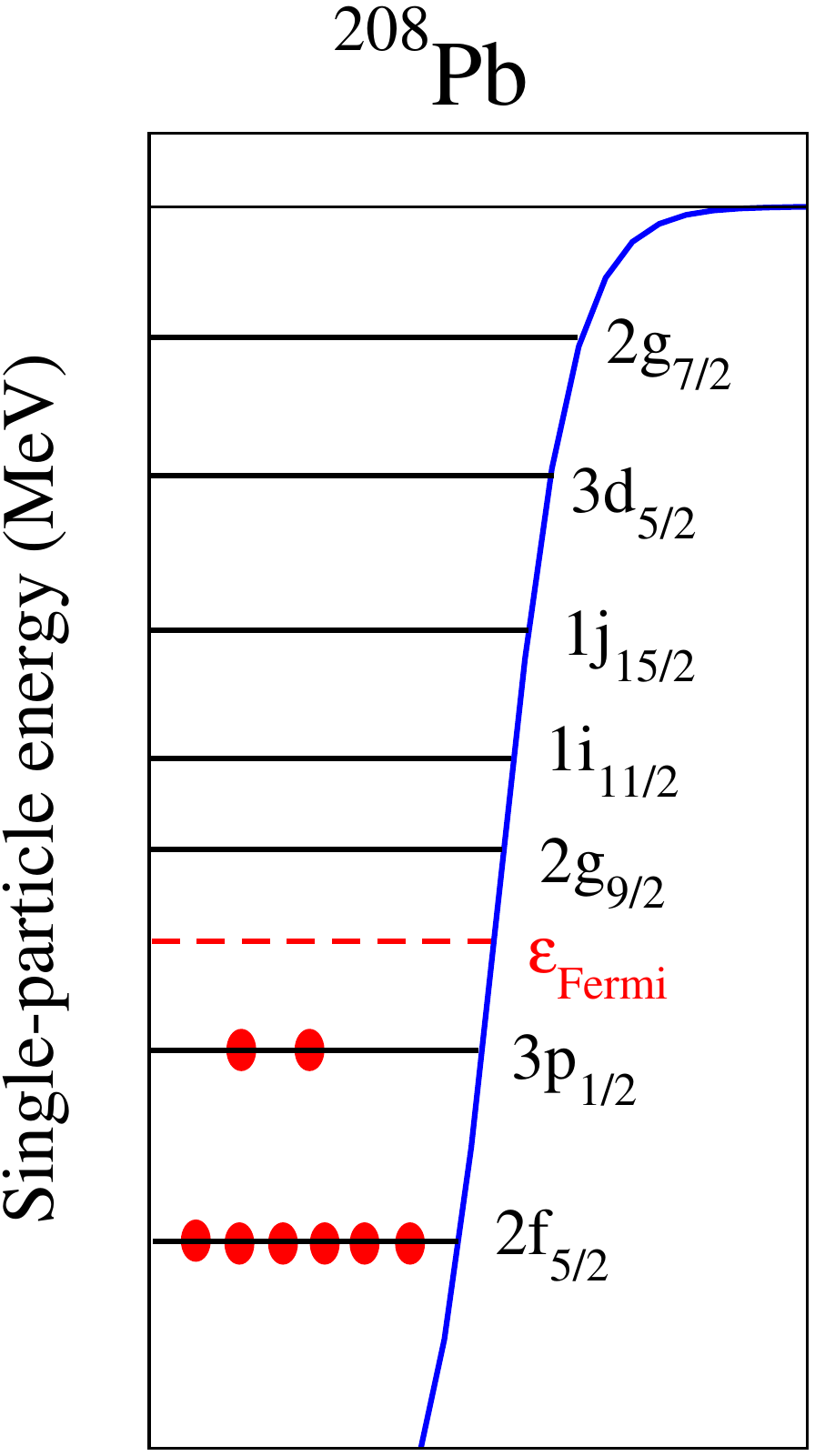} \end{center}
\end{minipage}
\begin{minipage}[t]{.55\linewidth}
\begin{center}\includegraphics[width=0.95\columnwidth]{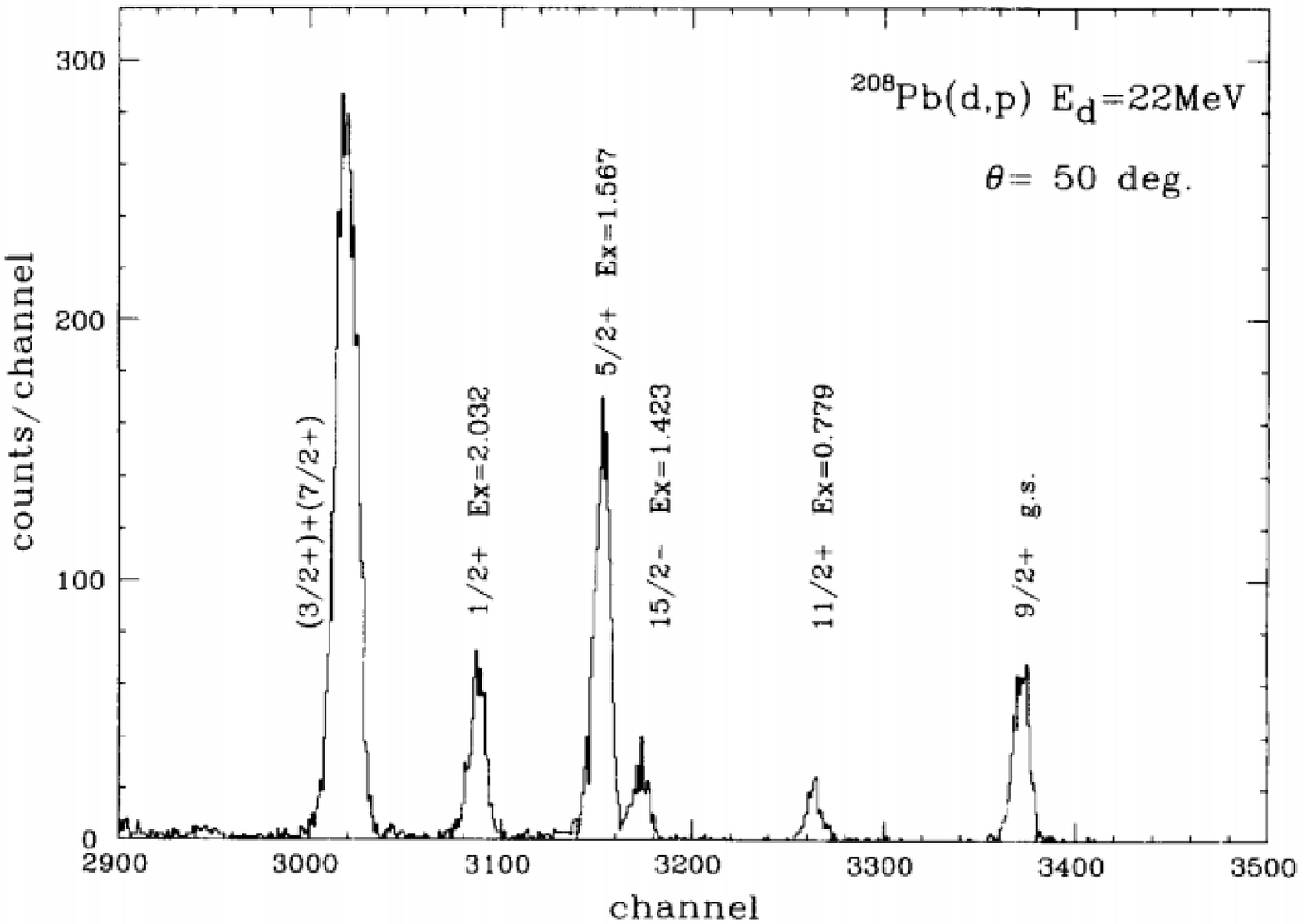} \end{center}
\end{minipage}
\caption{\label{fig:pb208dp}Left: Single-particle orbitals for $^{208}$Pb. Right: Experimental energy spectrum of protons for the $^{208}$Pb($d$,$p$) reaction at $E_d=22$~MeV at $50^\circ$.  Quoted from ref.~\cite{Hir98} with permission from Elsevier.} 
\end{figure}

The integral in the target coordinates $\xi$ involve the  overlap function between the target $A$ and the residual nucleus $B$ wave functions. Resorting to  eqs.~(\ref{eq:over_ba}), (\ref{eq:c2s}) and (\ref{eq:Camp})  we end up with the transition amplitude:
\be
{\cal T}^\mathrm{DWBA}_{d,p} =
C^B_{An}
\int \int \chi_{p}^{(-)*}(\bK_p, \bR') 
 \tilde{\varphi}^{\ell j I}_{An}(\brp)^{*} ~V_{pn}(\br) ~ \chi^{(+)}_{d}(\bK_d,\bR) \, \varphi_{d}(\br) d\brp d\bR' ,
\ee
and the corresponding differential cross section
\be
\label{eq:dsdw_dwba}
\left (\frac{d\sigma}{d\Omega} \right)_{\beta,\alpha} =
\frac{\mu_\alpha \mu_\beta}{(2 \pi \hbar^2)^2}
S^B_{An} 
\left | \int \int \chi_{p}^{(-)*}(\bK_p, \bRp) \,
 \tilde{\varphi}^{\ell j I}_{An}(\brp)^{*} ~V_{pn}(\br) ~ \chi^{(+)}_{d}(\bK_d,\bR) \, \varphi_{d}(\br) d\brp d\bRp \right |^2 .
\ee

The presence of the overlap function $S^B_{An}  \tilde{\varphi}^{\ell j I}_{An}(\brp)$ tells us that the transfer cross section contains information about the single-particle content of a given state of the residual nucleus. States with a strong single-particle character (large spectroscopic factor) will be strongly populated. A extreme case is that of a single nucleon outside a closed-shell core, such as in the $^{208}$Pb($d,p$)$^{209}$Pb reaction. This is shown in fig.~\ref{fig:pb208dp}. Low-lying states above the Fermi level in $^{209}$Pb are populated and this allows for a natural spin assignment according to a simple shell-model picture.

\begin{figure}
\begin{minipage}[t]{.45\linewidth}
\begin{center}\includegraphics[width=0.85\columnwidth]{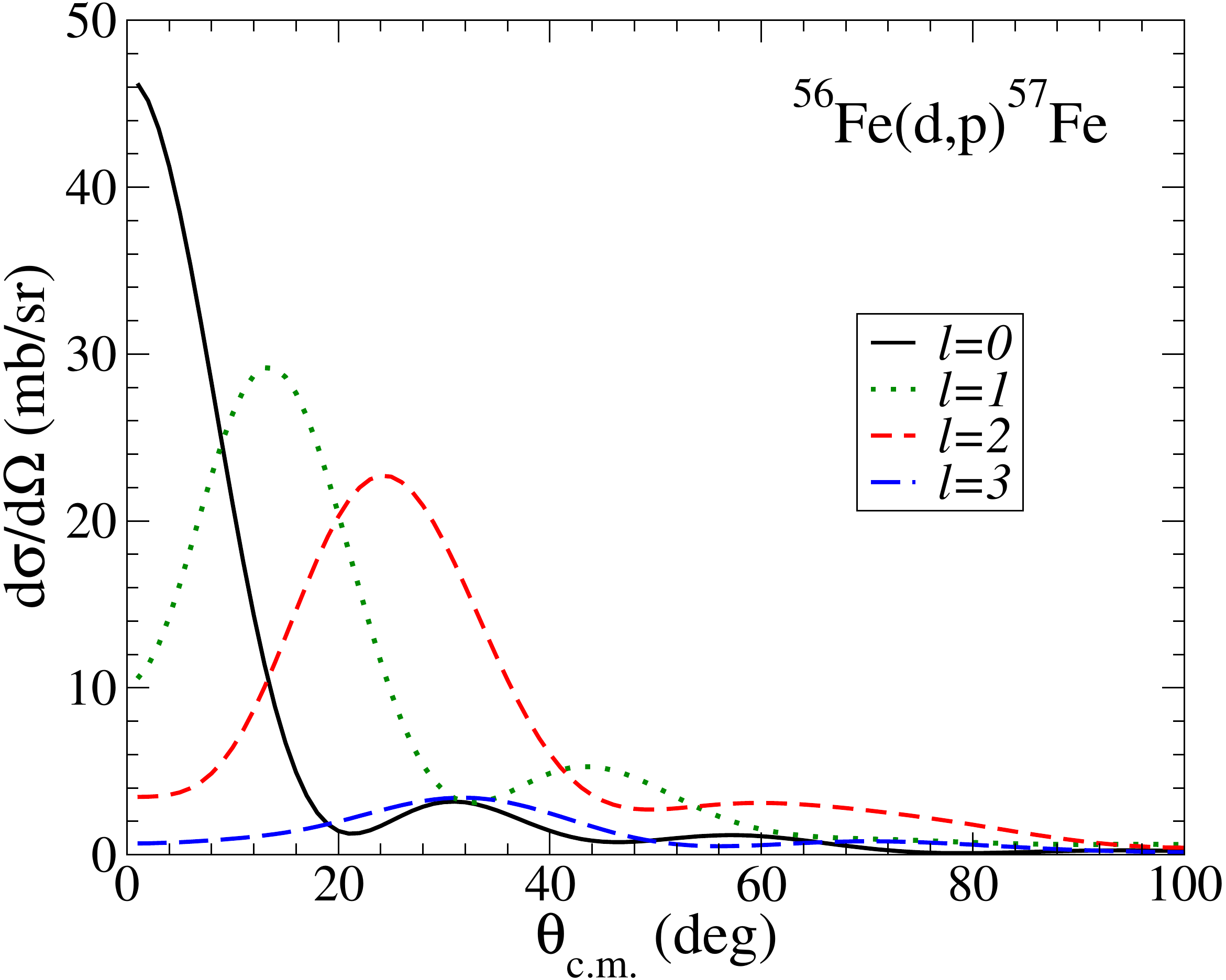} \end{center}
\end{minipage}
\begin{minipage}[t]{.55\linewidth}
\begin{center}\includegraphics[width=0.85\columnwidth]{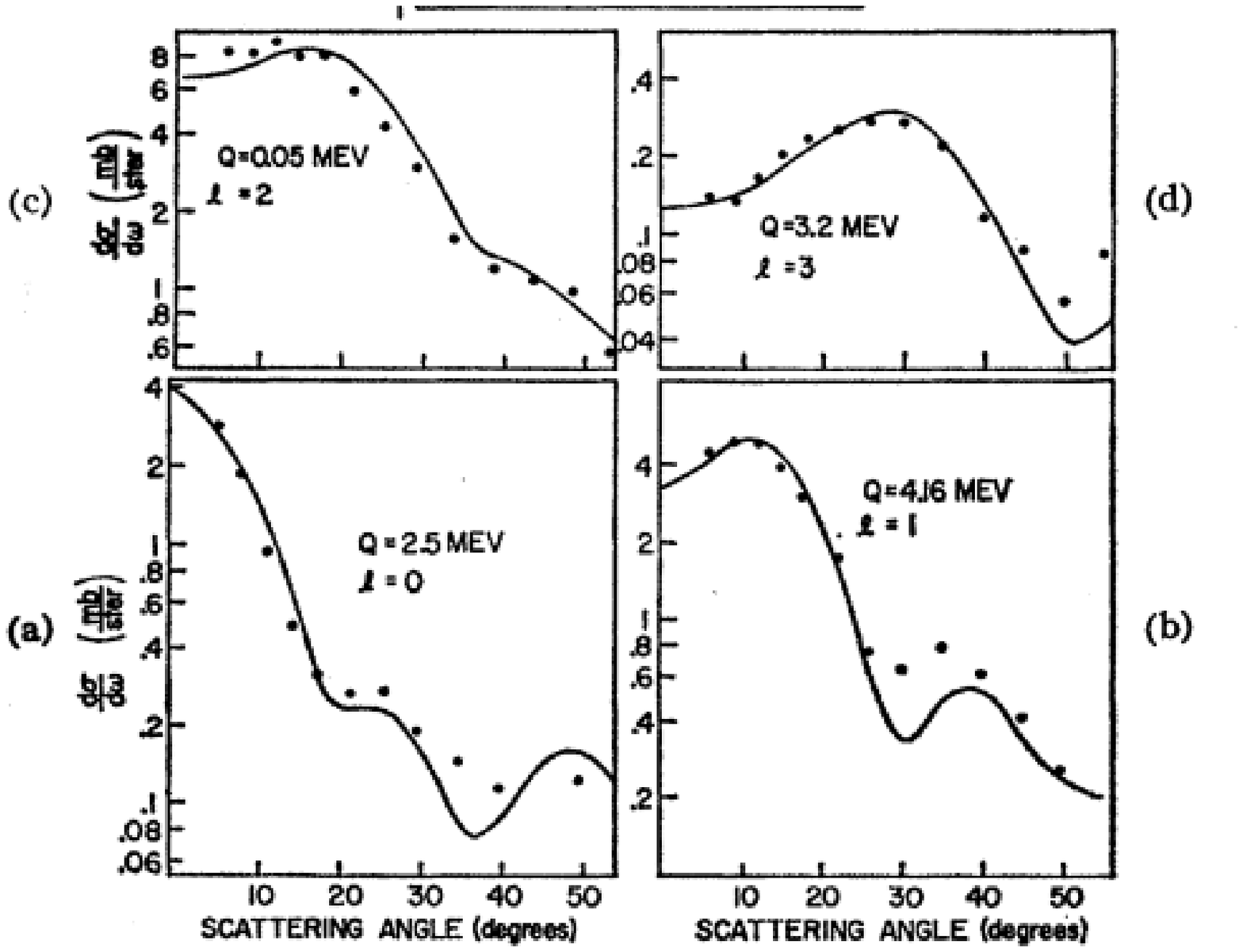} \end{center}
\end{minipage}
\caption{\label{fig:fe56dp_ldep}Left: DWBA calculations for $^{56}$Fe($d$,$p$)$^{57}$Fe reaction at $E_d=15$~MeV  assuming several values of the orbital angular momentum of the transferred neutron in the final nucleus. Right: Comparison with experimental data. From Cohen {\it el al.} \cite{Coh62}.}
\end{figure}

In a more general case, the determination of the angular momentum of the populated states is done with the assistance of the angular distribution of these states. This stems from the fact that the angular dependence of eq.~(\ref{eq:dsdw_dwba}) is strongly dependent on the orbital angular momentum $\ell$ so, upon comparison with experiment, the value of $\ell$ can be determined. An example is shown in fig.~\ref{fig:fe56dp_ldep}, corresponding to the $^{56}$Fe($d$,$p$)$^{57}$Fe reaction at $E_d=15$~MeV. The left panel shows DWBA model calculations for a hypothetical final state in $^{57}$Fe with different values of $\ell$. It is seen that, the larger the value of $\ell$, the larger the value of the angle corresponding to the first peak. In particular, for $\ell=0$ the distribution peaks at $\theta=0$. The right panel shows the comparison of experimental data for this reaction with DWBA calculations  \cite{Coh62}, where the sensitivity of the angular distributions on $\ell$ is clearly seen.

Finally, under the assumption of the validity of the DWBA approximation, the absolute magnitude of the spectroscopic factor for the decomposition $B \rightarrow A + n$ can be determined by comparing the magnitude of the experimental and calculated cross sections. 

 \begin{figure}
\begin{center}
\includegraphics[width=0.75\textwidth]{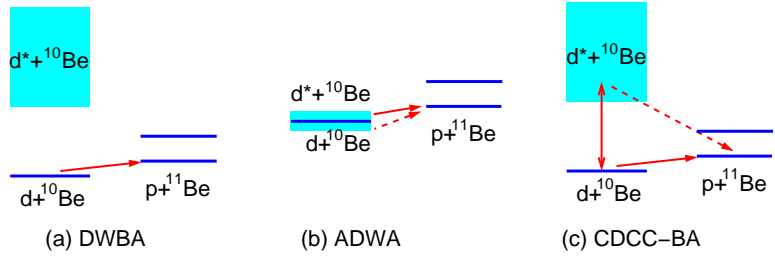}
\end{center}
\caption{\label{be10dp_schemes} Schematic representation of DWBA, ADWA and CDCC-BA approaches for a $(d,p)$ transfer reaction.}
 \end{figure}

\subsection{Influence of breakup channels on transfer: the ADWA method}
In the DWBA method, the three-body wave function appearing in the transition amplitude  
(\ref{eq:T-trans_exact}) is approximated by the elastic component ($\Psi^\mathrm{(+)} \approx \chi^{(+)}_{d}(\bK_d,\bR) \varphi_{d}(\br)$). This approximation is motivated by the assumption that the elastic channel dominates the reaction. However, this choice, albeit intuitively plausible, deserves some caution.
First, the phenomenological DWBA approach relies on the use of optical potentials, usually taken as local, angular momentum-independent potentials, chosen to reproduce
elastic scattering. This only means that the optical potentials will reasonably reproduce the phase shifts, for all partial waves, in the elastic channel.
In other words, the standard DWBA approach reproduces the elastic wave function asymptotically, at large projectile-target distances.  It is not
obvious that this elastic wave function reproduces correctly the elastic component of the wave function in the
radial range relevant for the transfer T-matrix elements. Second, the preponderance of the elastic cross section means that the elastic channel dominates the full wave function asymptotically, but this does not necessarily imply that the elastic channel is also dominant at shorter distances, where the transfer process takes place. Inspection of eq.~(\ref{eq:Tdwba-dp}) shows that the transfer cross section is dominated by small $p$-$n$ separations (for which $V_{pn}$ is  non-negligible). Consequently, an accurate evaluation of (\ref{eq:Tdwba-dp}) requires a good approximation of $\Psi^\mathrm{(+)}(\br,\bR)$  within the range of $V_{pn}$. This includes not only elastic channel component (i.e.\ with $p$-$n$ forming a deuteron) but also excited components, in which the $p$-$n$ system is unbound. Stated otherwise, this requires an approximation of  $\Psi^\mathrm{(+)}$ including the deuteron breakup channels. 

The problem has been addressed at length by R.C.\ Johnson and co-workers, who have provide some practical solutions within the so-called {\it adiabatic approximation}. The simpler of these solutions, proposed by Johnson and Soper \cite{Ron70}, 
was originally formulated for $(d,p)$, or $(d,n)$ reactions, although it can be applied to other weakly bound composite systems.
It is based on the fact that the composite projectile has a relatively low binding energy (2.22 MeV in the case of the deuteron), and so, if 
the collision energy is relatively high, we can expect that, during the collision process, the relative proton-neutron coordinate does not change 
significantly;  it is ``frozen''. 

Note that, 
even if the  $p$-$n$ wave function $\varphi_{d}(\vecr)$ has a relatively long range (which is also the case of halo nuclei),  $V_{pn}(r)$ has 
a much shorter range. Therefore, for the purpose of evaluating the transfer matrix element, one can calculate the adiabatic wave function using the potential
evaluated at $\vec r = 0$. This leads to the Johnson--Soper approximation \cite{Ron70}, in which
\be
\Psi^{(+)}(\vec R, \vec r) \simeq \chi_{JS}^{(+)}(\vecR) \varphi_{d}(\vec r) ,
\ee
where $\chi_{JS}^{(+)}(\vec R)$ is the solution of a two-body scattering problem, on the coordinate $\vec R$, in which the interaction is given by
\be
U_{JS}(R) = U_{pA}(R)+U_{nA}(R)  .
\label{eq:JS}
\ee

We see that, in this limit, the adiabatic theory of the transfer amplitude adopts a form akin to that found in the DWBA theory, but this analogy is misleading because the function $\chi_{JS}^{(+)}(\vecR)$ includes contributions from breakup and the potential  $U_{JS}(R)$ may have little to do with the optical potential describing the deuteron elastic scattering.  Due to the adiabatic approximation, the JT theory is not expected to be accurate at low incident energies. 

The adiabatic approximation is equivalent to neglect the excitation energy of the states of the projectile \cite{Ron70}. The adiabatic wave function takes into account the excitation to breakup channels, assuming that these states are degenerate in energy with the projectile ground state, as illustrated in  fig.~\ref{be10dp_schemes}(b). Therefore,  the ADWA approach 
takes into account, approximately, the effect of deuteron break-up on the transfer cross section, within the adiabatic approximation. So, it should be well suited to describe deuteron
scattering at high energies, around 100 MeV per nucleon. Systematic studies \cite{Har71,Sat71,Wal76} have shown that ADWA is superior to standard DWBA for $(d,p)$ scattering at these relatively high energies.

Although the zero-range adiabatic model of Johnson and Soper provides a systematic improvement over the conventional DWBA, there are situations in which the former fails to reproduce the experimental data \cite{Ste86,Ron89}. Models which go beyond the zero-range and adiabatic approximations are therefore needed. One of such models is the Weinberg expansion method of Johnson and Tandy \cite{JT74}. The idea  is to expand $\Psi^\mathrm{(+)}$ in terms of a set of functions which are complete within the range of $V_{pn}$. A convenient choice is the set of Weinberg states (also called Sturmians), 
\begin{equation}
\label{eq:Psi-W}
 \Psi^{(+)}(\bR,\br)=\sum_{i=0}^{N} \phi^W_{i}(\br)\chi^W_{i}(\bR) \, ,
\end{equation}
where $\phi^W_i(\xi)$ are the Weinberg states, which are solutions of the eigenvalue equation
\be
[ \hat{T}_\vecr  + \alpha_i V_{pn} ] \phi^W_i(\vecr) = - \varepsilon_d \phi^W_i(\vecr) ,
\ee
where $\varepsilon_d=2.225$~MeV is the deuteron binding energy and where $\alpha_i$ are the eigenvalues, to be determined along with the eigenfunctions. Beyond the range of the potential all the Weinberg states decay exponentially, like the deuteron ground-state wave function. For  $i=0$, $\alpha_0=1$ an-d so $\phi^W_0(\vecr)$ is just proportional to the deuteron ground state. As $i$ and $\alpha_i$ increase, they oscillate more and more rapidly at short distances. The Weinberg states form a complete set of functions of $\vecr$ inside the range of the potential $V_{pn}$. They are well suited to expand $ \Psi^{(+)}$ in this region, as it is required by the amplitude (\ref{eq:Tdwba-dp}). They do not satisfy the usual orthonormality relation but the less conventional one
\be
\langle \phi^W_i | V_{pn} | \phi^W_j \rangle = - \delta_{ij} .
\ee
If we retain in  (\ref{eq:Psi-W})  only the leading term, $\Psi^{(+)}(\bR,\br) \approx \phi^W_0 \chi^W_{0}(\bR)$, one finds \cite{JT74} that $\chi^W_{0}$ verifies the single-channel equation
\be
[\hat{T}_\vecR  + U_{JT}(\vecR) - E_d ]  \chi^W_{0}(\vecR) =0 ,
\ee
with $E_d= E - \varepsilon_d$ and where the potential $U_{JT}$ is given by:
\be
\label{eq:UJT}
U_{JT}(R)=  \frac{\langle \varphi_{pn}(\br) | V_{pn} (U_{nA}+U_{pA}) | \varphi_{pn}(\br) \rangle}{\langle \phi_{pn}(\br) | V_{pn} | \phi_{pn}(\br) \rangle} .
\ee
The bra and ket in this equation mean integration over $\vecr$, with fixed $\vecR$. Interestingly, in the zero-range limit, $U^{JT}(R)$ reduces to the JS potential, eq.~(\ref{eq:JS}). Therefore, the zero-order result given by eq.~(\ref{eq:UJT}) can be regarded as a finite-range version of the adiabatic (JS) potential. These two models are globally referred to as Adiabatic Distorted Wave Approximation (ADWA). However, it is worth noting that the full Weinberg expansion makes no reference to the incident energy and, as such, does not involve the adiabatic approximation. This suggests that a stripping theory based on this Weinberg expansion can be used at low energies, where the adiabatic condition is not well satisfied.  The inclusion of higher order terms ($i>0$) in the Weinberg expansion has been investigated in refs.~\cite{Lai93,Pan13}. 

An appealing feature of the ADWA  is that its  ingredients are completely determined by experiments. These ingredients are the proton-target and neutron-target optical potentials, evaluated at half of the deuteron incident energy,  as well as the  well known proton-neutron interaction. 
 On the negative side, the ADWA approach does not consistently describe elastic scattering and nucleon transfer. Although, physically, 
elastic scattering, transfer and break-up should be closely related, so that the increase of flux in one channel should reduce the flux in the others,
this connection is not present in ADWA. Furthermore, the arguments leading to ADWA are strongly associated with the assumption that the transfer is governed by
a short range operator. So, it is not obvious that the approximations remain valid for other weakly bound systems, like \nuc{11}{Be}. Even in the case
of $(d,p)$ scattering, the transfer matrix element is determined not only by the $n$-$p$ interaction, but also by the proton-target and neutron-target 
interactions, that define the {\it remnant} term. The role of these terms, that would have contributions of three-body configurations in which proton and neutron are not so close together, is not clear {\it a priori}.  An alternative method, that avoids the presence of these remnant terms, has been proposed by Timofeyuk and Johnson \cite{Tim99}.

\begin{figure}
\begin{minipage}[c]{.25\textwidth}
\includegraphics[height=2.5cm]{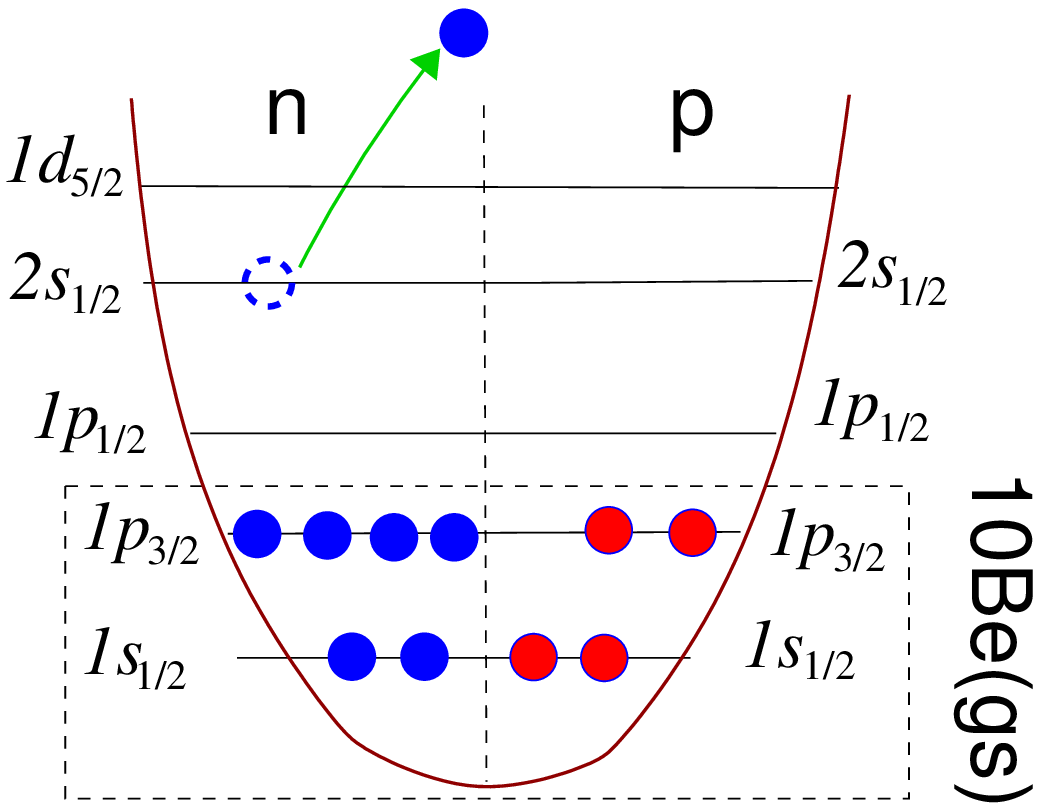} \\
\includegraphics[height=2.5cm]{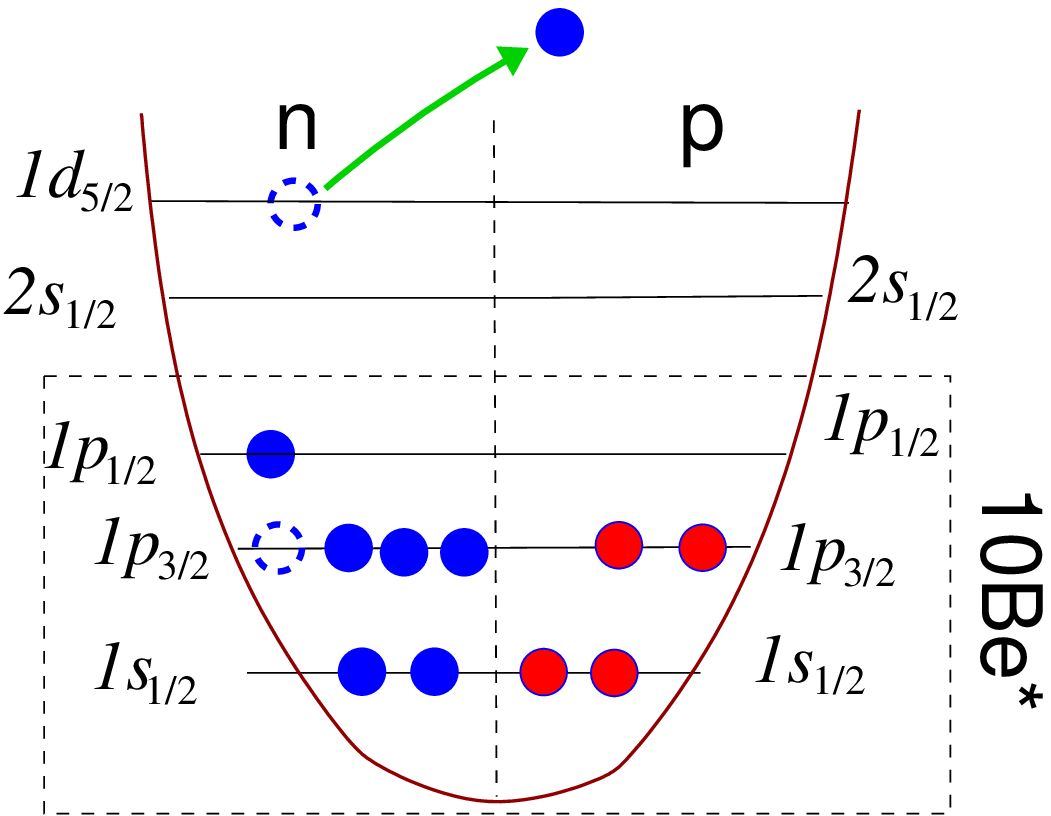} 
\end{minipage}
\begin{minipage}[c]{.65\textwidth}
\begin{center}
\includegraphics[height=5.5cm]{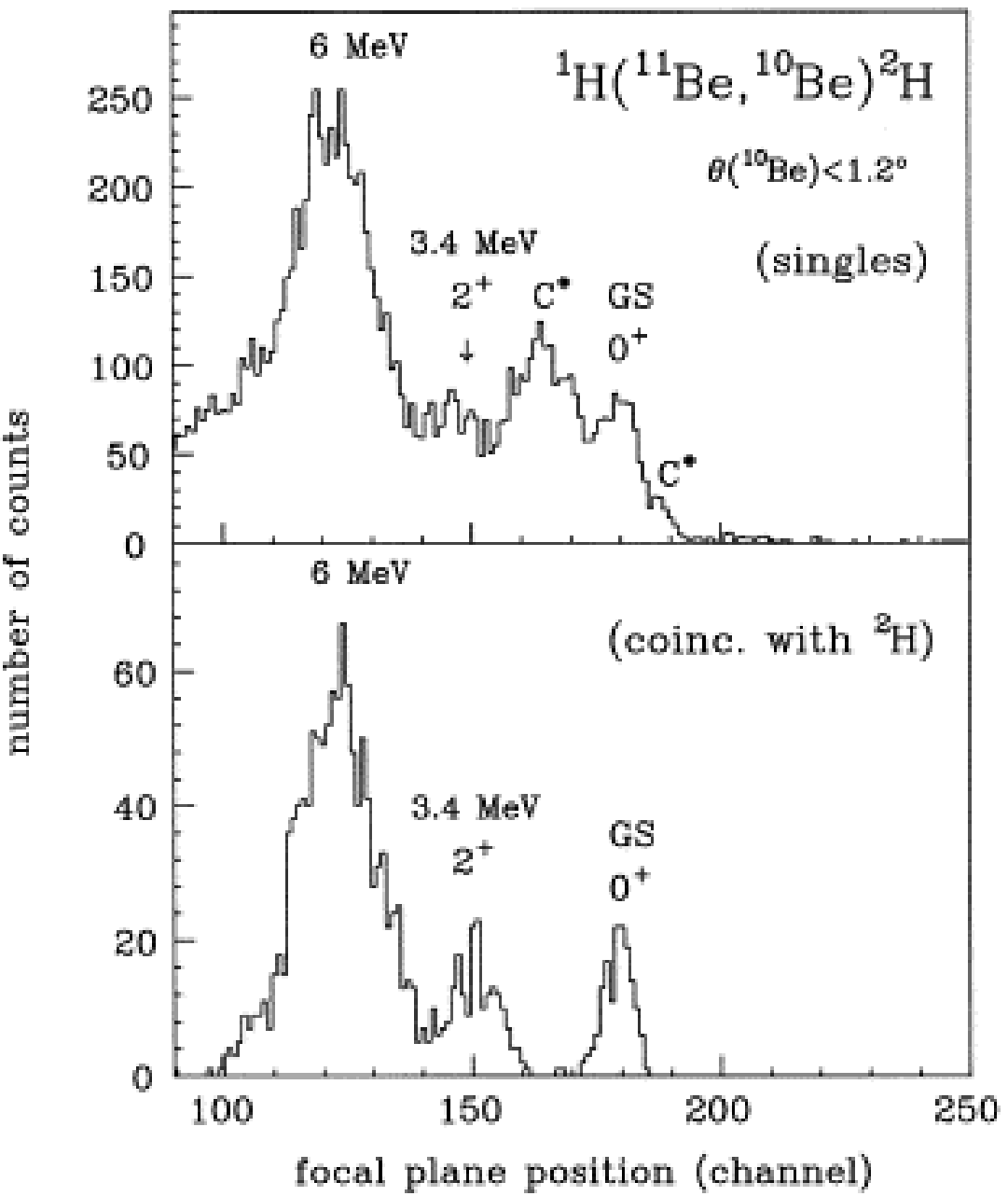} 
\includegraphics[height=5.5cm]{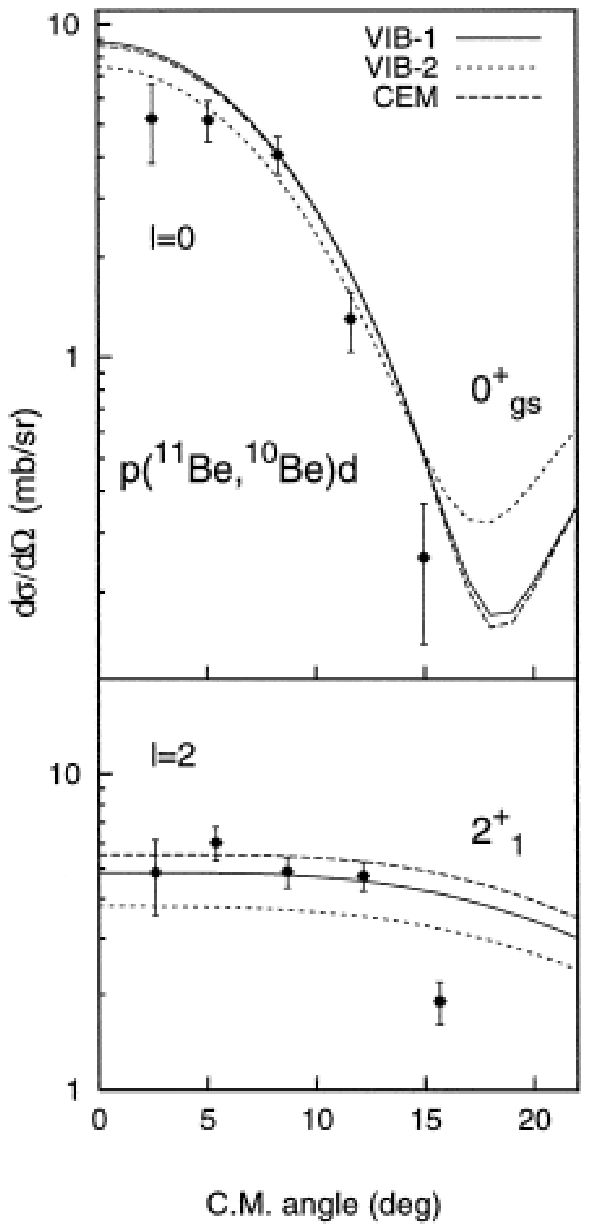} 
\end{center}
\end{minipage}
\caption{\label{fig:be11pd}Application of the ADWA method to the $^{11}$Be($p$,$d$)$^{10}$Be reaction. The left panels show a shell-model picture of the $^{11}$Be nucleus. The middle panels show the data from  Fortier {\it et al.} \cite{For99}, displaying the excitation energy function of $^{10}$Be events, where one can distinguish the peaks corresponding to the ground state and  the first excited state ($2^+$, $E_x=3.367$~MeV). The rightmost panels show the angular distributions corresponding to these two peaks compared  with ADWA calculations, assuming that they correspond to the neutron removal from the $2s_{1/2}$ and $1d_{5/2}$ single-particle configurations in $^{11}$Be, respectively.}
\end{figure}

As an example of the application of the ADWA model, we consider the reaction  $^1$H($^{11}$Be,$^{10}$Be)$^2$H measured at GANIL \cite{For99,Win01}. Considering the nucleus \nuc{11}{Be} as a neutron outside a \nuc{10}{Be} core we may write for the $^{11}$Be ground state 
\be
|^{11}{\rm Be} \rangle_\text{g.s.}  =  \alpha \mid \shalfzero \rangle+ 
            \beta  \mid \dhalftwo \rangle + \ldots
\ee
where only the two dominant configurations ($2s_{1/2}$ and $1d_{5/2}$) are indicated explicitly. By comparing with (\ref{eq:a_exp}), the $\alpha$ and $\beta$ coefficients are the spectroscopic amplitudes for these two configurations, respectively. The left cartoons in fig.~\ref{fig:be11pd} illustrate these two configurations of  $^{11}$Be, using a simple independent-particle-model picture. The $2s_{1/2}$ configuration is associated with the $^{10}$Be(g.s.). Thus, removing a neutron from the $2s_{1/2}$ orbital will produce $^{10}$Be in its ground state.  Conversely, if the neutron is removed from the $1d_{5/2}$ orbital, the $^{10}$Be will be produced in the first excited state ($E_x =3.4$~MeV). Note that these arguments assume that the reaction occurs in one-step, which is a reasonable approximation at this energy. The middle panel in fig.~\ref{fig:be11pd} shows the experimental yields of these $^{10}$Be states. In addition to the two  peaks corresponding to the  $^{10}$Be g.s.\ and first excited state, an even more prominent peak is observed for excitation energies of $\approx$6~MeV, which is  due to the stripping of neutrons from the $1p_{3/2}$ shell.  The rightmost panels in fig.~\ref{fig:be11pd}  show the comparison of the angular distributions of  the two first states of $^{10}$Be with ADWA calculations. This comparison allowed for a confirmation of the $\ell$ assignment, as well as a determination of the corresponding spectroscopic factors $\alpha^2$ and $\beta^2$ \cite{For99,Win01}.

\subsection{Continuum Discretized Coupled Channels Born Approximation CDCC-BA \label{sec:cdcc-ba}}
An alternative choice for the three-body wave function $\Psi^{(+)}$ to be used in the transfer amplitude (\ref{eq:T-trans_exact}) is the CDCC expansion, eq.~(\ref{PsiCDCC}
). When inserted into the transition amplitude we get, for the $(d,p)$ case, 
\begin{align}
{\cal T}^\mathrm{CDCC-BA}_{d,p}  & =
C^B_{An}
\langle \chi_{p}^{(-)}(\bK_p, \bR') 
 \tilde{\varphi}^{\ell j I}_{An}(\brp) \, V_{pn}(\br) \, \chi^{(+)}_{0}(\bK_d,\bR) \varphi_{d}(\br) \rangle 
\nonumber \\ 
& +
 C^B_{An} \sum_{i=1}^{N}
\langle \chi_{p}^{(-)}(\bK_p, \bR') 
 \tilde{\varphi}^{\ell j I}_{An}(\brp) \, V_{pn}(\br) \, \chi^{(+)}_{i}(\bK_d,\bR) \varphi^{i}_{pn}(\br) \rangle .
\end{align}
The first term in this expression corresponds to the {\it direct transfer}, in which the neutron is transferred  directly from the deuteron ground state, whereas the second term accounts for the {\it multi-step} transfer occurring via the breakup states of the $p$-$n$ system. These two types of processes correspond, respectively, to the solid and dashed lines in fig.~\ref{be10dp_schemes}(c) for the \nuc{10}{Be}($d$,$p$)\nuc{11}{Be} case. Clearly, the multi-step process going through the breakup channels are omitted in the DWBA calculation. At most, the DWBA considers the effect of these channels on the elastic scattering if a suitable choice of the entrance optical potential is made. The adiabatic approximation includes in principle both mechanisms, but under the assumption that the excited (breakup) channels of the projectile are degenerate with the ground state [fig.~\ref{be10dp_schemes}(b)]. The advantage of the CDCC-BA approach is that all relevant bound and continuum states in the $b+x$ system are explicitly included in the calculation. 

Some early  comparisons between these three methods can be found in refs.~\cite{Raw75,Ise83,Ama83,Kaw86a} and the main results are also summarized in refs.~\cite{Aus87,Gom14}. Due to numerical limitations, these first studies where done using a zero-range approximation of the $V_{bx}$ potential. Overall, it was found that the ADWA model describes well the direct transfer contribution. However, the multi-step contributions, which are completely absent in DWBA, are described very inaccurately by the adiabatic approximation. At low energies ($E_d < 20$~MeV) the discrepancy between the zero-range ADWA and CDCC-BA calculations can be understood because at these energies the adiabatic approximation itself is questionable. However, even at medium energies  ($E_d \approx 80$~MeV) there are situations in which transfer through breakup channels is found to be very significant, and therefore the ADWA method does not work well either.  In these situations, the CDCC-BA or the Weinberg expansion should be better used instead. The disadvantage of the CDCC-BA calculations is that, in principle, a large basis of internal states has to be included, making this approach much more demanding numerically.

Finite-range effects have been studied within the adiabatic approximation in refs.~\cite{Lai93,Ngu10} and were found to be small ( $<10\%$) at energies below 20-30 MeV/u but become more and more important as the incident energy increases. This limitation should be also taken into account in the analysis of experimental data. 

Along with deuteron breakup,  target excitations may affect the $(d,p)$ and $(p,d)$ transfer cross sections. The problem has been recently tackled in several ways. Deltuva {\it et al.} \cite{Del16b} used  a formulation of the momentum space Faddeev equations including {\it core} excitations, and studied the effect of $^{10}$Be excitations in  the $^{10}$Be($d$,$p$)$^{11}$Be reaction.  It is found that the cross sections are no longer proportional to the spectroscopic factor and the departure from this proportionality increases with increasing incident energies, reaching a maximum at a deuteron energy of $E_d\approx$60 MeV.  Similar results and conclusions were achieved in \cite{Gom17b} using two alternative methods. One uses an extended ADWA model, with a deformed adiabatic potential. The other uses the extended version of the CDCC method including target excitations discussed in sec.~\ref{sec:corex}, which is then used  in the transfer transition amplitude (\ref{eq:Tdist}). These works conclude that the main deviation from the pure three-body calculation with inert bodies  originates from the destructive interference of the direct one-step transfer and the two-step transfer following target excitation.

\subsection{Transfer reactions populating unbound states}
So far, we have considered transfer reactions as a tool to investigate bound states of a given nucleus. However, in a rearrangement process, the transferred particle can populate also unbound states of the final nucleus. This opens the possibility of studying and characterizing structures in the continuum, such as resonances or virtual states. 

As in the case of transfer to bound states, the simplest formalism to analyze these processes is the DWBA method. In this case, the bound wave function $\varphi^{\ell j I}_{An}(\brp)$ appearing in the final state in eq.~(\ref{eq:dsdw_dwba}) should be replaced by a positive-energy wave function describing the state of the transferred particle $n$ with respect to the  core $A$.  In principle, for this purpose one could  use a scattering state of the $n+A$ system at the appropriate relative energy. However, this procedure tends to give numerical difficulties in the evaluation of the transfer amplitude due to the oscillatory behaviour of both the final distorted wave and the wave function $\varphi^{\ell j I}_{An}(\brp)$.  To overcome this problem, several alternative methods have been attempted. We enumerate here some of them:

\begin{enumerate}[(i)]
\item The bound state approximation \cite{Cok74}. In the case of transfer to a resonant state, this method replaces the scattering state $\varphi^{\ell j I}_{An}(\brp)$ by a weakly bound wave function with the same quantum numbers $\ell$ and $j$. In practice, this can be achieved by starting with the potential that generates a resonance at the desired energy and increasing progressively the depth of the central potential until the state becomes bound. 

\item Huby and Mines \cite{Hub65} use a scattering state for $\varphi^{\ell j I}_{An}(\brp)$, but it is multiplied by  a convergence factor $e^{-\alpha r'}$ (with $\alpha$ a positive real number), which artificially eliminates its contribution to the integral coming from large $r'$ values, and then extrapolate numerically to the limit $\alpha \rightarrow 0$. The convergence factor can be physically justified taking into account that the incident and outgoing fragments are not characterized by well-defined linear momenta, but correspond instead to wave-packets. Thus, physical results depend on energy averages of stationary-state scattering amplitudes which, in the case of three-body breakup, destroy the asymptotic oscillations of the integrand in the same manner as the convergence factor \cite{Aus70}.   

\item Vincent and Fortune \cite{VF70}  questioned the bound state approximation arguing that, in general, the bound state and resonant form factors can be very different and, even in those cases in which the fictitious form factor gives the correct shape, they can lead to very different absolute cross sections. They suggest using the actual scattering state, but choosing an integration contour along the complex plane  in such a way that the oscillatory integrand is transformed into an exponential decay, thus improving the convergence and numerical stability of the calculation. 

\item In a real transfer experiment leading to positive-energy states, one does not have access to a definite final energy, but to a certain region of the continuum. That is to say, the extracted observables, such as energy differential cross sections, are integrated over some energy range which, at least, is of the order of the energy resolution of the experiment. This suggests a method of dealing with the unbound states consisting in 
discretizing the continuum states in energy bins, as in the CDCC approximation. 
\end{enumerate}  

\begin{figure}
\begin{center}\includegraphics[width=0.55\textwidth]{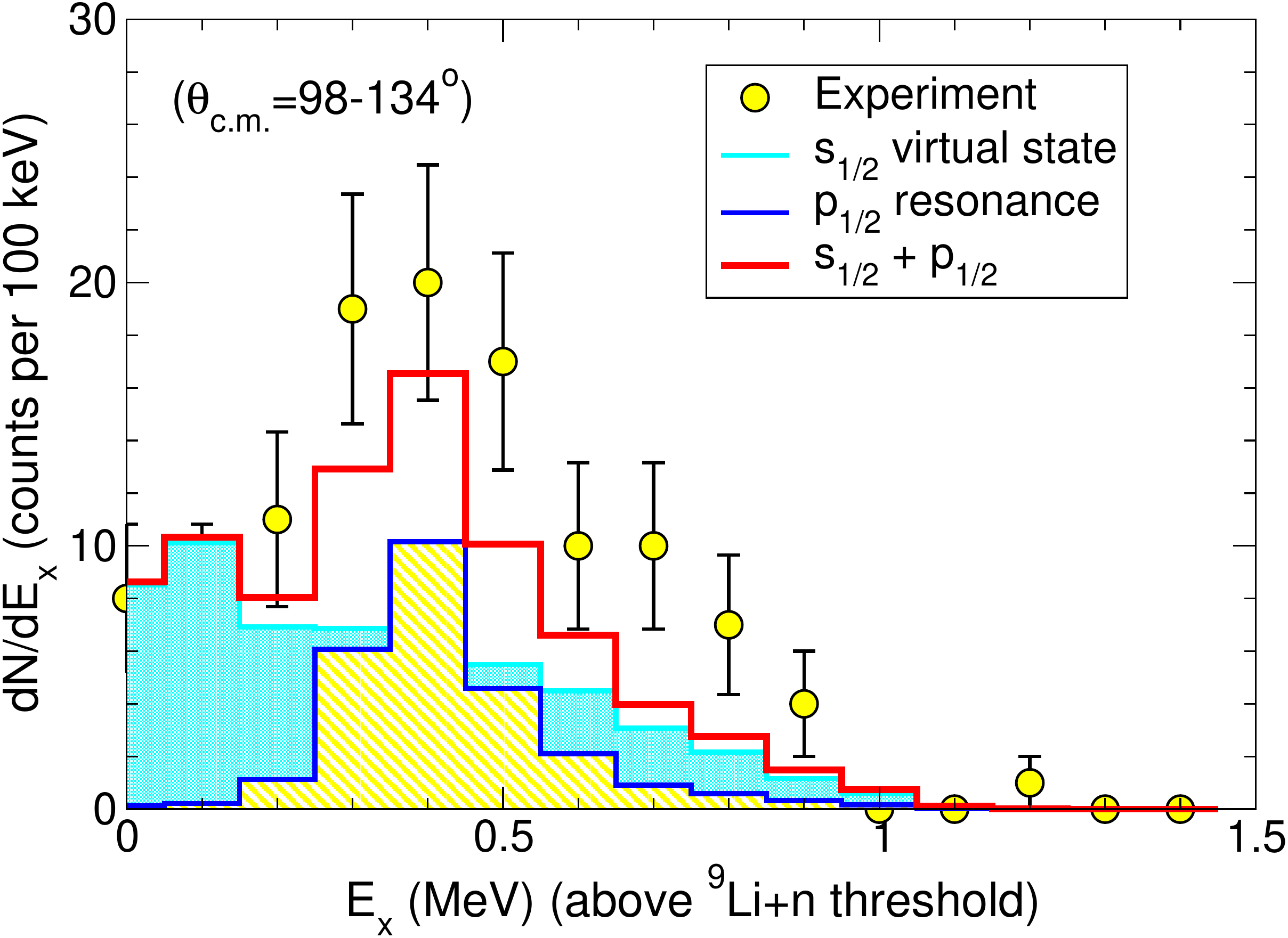}\end{center}
\caption{\label{li9dp}Experimental yield for the population of $^{10}$Li states for the reaction \nuc{2}{H}(\nuc{9}{Li},$p$)\nuc{10}{Li}$^{*}$ as a function of the  $n$+$^{9}$Li relative energy. The calculations consider the transfer of a neutron from the deuteron to a set of final states of the $^{10}$Li system, using a binning discretization method. Adapted from ref.~\cite{Jep06}.}
\end{figure}

An advantage of the  method (iv) is that it can be equally applied to both resonant and non-resonant continuum final states. An example is shown in fig.~\ref{li9dp}, which corresponds to the  differential cross section, as a function of the $n$-$^{9}$Li relative energy, for the reaction \nuc{2}{H}(\nuc{9}{Li},$p$)\nuc{10}{Li}$^{*}$ at 2.36~MeV/u measured at REX-ISOLDE \cite{Jep06}. The lines are the results of CDCC-BA calculations, including the transfer to  $^{10}$Li$^{*}$ continuum states, showing the separate contribution of the s-wave ($\ell=0$) continuum and $p$-wave ($\ell$=1) continuum. The strength of the measured cross section close to zero energy is due to the presence of a virtual state in the  \nuc{10}{Li}$^{*}$ system, whereas the peak around 0.4~MeV is due to a $p_{1/2}$ resonance. This is an example of how the use of transfer reactions can provide information of the continuum structure of weakly-bound or even unbound systems. 

\section{Final remarks}
We have reviewed the current status of the theoretical description of nuclear reactions, with emphasis in the case of reactions involving by weakly-bound nuclei. Although  formal  quantum mechanical scattering theory provides exact solutions to compute the scattering amplitudes and their associated cross sections for a general scattering problem, we have seen that important approximations must be applied in practice. These approximations try to reduce the extremely complicated  many-body scattering problem into a simplified, numerically tractable one. These approximations are usually tailored to specific types reactions, giving rise to a large variety of models. Although we have reexamined the most common ones, the models presented here  do not exhaust the variety of approaches available in the literature. Some of the omitted topics are the following (again, the list is not exhaustive):
 
\bi
\item {\it Two-particle transfer reactions}. In the case of transfer reactions, we have restricted ourselves to the case of one-particle transfers, and have omitted altogether the important case of two-particle transfer. These are perfect tools to study the so-called pairing rotations and vibrations, collective modes associated with a field (the pairing field) which changes the number of particles by two. The formal description of two-particle transfer reactions dates back to the seminal works of Glendenning \cite{Glen65} and Baynman \cite{Bay68} and has received renewed attention in recent years \cite{Pot13a,Pot13b}.

\item {\it Charge-exchange reactions.} The nucleon-nucleon interaction contains spin- and isospin-dependent terms which are able to induce spin and isospin changing transitions such as those triggered by the weak interaction in $\beta$-decay. For example, the presence of $\sigma \tau$ terms in the strong-interaction gives rises to spin-flip, isospin-flip processes analogous to the Gamow-Teller (GT) transitions.   In fact,
charge-exchange (CE) reactions, such as the $(p, n)$ or ($^3$He, $t$) reactions at intermediate beam energies, can selectively excite GT states up to high excitation energies in
the final nucleus. It has been found empirically that there is a close proportionality between
the cross-sections at $0^\circ$ and the transition strengths $B(GT)$ in these CE reactions. Therefore,
CE reactions are useful tools to study the relative values of $B(GT)$ strengths up to high excitation energies (see \cite{Fuj11} for a comprehensive review). Most recently, double charge-exchange reactions have been put forward as a promising tool to constrain the nuclear matrix elements involved in neutrino-less double-beta decay with potential access on the nature of the neutrino (Majorana versus Dirac) \cite{Cap15}.

\item {\it Ab-initio methods}. Our discussion has been mostly confined to macroscopic and few-body models. Much work is being done in recent years in the field of the so-called {\it ab-initio} approaches, whose final goal is to describe  both the structure    
and dynamics of nuclei starting from the nucleon-nucleon interaction. One of the most promising tools is the {\it no-core shell model} which enables to study bound and scattering states of many-body systems  (see \cite{Nav16} for a recent review). Due to the complexity of these calculations, applications have been restricted to light systems. Promising results have been obtained for a number of reactions, such as \nuc{4}{He}($n$, $n$)\nuc{4}{He} \cite{Nav16},  \nuc{3}{H}($d$,$n$)\nuc{4}{He}, \nuc{3}{He}($d$,$p$)\nuc{4}{He} \cite{Nav12}, \nuc{7}{Be}($p$,$\gamma$)\nuc{8}{B}, $^{10}$C(p,p)$^{10}$C, among others.

\ei

The curious reader is encouraged to go through the recommended references to discover by him/herself these and others recent developments and advances in the field of nuclear reaction modeling.

\acknowledgments
This paper is based on the lectures given during the  Course ``Nuclear Physics with Stable and Radioactive Ion Beams'' within the International  School of Physics Enrico Fermi, held at Varenna, Italy, Summer 2017. The author is deeply grateful to the  organizers for giving me the opportunity to enjoy such a stimulating environment with the other lecturers and attendants. This work is partially supported by the Spanish
 Ministerio de  Econom\'ia y Competitividad and FEDER funds under project 
 FIS2014-53448-C2-1-P  and by the European Union's Horizon 2020 research and innovation program under grant agreement No.\ 654002.

\thispagestyle{empty}

\bibliographystyle{varenna}
\bibliography{refer}

\end{document}